\documentclass[twoside,12pt]{article}
\usepackage{epsfig}

\usepackage{overpic}
\usepackage{axodraw}

\def\Journal#1#2#3#4{{#1} {#2} (#4) #3 }

\def\NPA{{\em Nucl. Phys.} A}

\def\PLB{{\em Phys. Lett.} B}

\def\PRL{\em Phys. Rev. Lett.}

\def\PREP{\em Phys. Rep.}

\def\PRD{{\em Phys. Rev.} D}
\def\PRC{{\em Phys. Rev.} C}

\newcommand{\be}{\begin{equation}}
\newcommand{\ee}{\end{equation}}
\newcommand{\ba}{\begin{eqnarray}}
\newcommand{\ea}{\end{eqnarray}}
\newcommand{\nn}{\nonumber}
\newcommand{\dsp}{\displaystyle}

\begin{document}
\vskip1cm
\begin{titlepage}
\begin{flushright}
LU TP 06-16\\
hep-ph/0604043\\
revised july 2006
\end{flushright}

\begin{center}
\vfill
{\Large \bf Chiral Perturbation Theory Beyond One Loop}\\[2cm]

{\bf Johan Bijnens}\\[1cm]

{Department of Theoretical Physics, Lund University\\
S\"olvegatan 14A, SE 22362 Lund, Sweden}

\vfill
\begin{abstract} 
The existing Chiral Perturbation Theory (ChPT) calculations
at order $p^6$ are reviewed. The principles of ChPT and how they are
used are introduced. The main part is a review of
the two- and three-flavour full two-loop calculations
and their comparison with experiment. We restrict the discussion to the
mesonic purely strong and semileptonic sector. The review concludes
by mentioning the existing results in finite volume, finite temperature
and partially quenched ChPT. 
\end{abstract}
\vfill
\end{center}
\end{titlepage}
\title{ \vspace{1cm} Chiral Perturbation Theory Beyond One Loop}
\author{Johan Bijnens\\
\\
Department of Theoretical Physics, Lund University\\
S\"olvegatan 14A, SE 22362 Lund, Sweden}
\maketitle
\begin{abstract} 
The existing Chiral Perturbation Theory (ChPT) calculations
at order $p^6$ are reviewed. The principles of ChPT and how they are
used are introduced. The main part is a review of
the two- and three-flavour full two-loop calculations
and their comparison with experiment. We restrict the discussion to the
mesonic purely strong and semileptonic sector. The review concludes
by mentioning the existing results in finite volume, finite temperature
and partially quenched ChPT. 
\end{abstract}

\section{Introduction}

Chiral Perturbation Theory (ChPT) is the low-energy effective field theory
of Quantum Chromo Dynamics (QCD)
where the degrees of freedom taken into account are the Goldstone
bosons from the spontaneous breakdown of the chiral symmetry and their
interactions. The subject grew out of the current algebra approach of the
1960ies. It was brought into its modern form by Weinberg, Gasser and Leutwyler
\cite{Weinberg,GL1,GL2}. Especially the work of Gasser and Leutwyler was
instrumental in the renaissance of effective field theory methods in low-energy
hadronic physics.
Introductions to ChPT
can be found in the lectures of
Refs.~\cite{Pich,Scherer1,Scherer2} as well as in
Sect.~\ref{sec:chpt}. Shorter introductions can be found in
Refs.~\cite{Eckerlecture,Gasserlecture}.
The lectures by Leutwyler~\cite{Leutwylerlecture}
stress the foundational aspects of ChPT and are a very recommended read.
Introductions to QCD with an emphasis on the low-energy aspects
and ChPT can also be found in the books by Donoghue, Golowich
and Holstein~\cite{DGHbook}, Georgi~\cite{Georgibook} and
Smilga~\cite{Smilgabook}.

Chiral Perturbation Theory (ChPT) is now a very large subject, no single
review can do the entire field justice.
Two main earlier reviews when the main part of ChPT was completed to
next-to-leading-order (NLO) or order $p^4$ are those by
Ecker \cite{eckerreview} and Mei\ss{}ner \cite{meissnerreview}.
This review concentrates on results at next-to-next-to-leading-order (NNLO)
or order $p^6$ in mesonic ChPT, restricted to purely strong or
electromagnetic processes. The wide field of applications beyond this,
weak mesonic decays \cite{eckerreview,Pich}, 
electromagnetic corrections to hadronic processes \cite{electro1,electro2},
non-relativistic approaches to hadronic atoms \cite{hadronic}
and the entire area of baryon and nuclear physics~\cite{nuclear}
are not treated.
The articles cited present an entrance to those fields.
This review includes the comparison with and prediction of experimental data
in connection with the existing order $p^6$ calculations.

In Sect.~\ref{sec:chpt}
ChPT itself is discussed. I start there by discussing the global symmetries
of QCD in Sect.~\ref{sec:chiral}. The consequences of a global symmetry
are embodied in the Ward identities involving
the Green functions of the theory. A very elegant method to automatically
produce Green functions obeying the Ward identities is the external field
method discussed in Sect.~\ref{externalfield}.
In QCD, the chiral symmetry is present in the Lagrangian but it is not visible
in the spectrum. The way to combine those two observations is by looking
at the concept of spontaneous symmetry breaking introduced here in
Sect.~\ref{spontaneous}. In fact, the chiral symmetry is not quite exact in
QCD. It is only valid when all quark masses are zero. But for the light quarks,
this can be treated as a perturbation. The way of dealing with this in general
is discussed in Sect.~\ref{spontaneousexplicit} and the spontaneous symmetry
breaking in QCD is briefly discussed in Sect.~\ref{spontaneousQCD}.
After this I proceed with ChPT at lowest order, built up from the Goldstone
bosons from spontaneous chiral symmetry breakdown in QCD.
The lowest order is introduced in Sect.~\ref{LOChPT} and its main applications
at that order are shown in Sect.~\ref{LOapplications}.
Nonrenormalizable effective field theories, as ChPT, have in principle
an infinite number of parameters. In order for them to be phenomenologically
useful, there has to be a way to order the various parts in order of
importance. This concept is called powercounting and discussed in
Sect.~\ref{powercounting}. Its interplay with renormalization is also
introduced there. Once the concept of powercounting is established, one has to
find out how to construct the explicit Lagrangians needed at each order
in the powercounting and find a way to do the renormalization in practice.
The main ideas behind both those subjects are treated in
Sects.~\ref{Lagconstruction} and \ref{renormalization} respectively.

After this brief introduction to ChPT and the principles behind it,
the main part of this review follows. It is split into three parts.
First a review of the existing calculations in infinite volume for the case of
two flavours. Here only the up and down quark are treated as light and
the relevant degrees of freedom are the pions only. The next part treats
the three-flavour case. Here up, down and strange quarks are all treated as
light and the full lowest mass pseudoscalar octet of pions, kaons and eta
are included as the relevant degrees of freedom. This part includes an overview
of all order $p^6$ calculations at infinite volume relevant in this domain.
The third part consists of those order $p^6$ calculations which do not belong
in either of the two previous categories. It includes finite volume
and finite temperature
calculations. I also briefly discuss what is known
for the partially quenched regime there.

For the two-flavour case I first discuss the early estimates of NNLO effects
using dispersive methods in Sect.~\ref{dispersive}.
The remaining subsections go through the various processes known to NNLO
order. $\gamma\gamma\to\pi^0\pi^0$ was the first case of a full NNLO
calculation and is discussed in Sect.~\ref{ggpipi}. The pion mass and
decay constant follow in Sect.~\ref{mpifpi}. $\gamma\gamma\to\pi^+\pi^-$
and the pion polarizabilities are an area of active experimental work
and at present there seems to be a discrepancy between data and ChPT.
The relevant calculations are reviewed in Sect.~\ref{ggpppm}.
A major recent success of theory of low-energy hadronic physics is the
extremely accurate description of low-energy pion-pion scattering. The role
of ChPT and the relevant calculations are discussed in Sect.~\ref{pipi}.
We have also included the full NNLO ChPT formula here since it can be
beautifully expressed in terms of elementary functions.
The two remaining existing calculations are those
of the pion vector and scalar form-factors, reviewed in Sect.~\ref{piform}
and the pion radiative beta decay, $\pi\to\ell\nu\gamma$. The latter 
is discussed
in Sect.~\ref{pilnugamma}. The present best values of the low-energy constants
(LECs) at NNLO are given in Sect.~\ref{valueslbari}.

When one considers three light quark flavours, the number of processes
increases rather dramatically. Due to the presence of many different scales,
the calculational difficulty also increases. Nonetheless, a great many
processes are known to NNLO also in this sector.
This starts with the vector two-point Green functions,
Sect.~\ref{vectortwopoint}, which were used as a laboratory for doing
three-flavour NNLO calculations and some small phenomenological applications.
A very similar calculation is required for the scalar two-point functions
discussed in Sect.~\ref{scalartwopoint}. The simplest two-loop calculation
is the quark-antiquark condensate.
It is treated in Sect.~\ref{quarkcondensate}.
The first calculations in the three-flavour sector requiring proper or
irreducible two-loop integrals are the axial-vector two-point functions, the
pseudoscalar meson masses and the decay constants. These are reviewed in
Sect.~\ref{massdecay}. The next calculation is in fact one of the more
elaborate ones, for the decays $K\to\pi\pi\ell\nu$. The reason this was done
was that this process is one of the main sources of the order $p^4$ or NLO
LECs. Results are reviewed in Sect.~\ref{Kl4}. With those results in hand,
the basic set of parameters of ChPT could be determined to NNLO.
This lead to a full flurry of applications.
All electromagnetic, Sect.~\ref{vectorform} and scalar form-factors,
Sect.~\ref{scalarform} of the pseudoscalar mesons  have been worked out and
compared with existing data. The process $K\to\pi\ell\nu$ contains several
form-factors and is needed for the determination of the CKM-matrix element
$V_{us}$. The ChPT aspects of this are reviewed in Sect.~\ref{Kl3}.
This section proceeds with what is known for pion-pion scattering and
pion-kaon scattering and some possible consequences for the LECs $L_4^r$
and $L_6^r$. As very briefly indicated there, this is relevant for a possible
strong flavour dependence of spontaneous chiral symmetry breaking. 
We present some results here as well from the last remaining calculation,
$\pi,K\to\ell\nu\gamma$, in Sect.~\ref{Klng}.
We conclude with a few comments about the estimates of the NNLO or
order $p^6$ LECs. This is one of the main open questions in this field.
Note that there have been a few claims of two-loop calculations in the
literature which neglected the proper two-loop diagrams. None of these
are mentioned in this review.

The remaining part of this review is kept very short. I basically only mention
the calculations which have been done for finite temperature and volume and
give only a brief overview of the order $p^6$ work done for the
partially quenched case.

A few last remarks, a website with links to the actual formulas of
many of the papers reviewed here as well as some lectures on ChPT and
more general effective field theory is Ref.~\cite{formulas}.
There are also many calculations in the anomalous sector. Here the
order $p^6$ is only one-loop. A review is Ref.~\cite{anomalyreview}
and references to more recent work can be found in
the papers place where the order $p^6$ Lagrangian for this sector
was worked out~\cite{Lagp61,Lagp62}. I have done a reasonable effort to dig
out all relevant papers for this review.
With ChPT being such a large subject, I have
however most likely overlooked some directly relevant for the subject
considered.

\section{Chiral Perturbation Theory}
\label{sec:chpt}

\subsection{\it Chiral Symmetry}
\label{sec:chiral}

At the low energies discussed in this review only the lightest quarks
are relevant. The heavier quarks, charm, bottom and top play no role here.
We put the quarks together in a column vector
\be
\label{defq2}
q\equiv\left(\begin{array}{c}u\\d\end{array}\right)
\ee
for the two-flavour case or
\be
\label{defq3}
q\equiv\left(\begin{array}{c}u\\d\\s\end{array}\right)
\ee
for the three-flavour case. The conjugate row vector is analogously defined
via
\be
\overline q \equiv
\left(\begin{array}{cc}\overline u & \overline d \end{array}\right)
\quad\mbox{or}\quad
\overline q \equiv
\left(\begin{array}{ccc}\overline u 
  & \overline d &\overline s\end{array}\right)\,.
\ee
The generalization to $n_F$ flavours is obvious.

The gluons couple identically to all quark flavours. If all the masses are
equal for $n_F$ flavours we have a $SU(n_F)_V\times U(1)_V$ symmetry.
The $U(1)_V$ symmetry changes the phase of all the quark fields
simultaneously and corresponds to baryon number. The $SU(n_F)_V$
symmetry acts as
\be
\label{flavoursym}
q(x)\longrightarrow U q(x)\quad\mbox\quad U\in SU(n_F)_V\,.
\ee
The vector symmetry $SU(n_F)_V$ is known as isospin for the two-flavour case
and as the Gell-Mann-Ne'eman octet symmetry for the three-flavour case.

However, QCD has a larger symmetry structure. The QCD Lagrangian
is of the form
\be
\label{lagQCD1}
{\cal L}_{\mathrm{QCD}} = \sum_{i=u,d,s}
i \overline q_{iL} D\hskip-0.6em/\hskip0.2em q_{iL}
+i \overline q_{iR} D\hskip-0.6em/\hskip0.2em q_{iR}
- m_i \overline q_{iR} q_{iL}- m_i \overline q_{iL} q_{iR}+\cdots\,.
\ee
Here $ D\hskip-0.6em/\hskip0.2em$ is the covariant derivative
with the gluon field and the dots indicate the purely gluonic terms.
The sums over colours are understood and not explicitly written out.
The left and right handed quark fields are given by
\be
q_R = \frac{1}{2}\left(1+\gamma_5\right) q\quad\mbox{and}\quad
q_L = \frac{1}{2}\left(1-\gamma_5\right) q\,.
\ee

We define a quark mass matrix
\be
{\cal M} = \left(\begin{array}{ccc} m_u & & \\ & m_d & \\ & & m_s
\end{array}\right)\,
\ee
and left and right-handed column vectors $q_L$ and $q_R$
by replacing in (\ref{defq2}) and (\ref{defq3}) all quark fields
by their left and right-handed parts.
This allows us to rewrite the QCD Lagrangian as
\be
\label{lagQCD}
{\cal L}_{\mathrm{QCD}} =
i \overline q_L D\hskip-0.6em/\hskip0.2em q_L
+i \overline q_R D\hskip-0.6em/\hskip0.2em q_R
- \overline q_R {\cal M} q_L - \overline q_L {\cal M} q_R\,.
\ee
The form (\ref{lagQCD}) shows that there is in fact a larger symmetry
than the flavour rotations of (\ref{flavoursym}) whenever 
the quark masses are equal to zero, ${\cal M}=0$.
To be precise, we obtain the chiral symmetry group
\be
G=SU(3)_L\times SU(3)_R\times U(1)_V\times  U(1)_A\,,
\ee
with
\ba
q_R \longrightarrow g_R q_R && g_R \in SU(3)_R
\nn\\
q_L \longrightarrow g_L q_L && g_L \in SU(3)_L\,.
\ea
Under $U(1)_V$ all quarks have the same change in phase while under $U(1)_A$
the right and left-handed quarks have the opposite change in phase.
The symmetry is called chiral because it acts differently on the
left and right-handed quarks.

The $U(1)_A$ is only a symmetry of the classical action, not of the full
quantum theory of QCD. The divergence of the associated current does not vanish
due to the anomaly~\cite{ABJ}. It is nonzero by a total divergence but
instantons allow for this to have a physical effect~\cite{instantons}.
We will not consider its effects for the remainder of this paper.

The $U(1)_V$ symmetry corresponds to baryon number. We will only discuss
mesons in this review and hence we also drop this symmetry.
The final chiral symmetry of QCD in the limit where all quarks are massless
is thus
\be
G_\chi = SU(3)_L\times SU(3)_R\,.
\ee

\subsection{\it External field method}
\label{externalfield}

The consequences of a global symmetry are most clearly
expressed through the Ward identities for Green functions. These can
be derived using the methods given in most field theory books
but a particularly elegant method to obtain Green functions that obey the
Ward identities is the external field method. This method also allows
to show clearly how the knowledge of the Green functions of 
QCD in the chiral limit is sufficient also to describe the Green
functions away from the chiral limit. The particular version described here
was introduced by Gasser and Leutwyler~\cite{GL1}.

For the mesonic physics we discuss in this paper we look at 
Green functions or correlation
functions defined by vector and axial-vector currents, and scalar and
pseudoscalar densities. These are referred to together as external currents.
The currents are defined by
\ba
\label{currents}
V_\mu^{ij}(x) &=& \overline q_i(x) \gamma_\mu q_j(x)\,,
\nonumber\\
A_\mu^{ij}(x) &=& \overline q_i(x) \gamma_\mu\gamma_5 q_j(x)\,,
\nonumber\\
S^{ij}(x) &=& -\overline q_i(x)  q_j(x)\,,
\nonumber\\
P^{ij}(x) &=& \overline q_i(x) i\gamma_5 q_j(x)\,.
\ea
The indices $i,j$ run over the quark flavours $u,d$ or $u,d,s$
and the sum over colours is implicitly understood.

Green functions or correlation functions can in general be introduced via
the inclusion of sources in the Lagrangian. This is described in most
books on Quantum Field Theory, see e.g. chapter 9 in Ref.~\cite{Peskin}.
These sources are referred to as external sources or external fields.

The Lagrangian of massless QCD extended by the external fields is written as
\be
\label{lagQCD3}
{\cal L}_{\mathrm{QCD}}^{\mathrm{ext}} =
i \overline q_L D\hskip-0.6em/\hskip0.2em q_L
+i \overline q_R D\hskip-0.6em/\hskip0.2em q_R
- \overline q_R \left(s+ip\right) q_L - \overline q_L \left(s-ip\right) q_R
+ \overline q_L \gamma^\mu l_\mu q_L
+ \overline q_R \gamma^\mu r_\mu q_R\,.
\ee
The external fields $s$, $p$, $l_\mu$ and $r_\mu$ are space-time dependent
$n_F\times n_F$ matrix functions. The vector and axial-vector
fields, $v_\mu$ and $a_\mu$, are included via
\be
l_\mu \equiv v_\mu-a_\mu\,,\quad r_\mu \equiv v_\mu+a_\mu\,.
\ee
These external fields are all Hermitian matrices.

By taking functional derivatives with respect to the external sources $s$, $p$,
$v_\mu$ and $a_\mu$ from the functional integral we can then build up
all the wanted Green functions with the currents defined in (\ref{currents})
of massless QCD. To get an insertion of the current $V^{ij}_\mu(x)$
one needs to take the functional derivative w.r.t. $v_{ji}^\mu(x)$.

With these additional sources the global chiral symmetry described in the
previous section can be extended into a local chiral symmetry.
A local symmetry transformation is given by an element of the symmetry group,
$g_L\times g_R\in SU(n_F)_L \times SU(n_F)_R$
where $g_L$ and $g_R$ are now functions of the space-time point $x$.
The transformations under the {\em local} symmetry are
\ba
\label{externaltransformation}
q_L &\to& g_L\,q_L
\nonumber \\
q_R &\to& g_R\,q_R
\nonumber \\
\hat{\cal M}\equiv  \left(s+ip\right)&\to& g_R\,\hat{\cal M}\,g_L^\dagger,
\nonumber \\
l_\mu\equiv v_\mu-a_\mu&\to & g_L\,l_\mu\,g_L^\dagger
-i \partial_\mu\,g_L\,g_L^\dagger,
\nonumber \\
r_\mu\equiv v_\mu+a_\mu&\to & g_R\,r_\mu\,g_R^\dagger
-i \partial_\mu\,g_R\,g_R^\dagger.
\ea
For most of this paper, we will consider these symmetries as exact also
at the quantum level. They are anomalous but that effect is fully taken into
account by the Wess-Zumino-Witten\cite{WZ,Witten}
term and that one can be taken explicitly
into account as described in~\cite{GL1,GL2}.

The function that gives all the Green functions directly when
the functional derivative w.r.t. to the external fields is taken,
is called the generating functional, $G$
and it is given by the functional integral
\be
G(l_\mu,r_\mu,s,p)\equiv e^{\dsp i \Gamma(l_\mu,r_\mu,s,p)} = 
\frac{\dsp \int [dqd\overline qdG] 
e^{\dsp i\int d^4x {\cal L}_{\mathrm{QCD}}^{\mathrm{ext}}} }
{\left(\dsp \int [dqd\overline qdG] 
e^{\dsp i\int d^4x {\cal L}_{\mathrm{QCD}}^{\mathrm{ext}}}
\right)_{l_\mu,r_\mu,s,p=0} }\,.
\ee
The function  $\Gamma(l_\mu,r_\mu,s,p)$ is called the effective action.
$\int[dqd\overline qdG]$ indicates the functional or Feynman path integral
over all possible quark, anti-quark and gluon paths or configurations.

With this one can see how the generating functional for QCD with nonzero
masses is related to the one in the chiral limit. We have indicated the
two different cases here by the superscript $m_q\ne0$ and $m_q=0$.
Comparing the Lagrangians with and without the quark masses leads to the
relation
\be
\label{generatingmqne0}
G^{m_q\ne0}(l_\mu,r_\mu,s,p) = 
\frac{\dsp G^{m_q=0}(l_\mu,r_\mu,{\cal M}+s,p)}
{G^{m_q=0}(0,0,{\cal M},0)}\,.
\ee

In a similar fashion the response to external electroweak vector fields
can be included\footnote{We use here the word external since when
the exchanges of electroweak bosons between strongly interacting particles
is needed, the formalism needs to be extended.}.
The couplings of photons to quarks is described by a term in the Lagrangian
of the type
\be
e A_\mu \left(\overline q_L Q_L \gamma^\mu q_L
+\overline q_R Q_R \gamma^\mu q_R\right)\,,
\ee
with
\be
Q_L = Q_R = Q \equiv 
\left(\begin{array}{ccc}2/3 & & \\ & -1/3 & \\ & & -1/3\end{array}
\right)\,.
\ee
$A_\mu$ is the photon field.
The interactions with photons can thus be included
by changing
\be
\label{includegamma}
l_\mu \to l_\mu+e A_\mu Q\,, \quad r_\mu\to r_\mu+e A_\mu Q\,.
\ee
This should be understood 
in a way similar to the inclusion of quark masses by changing
$s$ to $s+{\cal M}$ as done via Eq.~(\ref{generatingmqne0}). 
In particular, the Green functions in the presence
of external electromagnetism and quark masses are up to a normalization
given by
$G(l_\mu+e A_\mu Q,r_\mu+e A_\mu Q,s+{\cal M},p)$.

The couplings to $W^\pm_\mu$ and $Z_\mu$ bosons can be included in a similar
fashion by looking at the standard model Lagrangian and seeing how
those couplings can formally be written as parts of $l_\mu$ and $r_\mu$,
just as we could write ${\cal M}$ formally as a part of $s$.
To be precise, in terms of the Cabibbo angle, $\theta_C$,
the charged $W^\pm$ couplings
are include by changing
\be
\label{includeW}
l_\mu\to l_\mu-\frac{g}{\sqrt{2}}\left(\begin{array}{ccc}
0 & \cos\theta_C W^+_\mu & \sin\theta_C W^+_\mu\\
\cos\theta_C W^-_\mu & 0 & 0\\
\sin \theta_C W^-_\mu & 0 & 0\end{array}\right)\,,
\ee
with $g= e/\sin\theta_W$, the Weinberg angle.

As described in this subsection, currents, external sources,
external fields, electroweak gauge bosons are different objects but are treated
in the end via the external fields $l_\mu$ and $r_\mu$. Many papers
tend to be somewhat cavalier with the choice of words.

We have described the Green functions here as calculated by using functional
derivatives. This is 
of course completely equivalent to calculating them directly
using Feynman diagrams. Similarly, amplitudes calculated directly with
Feynman diagrams are equivalent to those calculated from the Green functions
using standard LSZ reduction. The main reason for using this external field
formalism is that it allows for calculations maintaining chiral symmetry
throughout the entire calculation and allows for a simpler way of
classifying terms in ChPT as described below.

\subsection{\it Spontaneous symmetry breaking}
\label{spontaneous}

In Sect.~\ref{sec:chiral} we introduced the chiral symmetry of QCD
in the massless limit. This symmetry is however clearly not present in the
spectrum of hadronic states. If it was a good symmetry realized in
the same way as Lorentz or rotation symmetry the spectrum of strongly
interacting particles would look quite different. In particular,
one would have parity doublets. For every particle with a given spin and parity
there would be another one with the same spin but the opposite
parity. In the presence of small quark masses we would still expect
this to be approximately true, just like we find isospin doublets and triplets
as well $SU(3)$ octets and decuplets.
If one looks at the mass spectrum this is clearly not the case. There is
obviously no partner with a mass close to the pion, the rho or the proton.
The possible candidates, $a_0(980)$, $a_1$ and the Roper are very far away in
mass and have in general very different properties. Clearly, the chiral
symmetry, $G_\chi$, is not directly visible in nature.

Since QCD describes a very large collection of phenomena at high energies
extremely well, there must thus be another way to include this symmetry
in the real world. This was found by Goldstone~\cite{Goldstone}
and is often called the Nambu-Goldstone mode, while a direct realization
is referred to as the Wigner or Wigner-Eckart mode. Nambu's papers for this are
Ref.~\cite{Nambu}.

\begin{figure}[t]
\begin{center}
\begin{minipage}{16.5cm}
\begin{minipage}[t]{0.5\textwidth}
\begin{center}
\begin{overpic}[width=8cm]{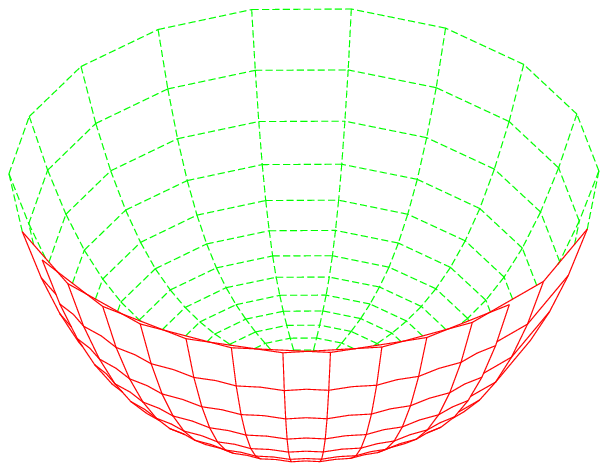}
\end{overpic}
\end{center}
\caption{The potential $V(\phi)$ for an unbroken symmetry.}
\label{figunbroken}
\end{minipage}
\begin{minipage}[t]{0.5\linewidth}
\begin{center}
\begin{overpic}[width=8cm,unit=0.5pt]{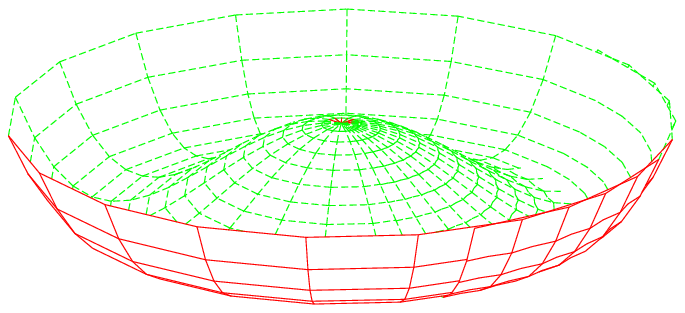}
\SetScale{0.5}
\SetWidth{3.}
\LongArrow(100,-10)(370,35)
\end{overpic}
\end{center}
\caption{The potential $V(\phi)$ for a spontaneously broken symmetry.
The arrow indicates a possible choice of vacuum.}
\label{figbroken}
\end{minipage}
\end{minipage}
\end{center}
\end{figure}

Let us first describe this mode for a simpler model. A complex scalar field
with Lagrangian
\be
\label{U1lagrangian}
{\cal L} = \partial^\mu\phi^*\partial_\mu\phi-V(\phi)\,.
\ee
We first look at a potential of the type shown in Fig.~\ref{figunbroken}
with a standard form of the type
\be
V(\phi) = \mu^2\phi^*\phi+\lambda\left(\phi^*\phi\right)^2\,.
\ee
We choose here $\lambda>0$ to have a stable theory. This Lagrangian has a
$U(1)$ symmetry under the phasetransformation
\be
\label{U1symmetry}
\phi\to e^{-i\alpha}\phi\,.
\ee
This transformation is rotation around the z-axis in Figs.~\ref{figunbroken}
and \ref{figbroken}.

If we choose $\mu^2>0$, the potential $V(\phi)$ has the form shown in
Fig.~\ref{figunbroken}, where the horizontal axes are the real and imaginary
part of $\phi$ while the vertical axis are $V(\phi)$.
In order to have a full theory we have to determine first the
vacuum, or lowest energy state, of the system. The contribution of the
kinetic term, $\partial^\mu\phi^*\partial_\mu$,
is minimized by a constant and spatially homogenous field $\phi_0$. 
From the form of the potential,
we can see that the total energy is thus minimized for a value of $\phi_0=0$.
I.e. $\langle\phi\rangle=0$. Excitations around the vacuum, which give the
particle spectrum, have only massive modes with a mass $m=\mu$.
Things to remark here: The vacuum is unique, i.e.
there is only one possible choice
of $\langle\phi\rangle$. There are two massive real modes in the spectrum
corresponding to the real and imaginary part of $\phi$. The interactions
of these particles are simply the four boson vertex directly present
in the Lagrangian (\ref{U1lagrangian}). This mode corresponds to the most
standard realization of symmetries like the realization of rotation
symmetries in standard quantum mechanics. States thus fall in
multiplets of the symmetry group and amplitudes obey the relations
of the Wigner-Eckart theorem.

However, when we choose the potential with the same form but take
$\mu^2<0$ the potential looks differently as depicted in Fig.~\ref{figbroken}.
The potential is still invariant under the symmetry~(\ref{U1symmetry}),
but now we have more than one option to choose from for the lowest-energy
state. The different choices are related by a symmetry transformation.
In order to start determining the particle spectrum we need to choose
a particular vacuum state (the lowest total energy is still given
by $\phi$ constant due to the kinetic term otherwise giving a positive
contribution). The arrow in Fig.~\ref{figbroken} indicates a possible
choice for $\langle\phi\rangle=\phi_0$. All possible choices are of
the form
\be
\label{vevphi}
\langle\phi\rangle=\phi_0= \frac{v}{\sqrt{2}} e^{i\alpha_0}
\ee
with
\be
v = \sqrt{-\mu^2/\lambda}\,.
\ee
Different choices of $\alpha_0$ lead to the same physics.
We can now try to get at the excitations around the vacuum. For that we need
to parameterize the changes of $\phi$ around $\phi_0$. This parameterization
can be done in many ways but let us choose here the form
\be
\label{expandphi}
\phi(x) = \left(v+\eta(x)\right) e^{i(\alpha_0+\pi(x)/v)}\,.
\ee
Putting this into the Lagrangian (\ref{U1lagrangian}) we obtain
\ba
\label{newU1lagrangian}
{\cal L} &=& \frac{1}{2}\partial^\mu\eta\partial_\mu\eta
+\frac{1}{4}\mu^2\eta^2-\lambda v \eta^3-\frac{1}{4}\lambda\eta^4
-\frac{1}{2}\mu^2 v^2-\frac{1}{4}\lambda v^4
+\frac{1}{2}\partial^\mu\pi\partial_\mu\pi
+\frac{1}{2}\left(2 v \eta +\eta^2\right)\partial^\mu\pi\partial_\mu\pi\,.
\ea
What do we see now, the Lagrangian has no obvious remainder
of the symmetry we originally had in Eq.~(\ref{U1symmetry}).
We say that the symmetry is {\em spontaneously broken}.
The fact that we needed to make a particular choice of the vacuum state
means that the original symmetry is no longer visible in the spectrum nor
in the interactions.
There is a massive mode, $\eta$, with mass $m^2=-\mu^2/2$ and a massless
mode, $\pi$. For the former there is some reminder of the symmetry in the
relations of the cubic to the quartic coupling and for the latter
there is a more striking result. It only interacts in a way that vanishes
for zero momentum. This is called a low-energy theorem and the massless mode
is called the Goldstone boson. We also see that the physical results
do not depend on the precise choice of the vacuum, $\alpha_0$ has disappeared
from the Lagrangian directly describing the excitations around the vacuum
of Eq.~(\ref{newU1lagrangian}). 

The appearance of both phenomena, a massless
mode and the vanishing of its interactions follow from the fact that
we have to choose a vacuum in every spacetime point. A qualitative description
is as follows. The massless mode
corresponds to choosing slightly different vacuum states in each space time
point. This gives a small kinetic energy but also momentum. We can think of
this as rolling around in the bottom of the valley in Fig.~\ref{figbroken}.
The interaction must vanish at zero momentum since the absolute
choice of vacuum cannot matter, we can thus shift $\pi(x)$ by an arbitrary
amount and no physics results should change. The interactions can thus
only proceed via derivatives, hence the low-energy theorems.

It is often said that the symmetry is realized nonlinearly in the
Nambu-Goldstone mode. This can be seen from the fact that the
transformation on $\pi(x)$ under the original symmetry (\ref{U1symmetry})
is
\be
\pi(x)\to\pi(x)-iv\alpha\,.
\ee
Note that the new Lagrangian (\ref{newU1lagrangian}) is invariant under this.
We will also talk in the remainder about a nonlinearly realized symmetry but
will construct for the case relevant for ChPT objects that do transform
linearly since this simplifies constructing the Lagrangians.
\begin{table}
\begin{center}
\begin{minipage}[t]{16.5 cm}
\caption{The two different modes of symmetry realizations compared.}
\label{tab:modes}
\end{minipage}
\begin{tabular}{c|c}
\hline
&\\[-2mm]
Wigner-Eckart mode & Nambu-Goldstone mode\\[2mm]
\hline
&\\[-2mm]
Symmetry group $G$  & $G$ spontaneously broken to subgroup $H$\\[2mm]
Vacuum state unique & Vacuum state degenerate\\[2mm]
Massive Excitations & Existence of a massless mode\\[2mm]
States fall in multiplets of $G$
& States fall in multiplets of $H$\\[2mm]
Wigner Eckart theorem for $G$ & Wigner Eckart theorem for $H$\\[2mm]
 & Broken part leads to low-energy theorems\\[2mm]
Symmetry linearly realized & Full Symmetry, $G$, nonlinearly realized\\[2mm]
   & unbroken part, $H$, linearly realized\\[2mm]
\hline
\end{tabular}
\end{center}
\end{table}

I have used a very simple first model with the Lagrangian
of Eq.~(\ref{U1lagrangian}), to describe
the most important features here. Let me now shortly indicate which parts
generalize to other theories.
In general we have a continuous symmetry group $G$ which is generated by
a number of generators $T^a$. A generic group element can be schematically
written as $g=\exp(i \epsilon^a T^a)$. The choice of vacuum
leaves in general not all generators invariant. The subspace of generators,
 $T^b$ that leave the vacuum invariant generates the unbroken
subgroup $H$ with elements of the form $\exp(i \epsilon^bT^b)$.
The broken part of the symmetry group corresponds to those generators
that move the vacuum around, i.e.,
\be
T^c |0> \ne 0\,.
\ee
The space on which the Goldstone bosons live is the space of possible
vacua. This space has the structure $G/H$, the coset space of the group $G$
with its unbroken subgroup $H$. In our example, this coset space was
the bottom of the valley and could be simply parameterized by $\pi(x)$.
The effective Lagrangian after spontaneous symmetry breakdown can be made
fully invariant also under the full symmetry group but this in general
implies nonlinear transformations for the Goldstone bosons.

Alternative introductions to spontaneous symmetry breaking can be found in most
books on particle physics when the Higgs mechanism is introduced.
A review that discusses many more places where Goldstone bosons
and effective Lagrangian appear is Ref.~\cite{Burgess}. 
More mathematical descriptions can be found in the lectures by
Pich~\cite{Pich} and Scherer~\cite{Scherer1,Scherer2}.

\subsection{\it Spontaneous symmetry breaking in the presence of explicit
symmetry breaking}
\label{spontaneousexplicit}

\begin{figure}[t]
\begin{center}
\begin{overpic}[width=8cm,unit=0.5pt]{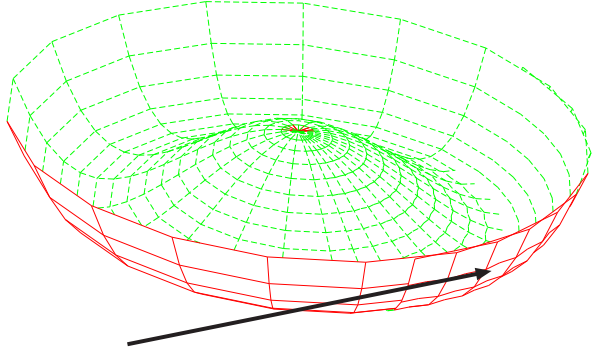}
\SetScale{0.5}
\SetWidth{3.}
\LongArrow(100,-10)(373,45)
\end{overpic}
\begin{minipage}{16.5cm}
\caption{The potential $V(\phi)$ for a spontaneously broken symmetry
in the presence of a small explicit symmetry breaking term.
The arrow indicates now the only possible choice of vacuum.}
\label{figtilted}
\end{minipage}
\end{center}
\end{figure}

In QCD, the chiral symmetry is not exact, the term with the quark masses
is not invariant under chiral symmetry. So what happens if there is
explicit symmetry breaking present at the same time as the spontaneous
symmetry breaking. We will again discuss it in the framework of our simple
model with a $U(1)$ symmetry and a complex scalar field.
We now add to the Lagrangian of Eq.~(\ref{U1lagrangian}) a term
of the form
\be
{\cal L}_{\mathrm{extra}} = -\beta(\phi+\phi^*)\,.
\ee
This extra term is not invariant under the symmetry transformation
(\ref{U1symmetry}) and the potential looks tilted as shown in
Fig.~\ref{figtilted}. What is important is that the tilting is small compared
to the height of the central bump so it can still be treated as a perturbation.
The vacuum in Eq.~(\ref{vevphi}) is no longer unique but only the
one with $\alpha_0=0$ is the lowest energy state.

The physics has also slightly changed. We expand around the vacuum again
by using (\ref{expandphi}) with $\alpha_0=0$ and obtain in addition
to the terms in Eq.~\ref{newU1lagrangian})
\be
{\cal L}_{\mathrm{extra}} = -\beta\sqrt{2}(v+\eta)
\left(e^{i\pi(x)/v}+e^{-i\pi(x)/v}\right)\,.
\ee
The linear term in $\eta$ can be removed by a small additional shift.
This happened because the lowest energy state is slightly shifted compared
to the value $v=\sqrt{-\mu^2/\lambda}$. But more importantly, when we expand
the exponentials, we now find that the $\pi(x)$-field has gotten a small
mass, small compared to the mass of the $\eta$-field, and no longer has only
derivative interactions.
The $\pi$ mass
\be
m_\pi^2 \approx \frac{2\sqrt{2}\beta}{v}\,.
\ee
is small and can be expanded in the small symmetry breaking parameter
$\beta$. The particle corresponding to it, is now called a
pseudo-Goldstone boson.
As long as the explicit symmetry breaking is small, we can still use
Goldstone's theorem as a first approximation and then add the corrections
systematically. This is precisely what we do in ChPT when the light
quark masses are explicitly included.

\subsection{\it Spontaneous symmetry breaking in QCD}
\label{spontaneousQCD}

We already argued in Sect.~\ref{spontaneous} that the chiral symmetry of
QCD cannot be realized in nature since the predicted parity doublets do not
occur. We thus expect the chiral symmetry to be realized in the Nambu-Goldstone
mode. What theoretical evidence do we have directly for this?

Most of the remainder of this paper is about the Goldstone bosons from the
spontaneous chiral symmetry breakdown and their properties. In this way,
all those properties are strong indications that the picture described
below is correct. However let us first give the full theoretical
arguments.
\begin{itemize}
\item
It has been proven that the chiral symmetry is spontaneously broken in the
limit of a large number of colours and assuming confinement~\cite{CW}.
\item
The vector symmetries remain unbroken in a vectorlike symmetry as
QCD~\cite{VW}.
\item
Assuming confinement, the anomalies in the effective low-energy theory
must match those for the underlying QCD theory. For two flavours,
this can be done but not for three or more flavours. We thus need
spontaneous symmetry breaking in order to have a correct anomaly matching for
three or more flavours~\cite{thooftcargese}.
\end{itemize}

We thus believe that the flavour symmetry $SU(n_F)\times SU(n_F)$
is spontaneously broken down to the diagonal subgroup
$SU(n_F)_V=SU(n_F)_{L+R}$ also for the realistic case of three flavours.
There are eight broken generators and we thus expect eight Goldstone boson
degrees of freedom. If we look at the hadron spectrum there are eight natural
candidates for this. The three pions, $\pi^0$, $\pi^\pm$, four kaons,
$K^\pm$, $K^0$, $\overline{K^0}$ and the eta, $\eta$.  Goldstone's theorem
thus explains their low mass compared to the other hadrons as well as the
fact that their interactions are relatively weak. The three main qualitative
predictions that follow are:
\begin{enumerate}
\item The masses are rather small.
\item The masses obey the Gell-Mann-Okubo relation with squared masses rather
than linearly.
\item $\pi\pi$ scattering is fairly small compared to proton-proton scattering
and related to pion decay.
\item There is a nontrivial relation between the pion decay constant,
the axial-vector coupling of the nucleon $g_A$ and the pion nucleon
coupling $G_\pi$, the Goldberger-Treiman relation~\cite{GT}. 
\end{enumerate}
All of these predictions are well borne out by experiment. The first
three will be discussed in detail below.

In Sect.~\ref{spontaneous} the $U(1)$ symmetry was broken by a vacuum
expectation value of the field $\phi$. In QCD, the vacuum is also not invariant
under the full chiral symmetry $G_\chi$ but the quantity that most
characterizes the noninvariance of the vacuum is a composite of the
fundamental fields, the quark-antiquark bilinear condensate.
The vacuum of QCD is thus characterized by
\be
\label{qbarqvev1}
\langle \overline q_j q_i\rangle \ne 0\,.
\ee
This is the standard picture which we now know is true for the two-flavour
case~\cite{CGL1}. Everything we know indicates that it is also true for
the three-flavour case but the argument is not fully closed yet~\cite{Stern2}.

That the vacuum expectation value (\ref{qbarqvev1}) breaks chiral symmetry
can be easily seen when we rewrite it into left and right handed components
and as a matrix in the flavour space via
\be
\label{qbarqvev2}
(V^q)_{ij} = \langle \overline q_{jL} q_{iR}\rangle\,.
\ee
Under a chiral symmetry transformation $g_L\times g_R \in SU(n_F)_L\times
SU(n_F)_R$ this transforms as
\be
V^q\to g_R V^q g_L^\dagger\,.
\ee
We now choose a particular vacuum expectation value,
\be
\label{qbarqvev3}
(V^q)_{ij} = \frac{1}{2}\delta_{ij}\langle\overline q q\rangle\,,
\ee
where we used the generic symbol $\langle\overline q q\rangle$ for
a vacuum expectation value of one quark species in the chiral limit.
With this choice, one sees that the vector subgroup, $g_L=g_R$, leaves
$V^q$ invariant but any transformation with $g_L\ne g_R$ does not,
so the axial part of the symmetry group is spontaneously broken.

It now remains to parameterize the space of possible vacua.
For a spontaneous breakdown of $G=SU(n_F)_L\times SU(n_F)_R$ to
$H=SU(n_F)_V$ the coset space $G/H$ has itself the structure of an
$SU(n_F)$ manifold. The simplest parameterization for describing
the Goldstone boson is thus choosing them as an $n_F\times n_F$ special
unitary matrix $U$. This parameterization, together with some minor extensions,
is used in the remainder of this paper. 

We argued in Sect.~\ref{spontaneous} that in some sense the Goldstone
bosons live in the space of possible vacua. The same is true here.
We can parameterize the space of vacua of $V^q$ by the same special
unitary matrix $U$ via
\be
\label{qbarqvev4}
V^q = \frac{1}{2}U\langle\overline q q\rangle\,.
\ee
The matrix $U$ thus transforms under the chiral symmetry group as
\be
\label{Utransformation}
U\to g_R^\dagger U g_L\,.
\ee

There exist many possible alternative parameterizations. The solution
for the two-flavour case was originally found by Weinberg~\cite{Weinberg0}.
It was generalized by Coleman, Wess and Zumino to arbitrary symmetry breaking
patterns $G\to H$~\cite{CWZ}. The inclusions of states other than the Goldstone
bosons was worked out in Ref.~\cite{CCWZ}. The latter shows also explicitly
that there is no need for the existence of states related by parity in order
to have a fully chirally symmetric theory when the chiral symmetry is
spontaneously broken. The last two references also showed that their
parameterization is fully general and remains valid when loop effects are
taken into account.

\subsection{\it Lowest Order ChPT}
\label{LOChPT}

Chiral Perturbation Theory is the low-energy effective field theory of QCD
where the degrees of freedom taken into account are the Goldstone
bosons from the spontaneous breakdown of the chiral symmetry and their
interactions. We showed in Sect.~\ref{spontaneousQCD} that the
resulting Goldstone boson manifold can be parameterized by a special unitary
matrix $U$ which transforms under the chiral symmetry group as in
Eq.~(\ref{Utransformation}). We also want to include the external fields
introduced in Sect.~\ref{externalfield} and we want it to be fully invariant
under the chiral symmetry as required by the arguments of~\cite{CWZ}.
Note that these arguments including the loop level are worked out in great
detail in Ref.~\cite{Leutwyler1}.

In the remainder a lot of notation will be introduced. In particular
many traces of $n_F\times n_F$ matrices will appear. In order to make
these traces easier to see we introduce the notation
\be
\langle A \rangle = \mathrm{tr}_F\left(A\right)\,,
\ee
where $\mathrm{tr}_F$ denotes the trace over flavour indices.

Without external fields and derivatives the only terms that can
be constructed are of the form
\be
{\cal L}_0 = 
\alpha_0\langle U^\dagger U\rangle+\alpha_1\det U+\alpha_1^*\det U^\dagger\,.
\ee
This is only an irrelevant constant since for a special unitary matrix
we have that
\be
U^\dagger U = 1\quad\mathrm{and}\quad\det U = 1\,.
\ee
This corresponds to the fact that Goldstone bosons cannot have interactions
without derivatives or explicit symmetry breaking.

At the next order, we find that $\partial_\mu U$ is not chirally invariant.
A building block that transforms nicely can be constructed by defining
a covariant derivative
\ba
\label{covariantderivative}
D_\mu U &=& \partial_\mu U -i r_\mu U + i U l_\mu\,,
\nonumber\\
D_\mu U^\dagger &=& 
\partial_\mu U^\dagger -i l_\mu U^\dagger + i U^\dagger r_\mu\,.
\ea
These transform simply under the chiral symmetry group as
\be
\label {DUtransform}
D_\mu U\to g_R D_\mu U g_L^\dagger\,\quad
D_\mu U^\dagger\to g_L D_\mu U^\dagger g_R^\dagger\,.
\ee

With the covariant derivatives
in hand we can now construct the Lagrangian at the first nontrivial
order
\be
\label{L2x1}
{\cal L}_2 = \beta_1 \langle D_\mu U^\dagger D^\mu U\rangle
+ \beta_2 \langle \hat{\cal M} U^\dagger + U \hat{\cal M}^\dagger\rangle
+ i\beta_3 \langle \hat{\cal M} U^\dagger - U \hat{\cal M}^\dagger\rangle
\,.
\ee
In this construction we have assumed that
\be
\langle r_\mu\rangle = \langle l_\mu\rangle = 0\,.
\ee
This together with $\det U = 1$ implies that
\be
\langle U^\dagger D_\mu U\rangle = 0\,.
\ee
We have used the fact that the Lagrangian can be changed by partial
integration and that partial integration can be done with the covariant
derivatives. 

Under parity we interchange left and right. The parity transformation on $U$
is thus $U\to U^\dagger$. $\hat{\cal M}$ goes similarly to its complex
conjugate. The term proportional to $\beta_3$ thus violates parity and can be
dropped because of that.

The Lagrangian (\ref{L2x1}) is usually written in the form
\be
\label{L2}
{\cal L}_2 = \frac{\hat F^2}{4}\langle D_\mu U^\dagger D^\mu U+\chi U^\dagger
+U\chi^\dagger\rangle\,,
\ee
with
\be
\chi = 2 \hat B {\cal M} = 2 \hat B \left(s+ip\right)\,.
\ee
The specific values of the parameters $\hat F$ and $\hat B$ depend
on the number of flavours and they are conventionally written
as $F,B$ for the two-flavour case~\cite{GL1} and $F_0,B_0$ for
the three-flavour case~\cite{GL2}.

The Lagrangian for lowest order ChPT was first given by Weinberg in
Ref.~\cite{Weinberg0} for the two-flavour case
and soon generalized. 
A discussion about its construction, including many of the subtleties
involving the overall phase of $U$ can be found in Sects.~3 and 4
in Ref.~\cite{GL2} and the references mentioned therein.

\subsection{\it A few consequences of lowest order ChPT}
\label{LOapplications}

In this section I will only talk about the three-flavour case, $n_F=3$,
and thus stick to the notation for that case.
Let us derive a few simple consequences from the Lagrangian
(\ref{L2}). This can be done using the machinery of the effective action
or by simply using Feynman diagram calculations. Both methods have to give
the same answers and calculations have been performed using both approaches.
In this section we will stick to the simplest one, tree level Feynman
diagrams and identifying masses by the terms in the Lagrangian.

First we need to parameterize the matrix $U$. This is done
in terms of a matrix of meson fields in the form
\be
U = e^{i \sqrt{2} M /F_0}\quad\mathrm{with}\quad
M = \left(\begin{array}{ccc}
\frac{1}{\sqrt{2}}\,\pi^0+\frac{1}{\sqrt{6}}\,\eta &\pi^+&K^+\\
\pi^- & -\frac{1}{\sqrt{2}}\,\pi^0+\frac{1}{\sqrt{6}}\,\eta & K^0\\
K^- & \overline{K^0} & -\frac{2}{\sqrt{6}}\,\eta
\end{array}\right)\,.
\ee
We have used here the isospin triplet field for the $\pi^0$ and the octet
component only for the $\eta$. This is fine as long as we work
in the isospin limit with
\be
\label{isolimi}
m_u=m_d\,.
\ee
In this limit we have two quark masses
\be
\label{defmhat}
\hat m \equiv \frac{1}{2}\left(m_u+m_d\right)
\ee
and $m_s$.
We use the relation (\ref{generatingmqne0}) and set $s$ in (\ref{L2})
equal to ${\cal M}$.
We put the parameterization of $U$ into (\ref{L2}) and expand the
exponentials. Looking at the terms up to second order
in the meson fields, we find
\ba
\label{kinetic}
{\cal L}_2 &=&
\frac{1}{2}\partial_\mu\pi^0\partial^\mu\pi^0
+\partial_\mu K^+\partial^\mu K^-
+\partial_\mu\overline{K^0}\partial^\mu K^0
+\partial_\mu\pi^+\partial^\mu\pi^-
+\frac{1}{2}\partial_\mu\eta\partial^\mu\eta
- B_0 \hat m \,\pi^0\pi^0- 2 B_0\hat m\, \pi^+\pi^-
\nonumber\\ &&
- B_0 (\hat m+m_s)\, \overline{K^0}K^0
- B_0 (\hat m+m_s)\, K^+ K^-
- B_0 \frac{\hat m+2m_s}{3}\,\eta\eta\,.
\ea
Here we see several things.
The pions have the same mass
\be
\label{mpiLO}
m_\pi^2 = 2 B_0\hat m\,.
\ee
Similarly, the kaons have the same mass,
\be
\label{mKLO}
m_K^2 = B_0(\hat m+m_s)\,.
\ee
A relation between the pion, eta and kaon masses exists.
This relation is the famous Gell-Mann-Okubo (GMO) relation:
\be
\label{GMO}
m_\eta^2 = \frac{4}{3}m_K^2-\frac{1}{3}m_\pi^2\,.
\ee
We see here that this relation should be satisfied by the masses squared.
A naive application of the Wigner-Eckart theorem
and the symmetry group $SU(3)_V$ would have led to
the same relation but with the masses present linearly. The fact that
$\pi$, $\eta$ and $K$ are pseudo-Goldstone bosons from the spontaneously
broken chiral symmetry explains why the relation should be with
the quadratic masses. This follows if we include the additional
assumption that the lowest order term gives the bulk of the observed
masses, see e.g. Ref.~\cite{Stern1}.

The determination (\ref{mpiLO}) of the pion mass actually contains
another famous relation. We can get the quark-antiquark bilinear condensate
by taking functional derivatives of the generating functional $G$.
\be
\langle \overline u u\rangle = 
-\left(\frac{\delta}{i\delta s_{11}(x)}G\right)\,,
\ee
evaluated at the point where the external fields are set to zero.
Doing this we obtain to lowest order
\be
\label{qbarLO}
\langle\overline q q\rangle=\langle \overline u u\rangle =
\langle \overline d d\rangle =
\langle \overline s s\rangle =
- B_0 F_0^2\,.
\ee
We thus obtain the celebrated Oakes-Renner relation
\be
m_\pi^2 = -\frac{\hat m \langle\overline u u + \overline d d\rangle}{F_0^2}\,.
\ee

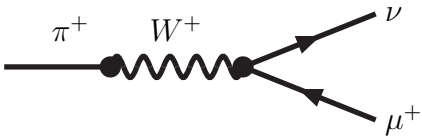
\begin{figure}
\begin{center}
\setlength{\unitlength}{1pt}
\begin{picture}(150,50)(-10,-5)
\SetScale{1.0}
\SetWidth{2}
\Line(-10,20)(30,20)
\Text(15,30)[b]{$\pi^+$}
\Photon(30,20)(80,20){4.}{5}
\Text(55,30)[b]{$W^+$}
\ArrowLine(80,20)(130,40)
\ArrowLine(130,0)(80,20)
\Text(134,40)[l]{$\nu$}
\Text(134,0)[l]{$\mu^+$}
\Vertex(30,20){4}
\Vertex(80,20){4}
\end{picture}
\begin{minipage}[t]{16.5cm}
\caption{The Feynman diagram responsible for the main pion decay
$\pi^+\to\mu^+\nu$.}
\label{figpidecay}
\end{minipage}
\end{center}
\end{figure}

A third example is the semileptonic decay of the pion.
This proceeds via the diagram shown in Fig.~\ref{figpidecay}.
The $\pi^+ W^-$ vertex can be derived from the Lagrangian (\ref{L2})
when we use the way to include the $W$-boson via (\ref{includeW}).
Alternatively, the coupling of the pion to the $W$-boson is regulated
by the pion decay constant defined by
\be
\label{defFpi}
\langle 0 | \overline d \gamma_\mu\gamma_5 u | \pi^+(p)\rangle
\equiv i \frac{F_\pi}{\sqrt{2}}\,p_\mu\,.
\ee
We can compare the two calculations or calculate the matrix-element 
(\ref{defFpi}) directly by taking a functional derivative with respect
to $a_\mu$. In both cases we reach the lowest order result
\be
\label{FpiLO}
F_\pi = F_0\,.
\ee

\begin{figure}
\begin{center}
\setlength{\unitlength}{1pt}
\begin{picture}(60,60)
\SetScale{1.0}
\SetWidth{2}
\Line(0,0)(60,60)
\Line(0,60)(60,0)
\Vertex(30,30){4}
\end{picture}
\begin{minipage}[t]{16.5cm}
\caption{The Feynman diagram responsible for pion-pion scattering
at lowest order. The lines are the pions.}
\label{figpipiLO}
\end{minipage}
\end{center}
\end{figure}
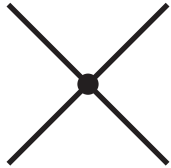

The final result we will allude to is $\pi\pi$ scattering.
This was first derived by Weinberg using current algebra
methods~\cite{Weinbergpipi}.
Here it follows from expanding the Lagrangian (\ref{L2}) to higher
orders in the meson fields. We then find vertices depicted schematically in
Fig.~\ref{figpipiLO}. In terms of the amplitude $A(s,t,u)$ defined
later in (\ref{defAstu}) the result he obtained is
\be
\label{pipiLO}
A(s,t,u) = \frac{1}{F_\pi^2}\left(s-m_\pi^2\right)\,.
\ee

\subsection{\it Powercounting and renormalization overview}
\label{powercounting}

The main purpose of this paper is to review higher order calculations
in ChPT. The Lagrangian in (\ref{L2}) is nonrenormalizable. This makes
it unfit to be used as a {\em fundamental} theory but produces no
problems for effective theories. We know that there is a well-defined
gauge theory, QCD, underlying ChPT. 
But still, in order to have a phenomenological usefulness, there must be
a way to limit the number of parameters that are present. In general,
nonrenormalizable theories have an infinite number of adjustable parameters.

What we will show in this subsection is that there exists a well-defined
way to order the various contributions of in terms of expansion parameters.
First, there are many quantities here, and we are not simply performing
an expansion in a small coupling constant as is done in
Quantum Electrodynamics. The expansion we have here is a long-distance
or a small momentum expansion, together with an expansion in the
quark masses.
We first call the magnitude of a typical momentum component $p$.
Since these come from derivatives, it is natural to also take
the external fields $l_\mu$ and $r_\mu$ of order $p$,
since these occur together
with the derivative in the covariant derivative 
of Eq.~(\ref{covariantderivative}).
On-shell particles have $p^2=m^2$. It is therefore natural from (\ref{mpiLO})
to take a quark-mass as order $p^2$. This in turn makes it natural to count
scalar and pseudoscalar external fields as order $p^2$ because
of the rule used in (\ref{generatingmqne0}).

With this counting we see that all terms in the Lagrangian (\ref{L2}) are
of order $p^2$, which is why we chose a subscript 2 there. This counting
can be generalized to all orders and is how we will order our series.
It was introduced by Weinberg in Ref.~\cite{Weinberg}.

\begin{figure}
\begin{center}
\begin{minipage}{5cm}
\unitlength=0.5pt
\begin{picture}(100,100)
\SetScale{0.5}
\SetWidth{1.5}
\Line(0,100)(100,0)
\Line(0,0)(100,100)
\Vertex(50,50){5}
\end{picture}
\hfill\raisebox{25pt}{$p^2$}\\[0.25cm]
\unitlength=0.5pt
\begin{picture}(100,30)
\SetScale{0.5}
\SetWidth{1.5}
\Line(0,15)(100,15)
\end{picture}
\hfill\raisebox{5pt}{$1/p^2$}\\[0.25cm]
$\int d^4p$\hfill$p^4$
\end{minipage}
\hskip2cm
\raisebox{0.cm}{
\begin{minipage}{7cm}

{ Some diagrams}\\[-1cm]
\begin{picture}(100,100)
\SetScale{0.5}
\SetWidth{1.5}
\Line(0,100)(20,50)
\Line(0,0)(20,50)
\Vertex(20,50){5}
\CArc(50,50)(30,0,180)
\CArc(50,50)(30,180,360)
\Vertex(80,50){5}
\Line(80,50)(100,100)
\Line(80,50)(100,0)
\end{picture}
\hskip-1cm\raisebox{25pt}
{$(p^2)^2\,(1/p^2)^2\,p^4 = p^4$}\\[0.25cm]
\unitlength=0.5pt
\begin{picture}(100,100)
\SetScale{0.5}
\SetWidth{1.5}
\Line(0,0)(50,40)
\Line(0,50)(50,40)
\CArc(50,70)(30,0,180)
\CArc(50,70)(30,180,360)
\Vertex(50,40){5}
\Line(50,40)(100,50)
\Line(50,40)(100,0)
\end{picture}
~~\raisebox{25pt}
{$(p^2)\,(1/p^2)\,p^4 = p^4$}
\end{minipage}
}
\begin{minipage}[t]{16.5cm}
\caption{The power-counting introduced by Weinberg illustrated on the
example of pion-pion scattering. See text for explanations.}
\label{figpower}
\end{minipage}
\end{center}
\end{figure}
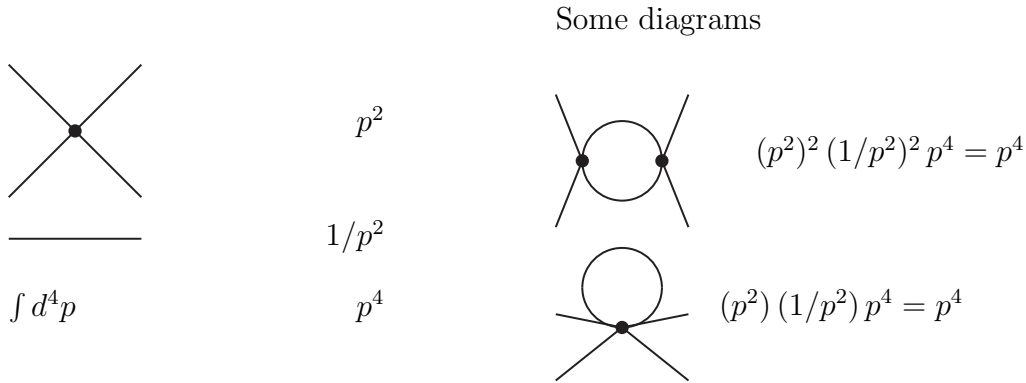

Let us first give it for a few simpler diagrams.
On the left hand-side in Fig.~\ref{figpower} the rules of counting
are shown. A vertex from the lowest order Lagrangian (\ref{L2})
counts as order $p^2$. A propagator is of order $1/p^2$ and a loop
integral is of order $p^4$. For dimensional reasons it must give
an extra four powers of momenta.
On the right-hand side we show two loop diagrams contributing
to pion-pion scattering. They are both order $p^4$ when counting the
number of loops, vertices and propagators. The two one-loop diagrams
are thus of the same order and also of the same order as a tree level
diagram with a vertex with four derivatives would be.

This type of observation is the underpinning of the expansion used in ChPT.
Let us know give this argument in general.
A generic diagram with $N_P$ propagators, $N_L$ loop integrations
and $N_n$ the number of vertices of order $p^n$. The total order of the
diagram is then
\be
\label{power1}
N_T = -2 N_P + 4 N_L +\sum_{n\ge2} n N_n\,.
\ee
Here we already used the fact that the lowest order Lagrangian is of
order $p^2$. We can rewrite this using the relation between the
number of internal lines, $N_I=N_P$, the number of vertices $N_V$ 
and the number of loops $N_L$,
\be
\label{power2}
N_I = N_L+N_V-1\,.
\ee
Since $N_V = \sum_n N_n$, Eq.~(\ref{power1}) can be rewritten as
\be
\label{power3}
N_T = 2+2 N_L+\sum_{n\ge2} (n-2) N_n\,.
\ee
Eq.~(\ref{power3}) is the basis of the perturbative expansion
of ChPT. The lowest order contribution to any process is given by a tree level
diagram with only vertices from ${\cal L}_2$. The next order, NLO, is formed
by one-loop diagrams with only vertices from ${\cal L}_2$ and
tree level diagrams with vertices from ${\cal L}_2$ and one-vertex from
the $p^4$ Lagrangian ${\cal L}_4$.

Eq.~(\ref{power3}) also shows the importance of Goldstone's theorem
for the existence of a perturbative expansion. Only because the lowest
order has derivatives or external fields do we have an expansion
where higher loops imply higher powers of $p$, i.e. Goldstone's
theorem is the source of the
requirement $n\ge 2$.

Note that the powercounting described here is closely related to the notion
of superficial degree of divergence described in most Quantum Field Theory
books see e.g. \cite{Peskin,ItzyksonZuber}.

The relation given in Eq.~(\ref{power2}) can most easily be understood
by induction. In a tree level diagrams all vertices need to be connected
by internal lines. The simplest diagram has one vertex and no internal line.
Keeping it at tree level but adding lines implies always adding one internal
line and one vertex, so far tree level diagrams we have
$N_I = N_V-1$. Every time we add an internal line, but no new vertex,
we create a new loop. We thus end up with (\ref{power2}).

Let me finish this section by giving the overview and general arguments
involved in ChPT and its construction and renormalization.
First we use Weinberg's conjecture~\cite{Weinberg}, see also the discussion
in~\cite{DHokerWeinberg}: {\em if one writes down the most general
possible Lagrangian, including {\em all} terms consistent with assumed
symmetry principles, and then calculates matrix elements with this
Lagrangian to any given order of perturbation theory, the result will simply be
the most general possible S-matrix element consistent with analyticity,
perturbative unitarity, cluster decomposition and the assumed symmetry
principles.}

Then we assume that the relevant degrees of freedom are the Goldstone
bosons from the spontaneous breaking of chiral symmetry and construct
the most general Lagrangian with them which has the full chiral
invariance. Using the results of Ref.~\cite{CWZ} this can then be brought into
a standard form. We have assumed here that we can use a {\em local} Lagrangian.
That this can be done was shown in Ref.~\cite{Leutwyler1}.
As said above, Goldstone's theorem implies that the lowest order Lagrangian
is of order $p^2$. We now use a regularization that conserves
chiral symmetry. Dimensional regularization~\cite{dimreg}
is the most standard choice. In a general Quantum Field Theory with
local vertices, all divergences that appear are local. Since we start with
a Lagrangian invariant under the symmetry and a fully invariant regularization,
the divergences are local but will have a structure that obeys the symmetry
structure. Since our constructed Lagrangian includes {\em all} possible
terms consistent with the symmetry, all divergences can thus be absorbed
into the coefficients of the Lagrangian. The total number of local terms
in the Lagrangian will be infinite, but since we can order the expansion
in terms of the order in powercounting in $p$, we have a well-defined
system with a finite number of parameters up to any given order in $p$.
A much more extensive version of this discussion where much attention is
paid to all the issues just mentioned here is Ref.~\cite{Leutwyler1}.

It is possible to use a regularization which is not chirally invariant.
One then needs to introduce also non-invariant counterterms and explicitly
enforce all Ward identities. Some of the problems involved are discussed
in the papers listed in Ref.~\cite{BG}.

Renormalization is a wide topic and can be found treated in most
Quantum Field Theory books, e.g.~\cite{Peskin,ItzyksonZuber}, one
concentrating on renormalization is ~\cite{Collins}.

\subsection{\it Construction of higher-order Lagrangians}
\label{Lagconstruction}

In Eq.~(\ref{L2}) we showed the lowest-order Lagrangian.
In Sect.~\ref{powercounting} we presented how ChPT can be systematically
extended to higher orders. A major part of this involves constructing
the most general Lagrangian at a given order in the powercounting in $p$.
This involves two steps. First we want to construct a {\em complete}
Lagrangian that includes all possible local terms that are invariant
under the full chiral symmetry. This is a rather elaborate exercise
but can be done in a fairly straightforward manner. However, this
procedure tends to end up with far too many terms. The more
challenging part is to find a {\em minimal} but still complete set of terms.

To construct a complete Lagrangian one first constructs a complete set of
quantities involving $U$, derivatives and the external fields that transforms
in a simple manner under the chiral symmetry. As an example,
for a quantity with one derivative there are three standard choices
\be
L_\mu = i U^\dagger D_\mu U\,\quad
R_\mu = i U D_\mu U^\dagger\quad\mathrm{and}\quad u_\mu\,.
\ee
The quantity $u_\mu$ needs a little more explanation. We write the
full matrix 
\be
U = u^2\,.
\ee
For a general chiral symmetry transformation
$g_L\times g_R\in SU(n_F)_L\times SU(n_F)_R$ there exists a matrix
$h\in SU(n_F)_V$ such that
\be
\label{defh}
u\to g_R u h^\dagger \equiv h u g_L^\dagger\,.
\ee
Eq.~(\ref{defh}) is the definition of $h$. The unitary matrix $h$ depends
nonlinearly on $g_L$, $g_R$ and $u$\,, but is unique. This is really
the general parameterization of \cite{CWZ,CCWZ} for the case of
$SU(n)\times SU(n)\to SU(n)$.
We now define
\be
\label{defu}
u_\mu = i\left\{
u^\dagger(\partial_\mu-i r_\mu)\,u -
u\,(\partial_\mu-i l_\mu)\,u^\dagger\right\}\,.
\ee

Under the symmetry, the three choices transform as
\ba
\label{utransform}
L_\mu &\to& g_L L_\mu g_L^\dagger\,,\nonumber\\
R_\mu &\to& g_R R_\mu g_R^\dagger\,,\nonumber\\
u_\mu &\to& h u_\mu h^\dagger\,,
\ea
which follow from Eqs.~(\ref{Utransformation},\ref{DUtransform},\ref{externaltransformation},\ref{utransform}).
The kinetic term of the lowest order Lagrangian can be written in terms of
all three using
\be
\langle L_\mu L^\mu\rangle = \langle R_\mu R^\mu\rangle
= \langle u_\mu u^\mu\rangle\,.
\ee
In this review we use the last choice but the others have also been
used, see e.g. \cite{DW,BBC,FS2}.

To get the order $p^4$ Lagrangian, we need the additional quantities
\ba
\label{defother}
\chi_\pm &=& u^\dagger\chi\,u^\dagger\pm u\,\chi^\dagger\,u,
\nonumber \\
f_\pm^{\mu\nu} &=& u\,F_L^{\mu\nu}\,u^\dagger\pm 
u^\dagger F_R^{\mu\nu}\,u,
\label{uquant}
\end{eqnarray}
where $F_L$ and $F_R$ denote the field strengths of the external fields 
$l$ and $r$, such that 
\be
F_L^{\mu\nu} = 
\partial^\mu l^\nu-\partial^\nu l^\mu-i\left[l^\mu,l^\nu\right]\,.
\ee
$F_R^{\mu\nu}$ is defined analogously in terms of $r$.
All the quantities in Eq.~(\ref{defother}) transform under chiral symmetry
as
\be
\label{othertransform}
X \to h X h^\dagger\,.
\ee
In (\ref{covariantderivative}) we defined a covariant derivative that
transforms simply. For objects transforming as (\ref{othertransform})
a covariant derivative can also be defined via
\be
\nabla_\mu X = \partial_\mu X + \Gamma_\mu X -  X \Gamma_\mu\,.
\ee
$\Gamma_\mu$ is the connection
\be
\Gamma_\mu = \frac{i}{2}
\left\{ u^\dagger\left(\partial_\mu-i r_\mu\right) u+ 
u\left(\partial_\mu -i l_\mu\right)u^\dagger\right\}\,.
\ee
Using the transformations defined earlier, it can be shown that
$\nabla_\mu X$ transforms as (\ref{othertransform}) as well.
One last relation which can be checked by putting in all definitions is:
\be
f_{-\mu\nu} = \nabla_\nu u_\mu-\nabla_\mu u_\nu\,.
\ee

The most general Lagrangian of order $p^4$ after using partial integrations
and all the identities mentioned above for the case of $n_F$ flavours is
\cite{GL2,BCE1}
\ba
\label{L4nf}
{\cal L}_4 &=& \sum_{i=0}^{12} {\hat L}_i X_i + \mbox{contact terms}
\nonumber\\  
&=& {\hat L}_0\,\langle u^\mu u^\nu u_\mu u_\nu \rangle 
+{\hat L}_1\,\langle  u^\mu u_\mu \rangle^2 
+{\hat L}_2\,\langle u^\mu u^\nu \rangle \langle u_\mu u_\nu \rangle
+{\hat L}_3\,\langle (u^\mu u_\mu)^2 \rangle
+ {\hat L}_4\,\langle u^\mu u_\mu \rangle \langle \chi_+\rangle 
\nonumber\\&&
+ {\hat L}_5\,\langle u^\mu u_\mu \chi_+ \rangle 
+ {\hat L}_6\,\langle \chi_+ \rangle^2 
+ {\hat L}_7\,\langle \chi_- \rangle^2
+ \frac{{\hat L}_8}{2}\,\langle \chi_+^2 + \chi_-^2 \rangle 
- i{\hat L}_9\,\langle f_+^{\mu\nu} u_\mu u_\nu \rangle 
+ \frac{{\hat L}_{10}}{4}\,\langle f_+^2 - f_-^2 \rangle 
\nonumber\\&&
+ i{\hat L}_{11}\,\left\langle \hat\chi_-\left( \nabla^\mu u_\mu - 
\frac{i}{2} \hat\chi_- \right) \right\rangle
+ {\hat L}_{12}\,\left\langle \left( \nabla^\mu u_\mu - 
\frac{i}{2} \hat\chi_- \right)^2 \right\rangle 
+ \hat H_1\,\langle F_L^2+F_R^2\rangle 
+ \hat H_2\,\langle\chi\chi^\dagger\rangle,
\end{eqnarray}
where the definition $\hat\chi_- \equiv \chi_- -\langle 
\chi_-\rangle/n_F$ has been applied. Furthermore, the lowest 
order equation of motion is given by
\begin{equation}
X_{\mathrm{EOM}} \equiv \nabla^\mu u_\mu - \frac{i}{2}\hat\chi_- = 0.
\end{equation}

The Lagrangian of Eq.~(\ref{L4nf}) contains three types of terms. 
The last type,
the terms proportional to $\hat{H_i}$ are contact terms which 
contain external fields only. Thus they are not relevant for low-energy 
phenomenology, but they are necessary for the computation of 
operator expectation values. Their values are determined by the precise 
definition used for the QCD currents, and they are conventionally 
labeled $h_i^r$ and $H_i^r$ for unquenched $\chi$PT with $n_f = 2$ and 
$n_f = 3$ quark flavors, respectively. 

The terms containing $\hat 
L_{11}$ and $\hat L_{12}$ are proportional to the equations of motion.
They can thus always always be reabsorbed into the Lagrangians
of higher orders.
A full proof can be found in Ref.~\cite{BCE1}, App.~A.
A discussion at lower level is Ref.~\cite{FS1}. The fact that these
can be put into the Lagrangians of higher order by field redefinitions
can be qualitatively understood by the following argument. The 
variation of the lowest order Lagrangian gives the equation of motion.
Using $U = u e^{i\xi} u$ for the variation of $U$ and varying $\xi$ gives
\be
\delta{\cal L}_2 \propto \langle\xi X_{\mathrm{EOM}}\rangle\,.
\ee
A term in a Lagrangian of the form $\langle A X_{\mathrm{EOM}}\rangle$ 
can thus removed
by a field redefinition of the type
$U\to u e^{i A} u$\,. The derivation with all factors correct and worked out to
all orders can be found in App.~A of Ref.~\cite{BCE1}\,.

The physically relevant terms are the remaining ones, containing
$\hat L_i$, $i=0,\ldots,10$. Just as for $\hat F$ and $\hat B$, the $\hat L_i$
are different for every value of the number of light flavours $n_F$.

For a specific number of flavours, additional relations exist, the
Cayley-Hamilton relations. These follow from the fact that any $n$-dimensional
matrix satisfies its own characteristic equation. This is described in Sect.~3
of Ref.~\cite{BCE1}. For the $p^4$ Lagrangian for three flavours
this allows for the removal of the term proportional to $\hat L_0$ via
\be
\langle u^\mu u^\nu u_\mu u_\nu\rangle = 
-2\langle u^\mu u_\mu u^\nu u_\nu\rangle
+\frac{1}{2}\langle u_\mu u^\mu\rangle\langle u_\nu u^\nu\rangle
+\langle u^\mu u^\nu\rangle\langle u_\mu u_\nu\rangle\,.
\ee
This is the same as Eq.~(7.24) in Ref.~\cite{GL2}.
The Cayley-Hamilton for two flavours is
\be
\left\{A,B\right\} = A\langle B\rangle+B\langle A\rangle+
\langle AB\rangle -\langle A\rangle\langle B\rangle
\ee
for arbitrary $2\times2$ matrices $A,B$. An additional contact term exists
for two-flavours as well at order $p^4$ since $\det\chi$ is invariant under
chiral $SU(n_F)\times SU(n_F)$ transformations and is order $p^4$ for two
flavours. As a result there are 7 $l_i$ and 3 $h_i$ parameters at order
$p^4$ for he two-flavour case~\cite{GL1}. The correspondence with
the general number of flavours Lagrangian is via
\ba
\label{li}
l_1 &=& -2\,\hat L_0 + 4\,\hat L_1 + 2\,\hat L_3\,, 
\nonumber \\
l_2 &=& 4\,\hat L_0 + 4\,\hat L_2 \,,
\nonumber \\
l_3 &=& -8\,\hat L_4 - 4\,\hat L_5 + 16\,\hat L_6 + 8\,\hat L_8\,, 
\nonumber \\
l_4 &=& 8\,\hat L_4 + 4\,\hat L_5 \,,
\nonumber \\
l_5 &=& \hat L_{10} \,,
\nonumber \\
l_6 &=& -2\,\hat L_9 \,,
\nonumber \\
l_7 &=& -16\,\hat L_7 - 8\,\hat L_8\,.
\ea
We can for the two-flavour case write the Lagrangian in the form
(\ref{L4nf}) but only the combinations in (\ref{li}) will show up
in experimentally relevant quantities.

All the same principles apply to the construction of the Lagrangian at
order $p^6$. However, since many more combinations are possible, it
becomes much harder to find a minimal set. This was accomplished in
Ref.~\cite{BCE1} after the first attempt of Ref.~\cite{FS2}.
We will not show the full Lagrangian here, it is given in the appendices
of Ref.~\cite{BCE1}. There also all the Cayley-Hamilton relations that
were used to obtain the minimal set for the two- and three-flavour case
can be found.  In Tab.~\ref{tabLEC} the number of independent parameters
at each order in the Lagrangian is summarized for the cases relevant
in this review.

\begin{table}
\begin{center}
\begin{minipage}{16.5cm}
\caption{\label{tabLEC}
The relevant sets of LECs, where the $i+j$ notation 
denotes the number of physically relevant ($i$) and contact ($j$) 
terms in the respective Lagrangians. At NLO, the latter ones are conventionally
denoted $h_i$ for $n_f = 2$ and $H_i$ for $n_f = 3$.
}
\end{minipage}
\vspace{.3cm}
\begin{tabular}{c|c|c|c|c|c}
 & \,\,$\chi$PT\,\, & \,\,$\chi$PT\,\, & $\chi$PT 
& \,\,PQ$\chi$PT\,\, & \,\,PQ$\chi$PT\,\, \\[1mm]
$n_f$ &  2 & 3 & $n$ & 2 & 3\\ 
 & & & & & \vspace{-.2cm} \\
\hline
 & & & & & \vspace{-.2cm} \\
LO & $F,B$ & $F_0,B_0$ & $\hat F_0, \hat B$ & $F,B$ & $F_0,B_0$\\
 & & & & & \vspace{-.2cm} \\ \hline
 & & & & & \vspace{-.2cm} \\
NLO & $l_i$ & $L_i$ &$ \hat L_i$
 &$ L_i^{(2pq)}$ &$ L_i^{(3pq)}$\\
$i+j$ & 7\,+\,3 & 10\,+\,2 & 11\,+\,2 & 11\,+\,2 & 11\,+\,2 \\
 & & & & & \vspace{-.2cm} \\ \hline
 & & & & & \vspace{-.2cm} \\
\,NNLO\, & $c_i$ & $C_i$ & $K_i$ & $K_i^{(2pq)}$ & 
$K_i^{(3pq)}$\\
$i+j$ & \,53\,+\,4\, & \,90\,+\,4\, & \,112\,+\,3\, & \,112\,+\,3\, 
& \,112\,+\,3\,
\end{tabular}
\end{center}
\end{table}

The Lagrangians at order $p^6$ contain very many terms and the relevant
operators can be found in Refs.~\cite{BCE1,BCE2}.
The parameters are labeled $K_i$ for the general-$n_F$ flavour case,
$C_i$ for the three-flavour and $c_i$ for the two-flavour case.

The parameters in the Lagrangians are often referred to as low-energy
constants (LECs)
and the terms in the Lagrangians
are sometimes called counterterms even though the latter strictly only means
the additional divergent parts defined in Sect.~\ref{renormalization}\,.
We will use the terms parameters in the Lagrangian and LECs interchangeably.

\subsection{\it Renormalization in practice}
\label{renormalization}

In this review we will not treat renormalization in detail. A short overview
is given in Sect.~\ref{powercounting}. A more comprehensive discussion,
including more details, of renormalization in ChPT can be found
in Ref.~\cite{BCEGS2}. A general treatment of renormalization is the book
by Collins~\cite{Collins}.

In ChPT one uses in general dimensional regularization to regularize
the divergences that occur. The number of space time dimensions
becomes noninteger and is written as
\be
d = 4 -2\epsilon\,.
\ee
All integrals are expanded in a Laurent-series in $\epsilon$
and the divergences occur as inverse powers of $\epsilon$~\cite{dimreg}\,.
These so-called pole-terms are then absorbed by adding to the parameters
in the Lagrangian also terms divergent when $\epsilon$ is sent to zero
in such a way that the final result is finite. When one only adds terms
sufficient to precisely cancel the poles the procedure is called
minimal subtraction (MS).
But, there are several
extra pieces that always show up together with the poles~\cite{BBDM}.
These can be subtracted together by adding finite parts into the
additional terms~\cite{BBDM}, a procedure known as modified minimal subtraction
($\overline\mathrm{MS}$), corresponding to choosing $c\ne1$ in Eq.~(\ref{MSbar}) below.

To order $p^4$ at most single poles can appear and we define
\be
\label{MSbar}
\hat L_i(d) = \frac{(\mu c)^{-2\epsilon}}{(4\pi)^2}
\left\{-\frac{\hat\Gamma_i}{2\epsilon}
+\hat L_i^r(\mu,c,\epsilon)\right\}\,.
\ee
The usual choice in ChPT is~\cite{GL1}
\be
\ln c=-\frac{1}{2}\left[\ln 4\pi +\Gamma'(1)+1\right]\,.
\ee
Note that the renormalization procedure in Eq.~(\ref{MSbar}) has
also introduced a scale $\mu$. The usual choice of low-energy constants
is
\be
\hat L_i^r(\mu) = \hat L_i^r(\mu,c,0)\,.
\ee
One more subtlety is the choice of the order $\epsilon$ part in
\be
\hat L_i^r(\mu,c,\epsilon) = \hat L_i^r(\mu,c,0)+\delta_i\epsilon+\ldots\,,
\ee
in this review we choose $\delta_i\equiv0$. Different choices are possible
but the difference can always be reabsorbed in the values of the parameters
at higher orders~\cite{BCE2}.

At order $p^6$ double poles can occur and the equivalent of (\ref{MSbar})
reads~\cite{BCEGS2,BCE2}
\be
\label{MSbarp6}
K_i(d) = \frac{(c\mu)^{-4\epsilon}}{\hat F^2}
\left[ K_i^r(\mu,d)-\hat \Gamma_i^{(2)}\Lambda^2-\left(
\hat\Gamma_i^{(1)}+\hat\Gamma_i^{(L)}(\mu,d)\right)\Lambda\right]
\ee
with
\be
\Lambda = -\frac{1}{32\pi^2\epsilon}\,.
\ee
The coefficients $\hat\Gamma_i$ are known for the 2,3 and $N_F$-flavour
case~\cite{GL1,GL2,BCE2} and likewise the coefficients $\hat\Gamma_i^{(1)}$,
 $\hat\Gamma_i^{2)}$ and $\hat\Gamma_i^{(L)}$~\cite{BCE2}.
These coefficients can be calculated in general using heat-kernel techniques
and background field methods, see Ref.~\cite{JackOsborn}
for the methods and earlier references.

Note that the renormalized $K_i^r$ have been made dimensionless by the
extra factors of $\hat F$, while the $K_i$
have dimensions. This extra factor of $\hat F^2$ has not always been
treated consistently in the literature.

The renormalized coefficients obey renormalization group equations
and there are consistency conditions between the various coefficients
appearing  in (\ref{MSbar}) and (\ref{MSbarp6}). These were first discussed
by Weinberg~\cite{Weinberg} and are called Weinberg consistency conditions.
They were used in Ref.~\cite{Colangelo0,BCE3} to obtain first two-loop results.
In the notation introduced above they are
\ba
\label{renormalizationgroup}
\mu\frac{d \hat L_i^r(\mu)}{d\mu}&=& -\frac{\hat\Gamma_i}{16\pi^2}\,,
\nonumber\\
\mu\frac{d K_i^r(\mu)}{d\mu}&=&
\frac{1}{16\pi^2}\left[2\hat\Gamma_i^{(1)}+\hat\Gamma_i^{(L)}\right]\,,
\nonumber\\
\mu\frac{d\hat\Gamma_i^{(L)}(\mu)}{d\mu}&=&
-\frac{\hat\Gamma_i^{(2)}}{8\pi^2}\,.
\ea
The last equation in (\ref{renormalizationgroup}) implies that the
coefficient of the double pole can be derived from only one-loop
diagrams~\cite{Weinberg,Colangelo0,BCE3}. This has been extended to
higher orders in Ref.~\cite{Buchler1}\,.

The renormalization coefficients $\hat\Gamma_i$ needed in (\ref{MSbar})
for the cases of 2,3 and $n_F$ flavours are denoted by $\gamma_i$,
$\Gamma_i$ and $\hat\Gamma_i$. The two-flavour ones are~\cite{GL1}
\be
\gamma_1=\frac{1}{3};~\gamma_2=\frac{2}{3};~\gamma_3=-\frac{1}{2};~
\gamma_4=2;~\gamma_5=-\frac{1}{6};~\gamma_6=-\frac{1}{3};~\gamma_7=0\,.
\ee
The three-flavour ones are~\cite{GL2}
\ba
{\Gamma}_1={3 \over 32}\; ,& &
{\Gamma}_2 ={3 \over 16}\; , \nonumber\\
{\Gamma}_3={0}\; , & &{\Gamma}_4={1 \over 8}\; , \qquad 
{\Gamma}_5={3 \over 8}\; , \nonumber\\
{\Gamma}_6={11 \over 144}\; , & &{\Gamma}_7=0\; , \qquad
{\Gamma}_8={5 \over 48}\; , \nonumber\\
{\Gamma}_9={1 \over 4}\; , & &{\Gamma}_{10}=-{1 \over 4}\; \; .
\ea
while the $n_F$ flavour case is given by~\cite{GL2}
\ba
{\hat \Gamma}_0={n \over 48}\; , & &{\hat \Gamma}_1={1 \over 16}\; ,\qquad 
{\hat \Gamma}_2 ={1 \over 8}\; , \nonumber\\
{\hat \Gamma}_3={n \over 24}\; , & &{\hat \Gamma}_4={1 \over 8}\; , \qquad 
{\hat \Gamma}_5={n \over 8}\; , \nonumber\\
{\hat \Gamma}_6={n^2+2 \over 16 n^2}\; , & &{\hat \Gamma}_7=0\; , \qquad
{\hat \Gamma}_8={n^2-4 \over 16 n}\; , \nonumber\\
{\hat \Gamma}_9={n \over 12}\; , & &{\hat \Gamma}_{10}=-{n \over 12}\; \; .
\ea

For the two-flavour case numerical values are usually not quoted for $l_i^r$
at a given scale but instead in terms of the barred quantities
defined by~\cite{GL1}
\be
\label{deflbari}
\bar l_i = \frac{32\pi^2}{\gamma_i}\, l_i^r-\log\frac{M_\pi^2}{\mu^2}\,.
\ee
These are independent of the scale $\mu$.

\section{Two-flavour calculations}
\label{twoflavour}

\subsection{\it Using dispersive methods for $p^6$}
\label{dispersive}

The $S$-matrix is unitary. The $T$-matrix defined by $S=1+iT$,
thus satisfies the relation
\be
T^\dagger T = -i(T-T^\dagger)\,.
\ee
This leads to various relations between the real and imaginary part of
amplitudes. These can be brought into relations for each diagram
separately and several parts of diagrams can thus be constructed from
their imaginary parts. The latter are well defined via Cutkosky rules.
The first phenomenological applications of ChPT beyond order $p^4$
were done using methods of this type.
In Ref.~\cite{GM} this was used to determine parts of the $p^6$ corrections
to pion form-factors. The parts that can be determined this way are the
nonanalytic dependences on kinematical variables.
With kinematical variables here is meant the momentum transfer $q^2$
for form-factors, the Mandelstam variables $s,t,u$ for scattering processes
and their equivalents for other processes.

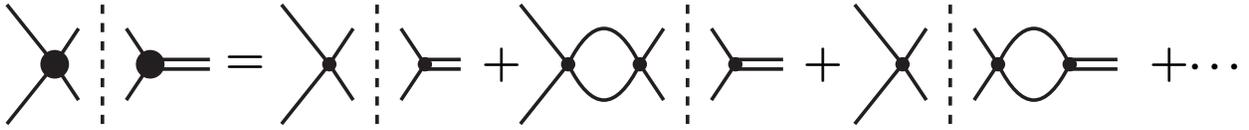
\begin{figure}
  \begin{center}
   \setlength{\unitlength}{0.9pt}
  \begin{picture}(525,50)
\SetScale{0.9}
  \SetWidth{1.5}
\Line(0,0)(20,25)
\Line(0,50)(20,25)
\Vertex(20,25){6}
\Line(20,25)(30,40)
\Line(20,25)(30,10)
\DashLine(40,0)(40,50){4}
\Line(50,40)(60,25)
\Line(50,10)(60,25)
\Vertex(60,25){6}
\Line(60,27)(85,27)
\Line(60,23)(85,23)
\Text(100,25)[]{\Large\boldmath$=$}
\Line(115,0)(135,25)
\Line(115,50)(135,25)
\Vertex(135,25){3}
\Line(135,25)(145,40)
\Line(135,25)(145,10)
\DashLine(155,0)(155,50){4}
\Line(165,40)(175,25)
\Line(165,10)(175,25)
\Vertex(175,25){3}
\Line(175,27)(190,27)
\Line(175,23)(190,23)
\Text(205,25)[]{\Large\boldmath ~$+$}
\Line(215,0)(235,25)
\Line(215,50)(235,25)
\Vertex(235,25){3}
\Curve{(235,25)(250,40)(265,25)}
\Curve{(235,25)(250,10)(265,25)}
\Vertex(265,25){3}
\Line(265,25)(275,40)
\Line(265,25)(275,10)
\DashLine(285,0)(285,50){4}
\Line(295,40)(305,25)
\Line(295,10)(305,25)
\Vertex(305,25){3}
\Line(305,27)(325,27)
\Line(305,23)(325,23)
\Text(340,25)[]{\Large\boldmath ~$+$}
\Line(355,0)(375,25)
\Line(355,50)(375,25)
\Vertex(375,25){3}
\Line(375,25)(385,40)
\Line(375,25)(385,10)
\DashLine(395,0)(395,50){4}
\Line(405,40)(415,25)
\Line(405,10)(415,25)
\Vertex(415,25){3}
\Curve{(415,25)(430,40)(445,25)}
\Curve{(415,25)(430,10)(445,25)}
\Vertex(445,25){3}
\Line(445,27)(465,27)
\Line(445,23)(465,23)
\Text(480,25)[l]{\Large\boldmath$+$\large$\cdots$}
\end{picture}  
\end{center}
\begin{center}
\begin{minipage}{16.5cm}
\caption{The imaginary part of the pion-form-factor to two-loop order.
It only needs the pion-pion scattering and the pion form-factor to
one-loop. The lines are pions, the double line is the insertion
of the external field. The thick dots indicate a general diagram,
the small dot a vertex from ${\cal L}_2$.
Adapted from Ref.~\cite{GM}}
\label{figdispersive1}
\end{minipage}
\end{center}
\end{figure}

The method works by evaluating the imaginary parts using the Cutkosky
cutting rules. The real parts can then be worked out using Cauchy's rule
when sufficient subtractions are made to make the dispersive integral
convergent. Due to the subtractions the analytical dependence
on the kinematical variables cannot be reconstructed in this way.
The principle is illustrated in Fig.~\ref{figdispersive1}.
We can see there that the knowledge of pion-pion scattering and the
form-factor to one-loop is sufficient to determine the imaginary part
of the form-factor
to two-loop order.

The methods used in \cite{GM} have been extended to other processes
as well. In particular, the structure of ChPT together with the Roy equations
was used to calculate the kinematical dependences of pion-pion scattering
to order $p^6$ in Ref.~\cite{Knechtpipi}. The observation made there,
which has since been generalized to many similar processes, is that the
imaginary
part up to order $p^6$ in scattering processes only depends on at most one
kinematical variable in a nonanalytic fashion at a time. This means that
up to order $p^6$ the amplitudes can be written in terms of single-variable
functions. This has been a very useful observation in simplifying many
of the full order $p^6$ calculations done afterwards. The same reference
also observed that all needed integrals for pion-pion scattering could
be performed analytically.

The third set of processes to be discussed fully in this way has been
the decay $\tau\to2\pi,3\pi$ in Ref.~\cite{CFU}.

\subsection{\it The process $\gamma\gamma\to\pi^0\pi^0$}
\label{ggpipi}

The first process to be fully calculated to 
two-loops
was in fact a rather difficult one. It was the process
$\gamma\gamma\to\pi^0\pi^0$. The interest in this process for ChPT
started because it was realized early on that, to order $p^4$, this
process did not depend on any of the order $p^4$ LECs~\cite{BC,DHL}.
It thus provided a clean prediction from ChPT.
This process was then measured by the Crystal Ball
Collaboration~\cite{ggpipiexp}. The overall size of the prediction
of Refs.~\cite{BC,DHL} was in good agreement with the data
but the rise with center-of-mass energy predicted by order $p^4$ ChPT
was not seen in the data.

\begin{figure}
\begin{center}
\includegraphics[width=10cm,angle=270]{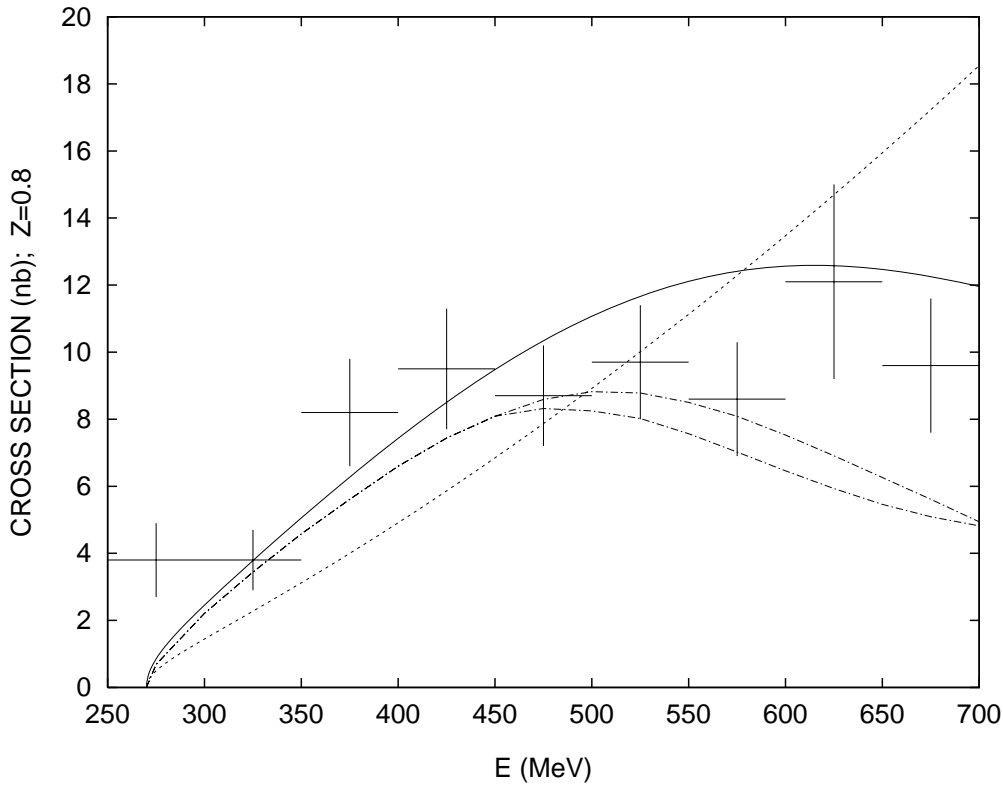}
\end{center}
\begin{center}
\begin{minipage}{16.5cm}
\caption{The cross-section for $\gamma\gamma\to\pi^0\pi^0$ with
$|\cos\theta|\le0.8$. The data points are from \cite{ggpipiexp}.
The dashed line is the order $p^4$ result from \cite{BC,DHL}. The full
line is the order $p^6$ calculation from \cite{BGS}. The dash-dotted lines
indicate the band of results from the dispersive
results\cite{MorganPennington}. The improvement of ChPT in going from $p^4$
to $p^6$ is remarkable. Figure from Ref.~\cite{BGS}.}
\label{figggpipi}
\end{minipage}
\end{center}
\end{figure}

This fact was repeatedly used to emphasize the inadequacy of ChPT,
see e.g. the discussion in Ref.~\cite{MorganPennington}. 
This prompted the authors of Ref.~\cite{BGS} to start the calculation
of this process to order $p^6$. It turned out that many of the
relevant integrals were in fact not known despite many years of
two-loop calculations in other circumstances. The necessary techniques
were developed by the authors of Ref.~\cite{GS}.
The full order $p^6$ calculation gave a significant improvement over
the $p^4$ calculation as is shown in Fig.~\ref{figggpipi} taken
from Ref.~\cite{BGS}. It should also be noted that the convergence of ChPT
is reasonable in the entire range below 600~MeV as can be seen from the
figure.

This calculation has since been redone with improved techniques for the
integrals~\cite{GIS1} where the results of the earlier calculation
have been essentially confirmed.

\subsection{\it The pion mass and decay constant}
\label{mpifpi}

The simpler observables, the pion mass and the decay constant were
calculated somewhat later. First in Ref.~\cite{Burgi1,Burgi2}
and later confirmed
by~\cite{BCEGS1,BCEGS2}. For the equal mass case, all relevant integrals
can be done explicitly. 
We quote here the result rewritten in terms of the physical
mass~\cite{BCT}.
\ba
\label{FpiF}
\frac{F_\pi}{F} &=& 1 + x_2 (l_4^r-L)
+x_2^2\Bigg[
\frac{1}{N} \left(-\frac{1}{2}l_1^r -l_2^r+\frac{29}{12}L\right)
  -\frac{13}{192}\frac{1}{N^2}
\nonumber\\&& + \frac{7}{4}k_1 + 
k_2 - 2 l_3^r l_4^r + 2(l_4^r)^2 - \frac{5}{4}k_4 + r_F^r\Bigg]+{\cal O}(x_2^3)
\, ,
\ea
and
\ba
\label{MpiM}
\frac{M_\pi^2}{M^2} &=& 1 + x_2 (2 l_3^r+\frac{1}{2}L)
+x_2^2\Bigg[\frac{1}{N}\left( l_1^r + 2 l_2^r  -\frac{13}{3}L\right)
  + \frac{163}{96}\frac{1}{N^2}
\nonumber\\&&
 -\frac{7}{2}k_1 - 2 k_2 - 4(l_3^r)^2 + 4l_3^r l_4^r - \frac{9}{4} k_3 
  + \frac{1}{4} k_4 +r_M^r\Bigg]+{\cal O}(x_2^3)\,.
\ea
The constants $r_F^r$ and $r_M^r$ denote the contributions from
the ${\cal O}(p^6)$ Lagrangian after modified minimal subtraction
and are given by~\cite{BCE2}
\ba
   r^r_F &=&   8c^r_{7} + 16c^r_{8} + 8c^r_{9}
\nonumber\\
   r^r_M &=&    - 32c^r_{6} - 16c^r_{7} - 32c^r_{8} - 16c^r_{9} 
 + 48c^r_{10} + 96c^r_{11} 
           + 32c^r_{17} + 64c^r_{18}\,. 
\ea
 
In Eqs. (\ref{FpiF}) and (\ref{MpiM}) we have used the quantities
\ba
\label{defvarious}
N&=&16\pi^2\,,\nonumber\\
x_2&=&\frac{M_\pi^2}{F_\pi^2}\,,\nonumber\\
L &=&\frac{1}{N}\log\frac{M_\pi^2}{\mu^2}\,,\nonumber\\
k_i&=& (4 l_i^r-\gamma_i L)L\,,\nonumber\\
M^2 &=& 2 B \hat{m}\,,
\ea
$M^2$ being the lowest order pion mass and $F$ the pion decay constant
in the chiral limit.
The $l_i^r$ are the finite part of the coupling constants $l_i$ in
${\cal L}_4$ after the $\overline{\mbox{MS}}$ subtraction
as given in Eq.~(\ref{MSbar}). The $k_i$ explicitly include the relations
between double logarithms and single logarithms that follows
from Weinberg's consistency conditions and were first introduced
in~\cite{BCE3}.

\subsection{\it The process $\gamma\gamma\to\pi^+\pi^-$ and polarizabilities}
\label{ggpppm}

The process $\gamma\gamma\to\pi^+\pi^-$ has been the focus
of a large amount of theoretical and experimental attention as well.
The order $p^4$ expression was worked out in Ref.~\cite{BC}
and lead to a clean prediction for the polarizabilities, see e.g.~\cite{DH}.

The agreement with the neutral pion cross-section and polarizabilities
is reasonable as discussed in~\cite{GIS1} and \cite{Filkov1},
 especially when the
order $p^6$ corrections corrections are taken into account.
References to earlier theoretical and experimental results on the
neutral pion polarizabilities can be found in those two papers.

For the charged pion polarizabilities, the comparison with experiment is
not that good. The order $p^6$ corrections were calculated in
Refs.~\cite{Burgi1,Burgi2} and were of the expected size.
With the new experiment \cite{ggpipiexp2} and dispersive derivations,
Ref.~\cite{Filkov2} and references therein, there is a significant disagreement
with the predictions of ChPT. The order $p^6$ calculation has been
redone recently~\cite{GIS2} and is essentially in agreement with the older
calculation as well.

\begin{figure}
\begin{center}
\includegraphics[width=10cm]{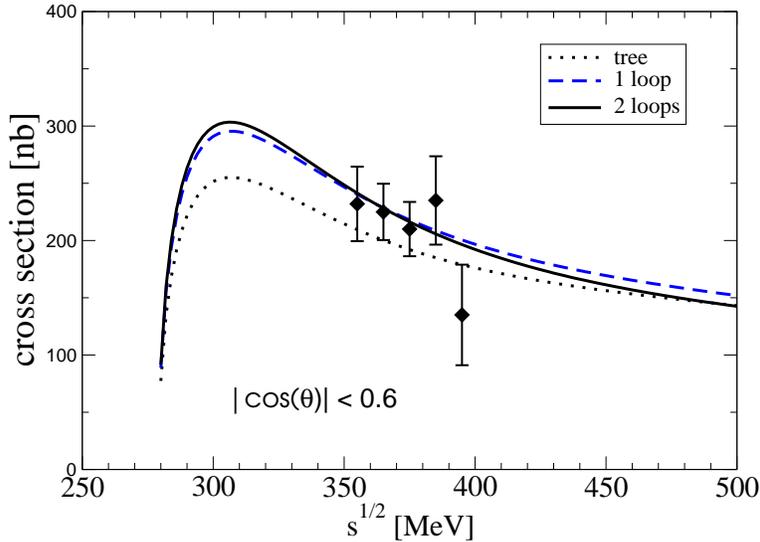}
\end{center}
\begin{center}
\begin{minipage}{16.5cm}
\caption{The cross-section of $\gamma\gamma\to\pi^+\pi^-$ with
$|\cos\theta|\le0.6$ in the regime where ChPT is fully valid.
The data are from Mark-II~+cite{dataggpipi}, the lowest order
result is equivalent to scalar QED, the order $p^4$ result
and order $p^6$ are from \cite{BC} and \cite{Burgi2,GIS2}.
Figure from Ref.~\cite{GIS2}.}
\label{figggpppm}
\end{minipage}
\end{center}
\end{figure}

The agreement with the measured cross-section data is quite good
and the ChPT series shows good convergence going from order $p^2$ to $p^4$
and $p^6$. This is shown in Fig.~\ref{figggpppm} where
the theoretical cross-section is compared with the data of
Ref.~\cite{dataggpipi}.

The values of the polarizabilities are not in good agreement with the ChPT
predictions, but there is a wide spread in the direct experimental values
and the dispersive estimates. A discussion with more references can be found
in \cite{GIS2} and \cite{Filkov2}. Here I only quote the final result
of order $p^6$ ChPT~\cite{GIS2}
\be
\left(\alpha_1-\beta_1\right)_{\pi^\pm} =
(5.7\pm1.0)\cdot 10^{-4}~\mathrm{fm}^3
\ee
and the result from the latest experiment \cite{ggpipiexp2}
\be
\left(\alpha_1-\beta_1\right)_{\pi^\pm} =
(11.6\pm1.5_\mathrm{stat}\pm3.0_\mathrm{syst}\pm0.5_\mathrm{mod})
\cdot 10^{-4}~\mathrm{fm}^3\,.
\ee
One sees that there is a clear, at present not understood, discrepancy.
The ChPT prediction is not expected to have large corrections,
it converged well and has good agreement 
with the direct data on $\gamma\gamma\to\pi^+\pi^-$.
The experiment of \cite{ggpipiexp2} has also large contributions
from the direct process $\gamma N\to \gamma N \pi$ and there might
be ununderstood effects in the separation from the pion pole contribution.

\subsection{\it Pion-pion scattering}
\label{pipi}

Pion-pion scattering has received an enormous amount of attention
both on the theoretical and experimental front. It is in some sense the
most pristine of hadronic processes, involving only the lightest
strongly interacting state. As mentioned earlier, the calculation by
Weinberg in current algebra was one of the great successes of that approach
in explaining the relative smallness compared to other strong interaction
cross-sections.

The amplitude for pion-pion scattering can be written as
\ba
\label{defAstu}
\langle \pi^d(p_4)\pi^c(p_3)\;\mbox{out}|\pi^a(p_1) \pi^b(p_2)\;\mbox{in}
\rangle &=&
\langle \pi^d(p_4)\pi^c(p_3)\;\mbox{in}|\pi^a(p_1) \pi^b(p_2)\;\mbox{in}
\rangle
\nonumber\\
&& +i
(2\pi)^4\delta^{4}(P_f-P_i)\left\{\delta^{ab}\delta^{cd}A(s,t,u)
+\mbox{cycl.} \right\}
\ea
where $s,t,u$ are the usual Mandelstam variables, expressed in units of the
physical pion mass squared $M_\pi^2$,
\be
s=(p_1+p_2)^2/M_\pi^2\,, \quad t=(p_3-p_1)^2/M_\pi^2 \,,\quad
u=(p_4-p_1)^2/M_\pi^2\label{eqmandel}\,.
\ee
Using these dimensionless quantities, the momentum expansion of the amplitude
amounts to a Taylor series in
\be
{\it x_2}=\frac{M_\pi^2}{F_\pi^2}\,,
\ee
where $F_\pi$ denotes the physical pion decay constant.

The lowest order result was found by Weinberg using current algebra methods
\cite{Weinbergpipi}, the order $p^4$ calculation was performed by
Gasser and Leutwyler \cite{GLpipi}.
The full calculation was done in \cite{BCEGS1,BCEGS2}.
The result obtained there is
 \ba 
A(s,t,u)&=& \hspace{.3cm}{\it
x_2}\left[s-1\right]
\nonumber\\
&&+{\it x_2}^2\left[b_1+b _ 2s + b_3 s^2 +
b_4 ( t - u )^2\right]\nonumber\\
&&+{\it x_2}^2\left[F^{(1)}(s) +G^{(1)}(s,t)+G^{(1)}(s,u)\right]\nonumber\\
&&+{\it x_2}^3\left[b_5s^3+b_6s(t-u)^2\right]\nonumber\\
&&+{\it x_2}^3\left[F^{(2)}(s)+G^{(2)}(s,t)
+G^{(2)}(s,u)\right]\,\!\nonumber\\
&&+O({\it x_2}^4)\,,
\label{amptot}
\ea
with
 \begin{eqnarray}
{ F^{(1)}}(\,{s}\,) &=& {\displaystyle \frac {1}{2}}\,
 \bar{J}(\,{s}\,)\,(\,{s}^{2} - 1\,)\,,\nonumber\\
{ G^{(1)}}(\,{s}, {t}) &=& {\displaystyle \frac {1}{6}}\,
 \bar{J}(\,{t}\,)\,(14\,-\,4\, s\,-\,10\, t\,+\,s\, t\,+\,2\, t^2\,)\,,\nonumber\\
{ F^{(2)}}(\,{s}\,) &=& \bar{J}(\,{s}\,) \left\{ {\vrule
height0.79em width0em depth0.79em} \right. \! \frac{1}{16 \pi^2}
\left(\! {\displaystyle
\frac {503}{108}}\,{s}^{3} - {\displaystyle \frac {929}{54}}\,{s}
^{2} + {\displaystyle \frac {887}{27}}\,{s} - {\displaystyle
\frac {140}{9}} \right) \nonumber\\
 &+&  {b_1}\,(\, 4\,{s}\, - 3) + {b_2}\,(\,
 {s}^{2} + 4\,{s}\, - 4) \nonumber\\
 &+& {\displaystyle \frac {{b_3}}{3}}\,\,(\,8\,{s}
^{3} - 21\,{s}^{2} + 48\,{s} - 32\,)
 + {\displaystyle \frac {{b_4}}{3}}\,
(\,16\,{s}^{3} - 71\,{s}^{2} + 112\,{s} - 48\,)
 \! \left. {\vrule height0.79em width0em depth0.79em}
 \right\} \nonumber\\
 &+ &\mbox{} {\displaystyle \frac {1}{18}}\,{ K_1}(\,{s}
\,)\, \left\{ \! \! \,20\,{s}^{3} - 119\,{s}^{2} + 210\,{s} - 135 -
{\displaystyle \frac {9}{16}}\,{ \pi}^{2}\,(\,{s} - 4\,)\, \!  \right\}  \nonumber\\
 &+ & \mbox{} {\displaystyle \frac {1}{32}}\,{ K_2}(\,{s}\,)\,
 \left\{ \! \,{s}\,{ \pi}^{2} - 24\, \!  \right\}  +
{\displaystyle \frac {1}{9}}\,{ K_3}(\,{s}\,)\,\left\{\,3\,{s}^{2} -
17\,{s} + 9\,\right\}\,, \nonumber\\
{ G^{(2)}}(\,{s}, {t}\,) &=& \bar{J}(\,{t}\,) \left\{
{\vrule height0.79em width0em depth0.79em} \right. \!
\frac{1}{16 \pi^2}
\left[
{\displaystyle \frac {412}{27}}\! -\! {\displaystyle \frac {s}{54}}
({t}^{2} + 5\,{t} + 159)
\! -\! t \left(\frac{267}{216}{t}^{2} - \frac{727}{108}{t} +
\frac{1571}{108} \right) \right] \nonumber\\
 &+&   {b_1}\,(2
 - {t})
+ {\displaystyle \frac {{b_2}}{3}}({t} - 4
)(2\,{t} + {s} - 5)
- {\displaystyle \frac {{b_3}}{6}}
({t} - 4)^{2}(3{t} + 2{s} - 8) \nonumber\\
&+& {\displaystyle \frac {{b_4}}{6}}\left(2{s}
(3{t} - 4)({t} - 4) - 32 t + 40t^2 - 11{t}^{3}\,\right)\!
  \left. {\vrule
height0.79em width0em depth0.79em} \right\} \mbox{} \nonumber\\
 & +&
{\displaystyle \frac {1}{36}}{ K_1}(\,{t}\,)
\left\{\,174 + 8\,{s} - 10\,{t}^{3}
 + 72\,{t}^{2} - 185\,{t} - {\displaystyle \frac {{\pi}^2}{16}}\,
(\,{t} - 4\,)\,(\,3\,{s}\! -\! 8\,)\, \!
 \right\}  \nonumber\\
 &+ & \mbox{} {\displaystyle \frac {1}{9}}\,{ K_2}(\,{t}\,)
\, \left\{ \! \,1 + 4\,{s} + {\displaystyle \frac {{\pi}^2}{64}}
\,{t}\,(\,3\,{s} - 8\,)\
\, \!  \right\}  \nonumber\\
 &+& {\displaystyle \frac {1}{9}}\,{ K_3}(\,{t}\,)\left\{
1 + 3{s}{t} - {s} + 3{t}^{2} - 9{t}\right\}
+ {\displaystyle \frac {5}{3}}\,{ K_4}(\,{t}\,)\,\left\{\,4 - 2\,{s} -
{t}\,\right\}\,.
\label{amptot1}
\end{eqnarray}
The loop functions $\bar{J}$ and $K_i$ are
\begin{eqnarray*}
\left(\begin{array}{l}\bar{J}\\
K_1\\
K_2\\
K_3\\
\end{array}\right)
=
\left(\begin{array}{cccc}
0&0&z&-4N\\
0&z&0&0\\
0&z^2&0&8\\
Nzs^{-1}&0&\pi^2(Ns)^{-1}&\pi^2\\
 \end{array} \right)
 \left(\begin{array}{c}
{h}^3\\
{h}^2\\
{h}\\
\displaystyle{-(2N^2)^{-1}}
\end{array}
\right)\,,
\end{eqnarray*}
and
\begin{eqnarray*}
K_4&=&\frac{1}{sz}\left(\frac{1}{2}K_1+\frac{1}{3}K_3+
\frac{1}{N}\bar{J}
+\frac{(\pi^2-6)s}{12N^2}\right)\,, \nonumber\\
\end{eqnarray*}
where
\[
{h}(s)=\frac{1}{N\sqrt{z}}\ln
\frac{\sqrt{z}-1}{\sqrt{z}+1} \quad ,\qquad z=1-\frac{4}{s} \; , \;
N=16\pi^2\,.
\]
The functions $s^{-1}\bar{J}$ and $s^{-1}K_i$ are analytic in the complex
$s$--plane (cut along
the positive real axis for $s \geq 4$), and they vanish
 as $|s|$ tends to infinity. Their real and
imaginary parts are continuous
at $s=4$.
The coefficients $b_i$ in the polynomial part
are given in App.~D of~\cite{BCEGS2} and their dependence on the $p^6$ LECs
can be found in \cite{BCE2}.

\begin{figure}
\begin{center}
\includegraphics[width=10cm]{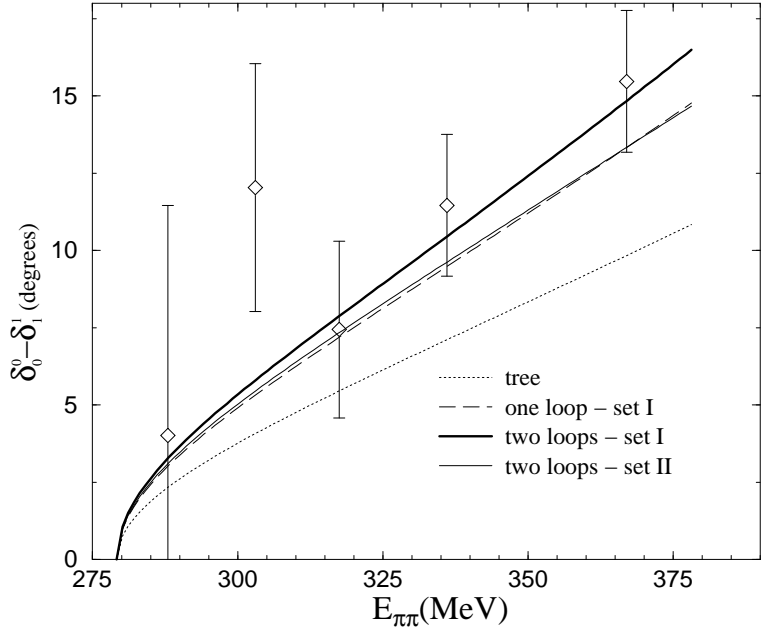}
\end{center}
\begin{center}
\begin{minipage}{16.5cm}
\caption{The pion-pion phase-shift difference $\delta_0^0-\delta_1^1$
as a function of the center-of-mass energy. Shown are the lowest order,
order $p^4$ and order $p^6$. The latter is shown for two sets of
input parameters together with the then most precise data~\cite{Rosselet}.
 Figure from Ref.~\cite{BCEGS2}.}
\label{figpipi1}
\end{minipage}
\end{center}
\end{figure}

A few comments are in order here. The result (\ref{amptot}) can be rewritten
in the form derived in \cite{Knechtpipi}. The relevant loop-integrals
can all be written in terms of elementary functions. This is a feature of
many of the results when all masses are equal but not for all.
The ChPT series in fact converges reasonably well as can be seen in
Fig.~\ref{figpipi1}.

This is a review on ChPT but some of the recent history in the theoretical
treatment of pion-pion scattering deserves mention.
The Roy equations \cite{Roy} have been reanalyzed in great detail in
Ref.~\cite{ACGL}. A similar analysis was also performed in \cite{Stern4}.
The results of this analysis have been combined with the results of
the order $p^6$ calculation mentioned here to get a series of very precise
predictions for the pion-pion scattering system in \cite{CGL2,CGL3}.
These predictions have been nicely confirmed by experiment
\cite{Pislak1,Pislak2} and \cite{DIRAC}. The relevance for the mechanism
of spontaneous symmetry breaking in QCD is discussed in \cite{CGL3}.
The case with a possible small value of the quark condensate \cite{Stern1}
is now ruled out.

The work of \cite{ACGL,CGL2,CGL3}
has been criticized in Refs.~\cite{PY1,PY2} on two grounds,
the high-energy input used as well as the fact that certain sum-rules
were not obeyed by the results of  \cite{ACGL,CGL2,CGL3}.
The criticisms in these papers have been answered in
Refs.~\cite{Caprini1,Caprini2}. They showed that the high-energy input used
in \cite{ACGL,CGL2,CGL3} also satisfied the requirements as well
as the one advocated in \cite{PY1,PY2}. They showed as well that
changing the high-energy input to the one of \cite{PY1,PY2} did not change
their result outside quoted errors. The sum-rules discussed in  \cite{PY1,PY2}
are not well convergent. Ref.~\cite{Caprini1,Caprini2} showed that more
convergent versions of the sum-rules are well satisfied by their results.

The successful prediction of pion-pion
scattering at low-energy from the combination of ChPT and Roy equations
is one of the great successes of theory in low-energy hadronic physics.

\subsection{\it Pion form-factors}
\label{piform}

The pion vector and scalar form-factors are also known analytically to
order $p^6$~\cite{BCT}. They are defined respectively by
\ba
\label{defFVFS}
\langle \pi^i(p_2)| \bar u u + \bar d d | \pi^j(p_1)\rangle &=&
 \delta^{ij} F_S(s_\pi)  \; ,
\nonumber\\
\langle \pi^i(p_2)|\frac{1}{2}\left( \bar u\gamma_\mu u - \bar d\gamma_\mu d
\right) | \pi^j(p_1)\rangle &=&
 i \varepsilon^{i3j} (p_{1\mu}+p_{2\mu}) F_V(s_\pi)\; ,
\ea
where $s_\pi = (p_2-p_1)^2$. The scalar form-factor is defined with an
isospin--zero scalar source. The isospin--one scalar form-factor can be
defined analogously but it only starts at ${\cal O}(p^4)$. 
The vector form
factor is here defined as the isovector part only,
and what we calculate here is its $I_z=0$
component. Similar definitions exist for the other isospin components.
In the limit of conserved isospin, these components are the relevant ones.
The tree level results have been long known and it is one for the
vector form-factor. The order $p^4$ result has been derived in Ref.~\cite{GL1}
and the full order $p^6$ expressions have been obtained by the
authors of Ref.~\cite{BCT}. Similar to the case of pion-pion scattering,
the expressions are fully given in terms of elementary functions.

\begin{figure}
\begin{center}
\includegraphics[width=10cm]{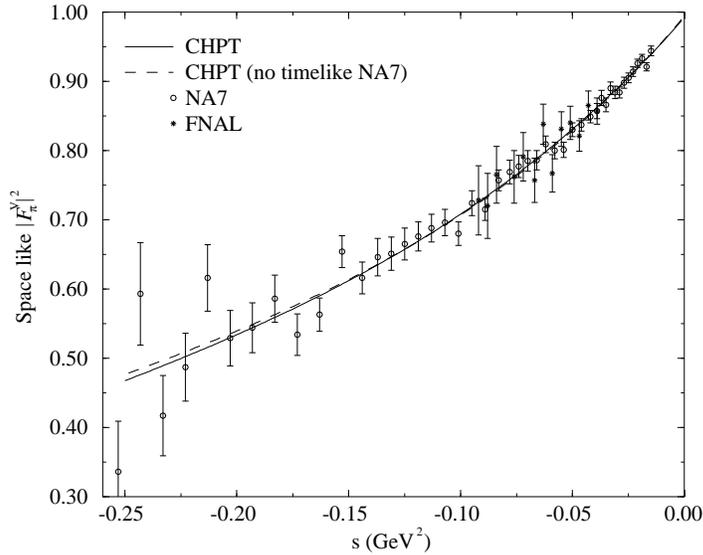}
\end{center}
\begin{center}
\begin{minipage}{16.5cm}
\caption{The spacelike vector form-factor of the charged pion and the best fit
of the ChPT order $p^6$ expression enhanced by a cubic analytic term.
Figure from Ref.~\cite{BCT}.}
\label{figFV}
\end{minipage}
\end{center}
\end{figure}

The two form-factors have a different behaviour. The vector form-factor
has relatively small corrections coming from the loop diagrams and has
the corrections dominated by the part given by the LECs of order $p^4$
and $p^6$. The fit to data is dominated by the spacelike measurements of
the pion form-factor at CERN by NA7. In Fig.~\ref{figFV} we have shown the
quality of the fit achieved. 

This fit also allowed for a good measurement of a $p^4$ and $p^6$ LEC.
The result is~\cite{BCT}
\ba
\label{valuel6}
\bar l_6 &=& 16.0\pm0.5\pm0.7\,,
\nonumber\\
r^r_{V2}(M_\rho) &=&  - 4c^r_{51} + 4c^r_{53} = (1.6\pm0.5)\cdot 10^{-4}\,.
\ea

For the pion scalar form-factor there is no direct experimental information.
The momentum dependence can be derived using dispersive methods~\cite{DGH}.
More recent determinations and discussion using the same method can
be found in \cite{Moussallam1,ABM,Caprini3}.
The result for the $p^4$ LEC that follows from a reasonable range for the
scalar radius is
\be
\label{valuel4}
\bar l_4 = 4.4\pm0.3
\ee
if one assumes a range for the scalar radius
\be
\langle r^2\rangle_S^\pi = 0.60\pm0.03~\mathrm{fm}^2\,.
\ee
An approximate value for an order $p^6$ constant could also be obtained from
this comparison, but its actual value was dependent on the other $p^4$
LECs input values used. Ref.~\cite{BCT} obtained
\be
r_{S3}^2 =  - 8c^r_{6} \approx 1.0\cdot 10^{-4}\,.
\ee

\subsection{\it $\pi\to\ell\nu\gamma$}
\label{pilnugamma}

The last full calculation in two-flavour ChPT to order $p^6$ I am aware of
is for the radiative decay of the pion in Ref.~\cite{BT1}.
The process $\pi^+\to\ell^+\nu\gamma$ at lowest order is nothing but the
QED correction to the pointlike $\pi\to\ell\nu$ decay.
First at order $p^4$ are there contributions from the pion structure
and the allow a sizable effect for $\pi^+\to e^+\nu\gamma$ since the
QED Bremsstrahlung contribution is helicity suppressed there.

The order $p^4$ contribution has been calculated in Ref.~\cite{GL1}.
The order $p^6$ calculation was added in Ref.\cite{BT1}. Again, a rather good
convergence from the order $p^4$ to the order $p^6$ result was seen.
The combination of order $p^4$ LECs that can be obtained from this calculation
is~\cite{BT1}
\be
\label{valuel5l6}
2 l_5-l_6 = 0.00315\pm0.00030\,.
\ee
This was derived from the value for the axial form-factor in the decay.
There have since been new data from the PIBETA collaboration~\cite{PIBETA}
which are compatible with the value of the axial form-factor used in \cite{BCT}
to determine (\ref{valuel5l6}).
Note that the general agreement of the data
with the distributions predicted by ChPT, or the
standard $V-A$ picture is not very good.
The variation of the form-factors
with momenta is expected to be small and the experimental
variation is significant, alternatively other form-factors can contribute.
This discrepancy is at the edge of statistical significance as discussed
in \cite{PIBETA}.

\subsection{\it Values of the low-energy constants}
\label{valueslbari}

The LECs of order $p^4$ were first determined in the original paper
\cite{GL1} using the then best values and the order $p^4$ expressions
that were derived there as well. All these quantities are now known to order
$p^6$ which in principle allows for a much more precise determination.
A problem that surfaces at this level is how to deal with the values
of the unknown order $p^6$ LECs. In all of the papers mentioned
earlier similar estimates using resonance saturation have been used.
Since quark-mass corrections in the two-flavour case are suppressed
by powers of $M_\pi^2$ the unknown parameters have typically a fairly
small effect but no general study of this has been undertaken.

The best estimate of $\bar l_1$ to $\bar l_4$ comes from the
combination of ChPT at order $p^6$ and the Roy equation analysis of
Ref.\cite{CGL3}. The other two known parameters are $\bar l_5$
and $\bar l_6$ as discussed earlier.
The best values at present are thus
\ba
\label{valueli}
\bar l_1&=&-0.4\pm 0.6\,,
\nonumber\\
\bar l_2 &=&4.3\pm0.1\,,
\nonumber\\
\bar l_3&=&2.9\pm2.4\,,
\nonumber\\
\bar l_4&=&4.4\pm0.2\,,
\nonumber\\
\bar l_6-\bar l_5 &=& 3.0\pm0.3\,,
\nonumber\\
\bar l_6 &=& 16.0\pm0.5\pm0.7\,.
\ea
The value for $\bar l_4$ is in complete agreement with the one in
Eq.~(\ref{valuel4}). The error is smaller due to the fact that the
work of \cite{CGL3} allowed to pin down some of the input better than in
\cite{BCT}. The remaining parameter $l_7$ is not known directly from
phenomenology. It can in principle be determined from lattice QCD evaluations
of the neutral pion mass in the presence of isospin breaking.
Its value was estimated to be about
\be
l_7\sim 5\cdot 10^{-3}
\ee
from $\pi^0$-$\eta$ mixing in \cite{GL1}.

Some combinations of order $p^6$ LECs are known as well.
Two have been derived from the vector and scalar form-factor of the
pion and have been given above. Ref.~\cite{CGL3} has derived some more
combinations which are implicit in the values of $b_5$ and $b_6$
quoted there.

\section{Three-flavour calculations}
\label{threeflavour}

The three-flavour calculations started soon after the first two-flavour
results had appeared. In the two-flavour case all calculations
have used the same renormalization procedure as discussed in
Sect.~\ref{renormalization} with the ChPT modified version of modified
minimal subtraction. This is unfortunately not true for the three-flavour
case where at least three different renormalization schemes have been
used. These schemes are related and can in principle be related to
each other. In practice it has made use of calculations of different
groups of authors difficult. The situation at present is that most
calculations performed by Bijnens and collaborators have been used
together with a similar scheme of analysis and the standard ChPT subtraction
scheme. Most of the other calculations have done only a rather small amount
of numerical analysis, often using a set of LECs determined at order $p^4$.

In this section I give a list of the calculations which have been
done and show a few typical numerical results. 
The section finishes with a discussion of how the order $p^6$ LECs have been
treated in the existing calculations and I also briefly discuss more recent
developments which have happened in this regard.

\subsection{\it The vector two-point functions}
\label{vectortwopoint}

The vector currents are defined as
\be
V_\mu^{ij}(x) = \overline{q}^i \gamma_\mu q^j\;,
\ee
where the indices
$i$ and $j$ run  over the  three light quark flavours,
$u$, $d$ and $s$.
Working in the isospin limit all SU(3) currents can be constructed
using isospin relations from
\ba
V_\mu^\pi(x) &=& \frac{1}{\sqrt{2}}\left(V_\mu^{11}(x)-V_\mu^{22}(x)\right)\,,
\nonumber\\
V_\mu^\eta(x) &=& \frac{1}{\sqrt{6}}\left(V_\mu^{11}(x)+V_\mu^{22}(x)
               -2V_\mu^{33}(x)\right)\,,
\nonumber\\
V_\mu^K(x) &=& V_\mu^{31}(x)\;.
\ea
These will be referred to as the isospin or pion, hypercharge or eta and
kaon vector currents respectively. All others can be defined from these
in the isospin limit.
E.g. the electromagnetic current corresponds to
\be
V_\mu^{{em}} =
\frac{e}{\sqrt{2}}V_\mu^\pi(x)+\frac{e}{\sqrt{6}}V_\mu^\eta(x)\;.
\ee
The two-point functions are defined in terms of the currents as
\be
\label{deftwopVV}
\Pi_{\mu\nu}^{Va}(q) \equiv
i\int d^4x\; e^{iq\cdot x}\;\langle 0|T(V_\mu^a(x)
V_\nu^a(0))^\dagger|0\rangle \,,
\ee
for $a=\pi,\eta,K$. All other vector two-point functions can be constructed
from these using isospin relations.
Lorentz-invariance allows to express them
in a transverse, $\Pi^{(1)}$, and a longitudinal, $\Pi^{(0)}$, part via
\be
\Pi_{\mu\nu}^{Va} = (q_\mu q_\nu-q^2 g_{\mu\nu})\Pi_{Va}^{(1)}(q^2)
 +q_\mu q_\nu \Pi_{Va}^{(0)}(q^2)\;.
\ee
For a conserved current, which is the case for the pion and eta current,
the longitudinal part vanishes because of the Ward identities.

\begin{figure}[htb]
\begin{center}
\includegraphics[width=8cm]{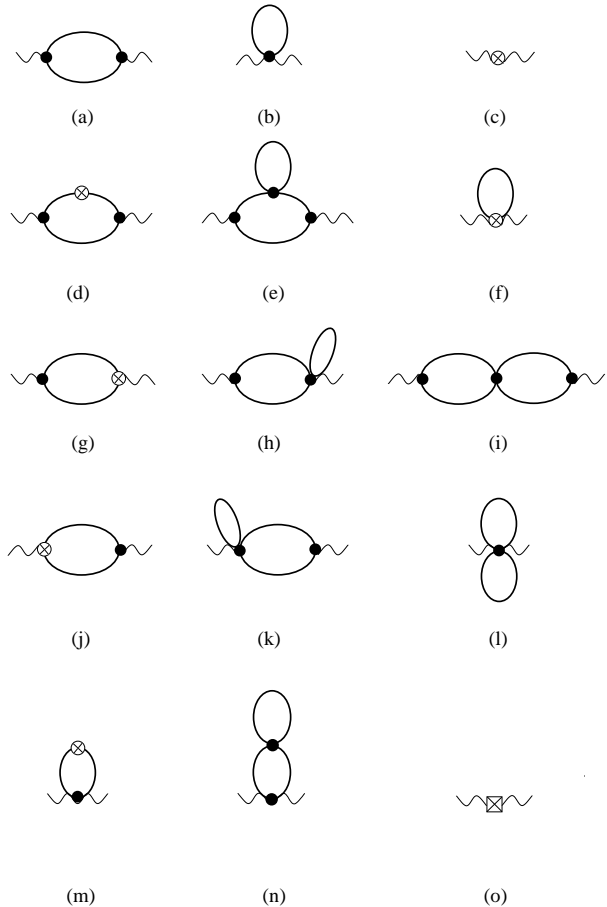}
\end{center}
\begin{center}
\begin{minipage}{16.5cm}
\caption{The Feynman diagrams needed to calculate the vector two-point
functions at order $p^6$.
The crossed circle stands for the order $p^4$ vertex insertion. Wiggly lines
are the external vector currents. Dots are order $p^2$ vertices
and a square is an order $p^6$  vertex. The solid lines are meson propagators.
Figure from Ref.~\cite{ABT1}.}
\label{figdiagsvector}
\end{minipage}
\end{center}
\end{figure}

These two-point functions were first evaluated in Ref.~\cite{GK1} for the
isospin and hypercharge, i.e. the diagonal, vector-currents.
The missing case for the vector-current with Kaon quantum numbers was
later added by Ref.~\cite{ABT1} and \cite{DK}.
The relevant diagrams to order $p^6$ are shown in Fig.~\ref{figdiagsvector}.
There is no order $p^2$ contribution, the order $p^4$ diagrams are the
first line, (a-c), and the remainder are the order $p^6$ diagrams.

There are no proper two-loop integrals needed to evaluate the vector
two-point functions to order $p^6$. They have not been used very much
for phenomenological purposes but could in principle be used as low-energy
constraints on sum-rule analyses in this channel. Some results were given
in Ref.~\cite{GK1}.
The papers \cite{ABT1} and \cite{GK1,DK} used a different subtraction scheme
so a full comparison has not been done, but the checked parts are in agreement
between the two independent evaluations.

The isospin breaking vector two-point function was calculated to order $p^6$
in Ref.~\cite{Maltman1} and were used as constraints in a sum-rule
analysis of isospin breaking in vector meson decay constants~\cite{Maltman2}.

\subsection{\it Scalar two-point functions}
\label{scalartwopoint}

The scalar two-point functions can be calculated from a similar set
of diagrams as the vector two-point functions by replacing the
insertion of the vector currents in Fig.~\ref{figdiagsvector}
by scalar currents. The main one used in this sector is
\be
\label{deftwopSS}
\Pi^S(q) \equiv
i\int d^4x\; e^{iq\cdot x}\;\langle 0|T(
(S^{uu}(x)+S^{dd}(x))
S^{ss}(0)^\dagger|0\rangle \,,
\ee
with the scalar densities $S^{ij}$ defined in Eq.~(\ref{currents}).
It was calculated to order $p^6$  by Moussallam~\cite{Moussallam2}
and used in an analysis to
obtain bounds on $L_6^r$.

\subsection{\it Quark condensates}
\label{quarkcondensate}

Another quantity which can be evaluated without proper two-loop integrals
at order $p^6$ is the quark condensate 
\be
\langle \overline q^i q^j\rangle.
\ee
 It has been evaluated in the
isospin limit in Ref.~\cite{ABT3} and the isospin breaking corrections
in Ref.~\cite{ABT4} for the three possible cases $i=j=u,d,s$.

The value of the quark condensate depends on the constant $H_1^r$ at order
$p^4$. This cannot be directly measured in any physical process. It's value
depends on the precise definition of the quark densities in QCD.
To order $p^6$ the local counterterms of that order also contribute
to $\langle \overline q^i q^j\rangle$.
The corrections when $L_4^r$ and $L_6^r$ are assumed to be zero are fairly
small as can be seen from the plots in Ref.~\cite{ABT3} for 
$\langle\overline u u\rangle$ and $\langle\overline d d\rangle$
but they were sizable for
$\langle\overline s s\rangle$.

As discussed shortly in Sect.~\ref{twoflavour}
it is now clear that the quark condensate in the two-flavour case remains large
also in the chiral limit where $m_u$ and $m_d$ are sent to zero.
The equivalent question for the three-flavour chiral limit remains open
as discussed in \cite{Moussallam1} and \cite{Stern2,Stern3}.
The situation here is analogous to the situation in the two-flavour case
before the latest results on pion-pion scattering threshold parameters,
all results seem to indicate that the standard picture as described in this
review is consistent but an alternative scenario is not ruled out.
The main remaining obstacle in the three flavour case is that the effects of
the order $p^6$ LECs have not been constrained in sufficient detail yet to see
how this question can be resolved. Work is ongoing whether the results
for $\pi\pi$ and $\pi K$ scattering can be sufficiently refined theoretically
to provide such a test.

\subsection{\it Axial-vector two-point functions, masses and decay-constants}
\label{massdecay}

These are the simplest quantities requiring proper two-loop integrals.
The axial-vector currents are defined as
\be
A_\mu^{ij}(x) = \overline{q}^i \gamma_\mu\gamma_5 q^j
\ee
where the indices
$i$ and $j$ run  over the  three light quark flavours,
$u$, $d$ and $s$.
Working in the isospin limit all SU(3) currents can be constructed
using isospin relations from
\ba
A_\mu^\pi(x) &=& \frac{1}{\sqrt{2}}\left(V_\mu^{11}(x)-V_\mu^{22}(x)\right)\,,
\nonumber\\
A_\mu^\eta(x) &=& \frac{1}{\sqrt{6}}\left(V_\mu^{11}(x)+V_\mu^{22}(x)
               -2V_\mu^{33}(x)\right)\,,
\nonumber\\
A_\mu^K(x) &=& V_\mu^{31}(x)\;.
\ea
These will be referred to as the isospin or pion, hypercharge or eta and
kaon axial-vector currents respectively. All others can be defined from these
in the isospin limit.

The two-point functions are defined in terms of the currents as
\be
\label{deftwopAA}
\Pi_{\mu\nu}^{Aa}(q) \equiv
i\int d^4x\; e^{iq\cdot x}\;\langle 0|T(A_\mu^a(x)
A_\nu^a(0))^\dagger|0\rangle \,,
\ee
for $a=\pi,\eta,K$. Using isospin relations,
all other axial-vector two-point functions
can be constructed
from these using isospin relations.
Lorentz-invariance allows to express them
in a transverse, $\Pi^{(1)}$, and a longitudinal, $\Pi^{(0)}$, part
\be
\Pi_{\mu\nu}^{Aa} = (q_\mu q_\nu-q^2 g_{\mu\nu})\Pi_{Aa}^{(1)}(q^2)
 +q_\mu q_\nu \Pi_{Aa}^{(0)}(q^2)\;.
\ee
The axial currents are only conserved in the chiral limit, i.e. when the
masses of all quarks in the currents involved vanish. Thus in general there
is both a longitudinal and a transverse part.

\begin{figure}
\begin{center}
\includegraphics[width=10cm]{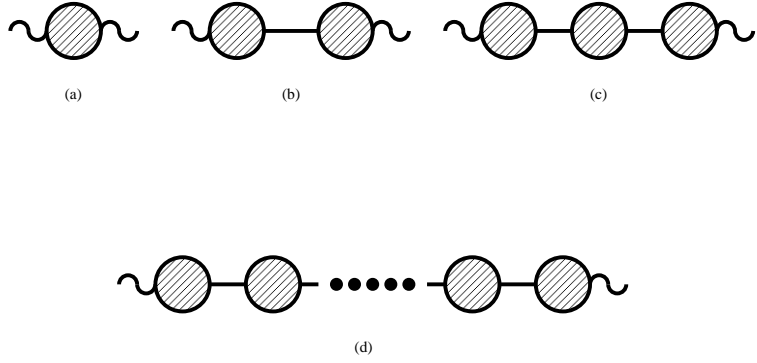}
\end{center}
\begin{center}
\begin{minipage}{16.5cm}
\caption{\label{figAA1PR} The diagrams contributing to the
axial-vector two-point function. The filled circles indicate
the one-particle-irreducible (1PI) diagrams. Solid lines are pseudoscalar meson
propagators and the wiggly lines indicate insertions of an
axial-vector current. For the inverse propagator the wiggly lines
are meson legs and for the decay constant the right wiggly line is a meson
leg while the left remains an axial current.
Figure from Ref.~\cite{ABT1}.}
\end{minipage}
\end{center}
\end{figure}

The axial-vector two-point functions have contributions from
one-particle-reducible diagrams, those where the cutting of one line makes the
diagram become disconnected. 
The full set is shown in Fig.~\ref{figAA1PR}.

\begin{figure}
\begin{center}
\includegraphics[width=8cm]{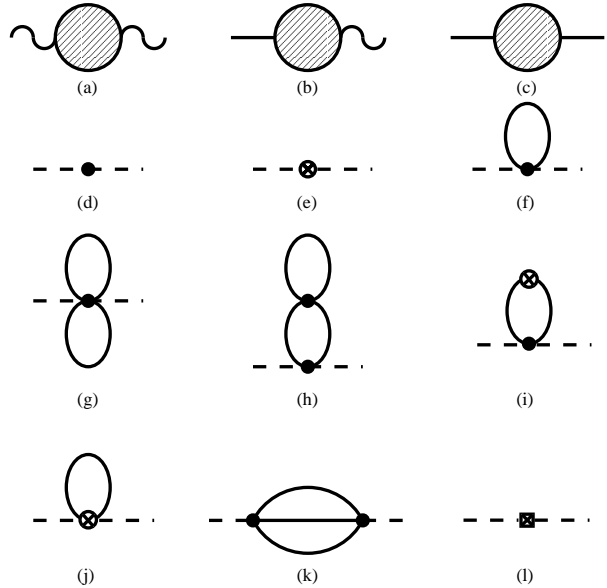}
\end{center}
\begin{center}
\begin{minipage}{16.5cm}
\caption{\label{figAA1PI} The set of diagrams contributing to the
1PI quantities. (a) axial-vector--axial-vector (b) axial-vector--pseudoscalar
(c) pseudoscalar--pseudoscalar. (d)-(l) the respective diagrams when the
dashed lines are replaced with the external legs of (a), (b) or (c).
A line is a meson propagator, 
a wiggly line an external source, a dot a vertex of
order $p^2$, a crossed circle a vertex of order $p^4$ and a crossed
box a vertex of order $p^6$. Figure from Ref.~\cite{ABT1}.}
\end{minipage}
\end{center}
\end{figure}

The filled circles in Fig.~\ref{figAA1PR} are the sum of all 1PI diagrams.
Those up to order $p^6$ are shown in Fig.~\ref{figAA1PI}.
The diagram in Fig.~\ref{figAA1PI}(k), the so-called sunset diagram is the
first diagram we encounter in this section that requires the evaluation
of proper two-loop integrals. Methods to perform these integrals
thus needed to be developed.
The methods derived in \cite{GS} need to be generalized to the case
with different masses in the loop. An efficient method to get
the sunset diagram at zero momentum was derived in Ref.~\cite{PostTausk}
using recursion relations between the various integrals. A different derivation
of the same relations was presented in Ref.~\cite{ABT1}.
The method for the momentum dependence of the sunsetintegrals of \cite{GS}
was extended to the different mass case in \cite{ABT1}. Ref.~\cite{GK2}
used a different variation on the sunset integral method of \cite{GS}
to perform their numerical analysis.

In Ref.~\cite{GK2} the axial-vector two-point function with pion and eta
quantum numbers was evaluated to order $p^6$.
They also used this to evaluate the pion
and eta masses and decay constants as well as a simple sum-rule
analysis~\cite{GK3}.
The axial-vector two-point functions were calculated also in Ref.~\cite{ABT1}
where the same quantities with kaonic quantum numbers were also evaluated
to order $p^6$. At this level all masses and decay constants were known
in the isospin limit.
Note that the masses can be calculated from the position of the pole in
the full propagator and the
decay constant by direct evaluation of the matrix element
\be
\langle 0 | A^a_\mu | M(p)\rangle = i \sqrt{2} F_M P_\mu\,.
\ee
$F_M^2$ can be determined as well from the residue of the meson pole
in the longitudinal part of the axial-vector two-point function.
That all these methods of calculating the masses and the decay constants
give the same answer was explicitly checked in Ref.~\cite{ABT1}.
 
What was found for the masses was that the order $p^4$ corrections
were small for the standard set of input parameters from fits
at order $p^4$, see Ref.~\cite{daphnereview} for this determination.
However, the order $p^6$ corrections were rather large for both
\cite{GK2} and \cite{ABT3}. It should be noted that there was analytical
agreement between those two references for the parts that could be
checked without relating the different renormalization schemes
and the different way of evaluating the integrals. At present there is no
obvious solution for the presence of these large corrections.
It is possible to choose the order $p^6$ LECs to cancel these corrections
but the LECs involved are those coming from the scalar sector and are the
most difficult ones to estimate. The naive estimates used in \cite{ABT1}
gave a very large range for the contribution from the order $p^6$ LECs,
from extremely large to zero.
In Sect.~\ref{scalarform} I will discuss a few more relevant results.

The masses and decay constants away from the isospin limit have also been
worked out. These calculations were reported in Ref.~\cite{ABT4}.

\subsection{\it $K_{\ell4}$}
\label{Kl4}

The decay $K\to\pi\pi\ell\nu$ was first treated using current algebra methods
to obtain the lowest order result.
The order $p^4$ calculation was performed by two groups
simultaneously~\cite{BijnensKl4,Riggenbach} with complete analytical agreement.
The reason for calculating this decay beyond lowest order was two-fold.
The most accurate determinations of the form-factors showed a significant
deviation from the lowest order prediction. It was expected that the order
$p^4$ calculation would allow to fit the experimental results and allow
for a more accurate determination of $L_i^r$ with $i=1,2,3$.

The corrections to the lowest order were of the expected order but not very
small. This prompted an investigation of higher orders using dispersive
methods. This was done in Ref.~\cite{BCG}. In that reference also the missing
form-factor $R$ defined below was evaluated to order $p^4$.
The higher order corrections
were of the expected order indicating a converging series for this decay.
The fit results of Ref.~\cite{BCG} have been the standard set of values
for the order $p^4$ LECs replacing the earlier full fit of Ref.~\cite{GL2}.

When the full work on pion-pion scattering in two-flavour ChPT was finished
it became interesting to check whether this calculation was compatible with the
values of the order $p^4$ LECs determined from $K_{\ell4}$.
To be able to do this $K_{\ell4}$ needed to be determined also to order $p^6$.
This calculation was performed by the authors of Ref.~\cite{ABT3,ABT2}.
I will now present the results from these references.

The $K_{\ell4}$ decay processes are
\ba
\label{processes}
K^+(p)&\to&\pi^+(p_+)\pi^-(p_-)\ell^+(p_\ell)\nu_\ell(p_\nu)\,,
\nonumber\\
K^+(p)&\to&\pi^0(p_+)\pi^0(p_-)\ell^+(p_\ell)\nu_\ell(p_\nu)\,,
\nonumber\\
K^0(p)&\to&\pi^-(p_+)\pi^0(p_-)\ell^+(p_\ell)\nu_\ell(p_\nu)\,.
\ea
The corresponding momenta are given inside the brackets after each particle.
Following
the original work \cite{cabibbo},
the $K_{\ell 4}$ decays are parameterized
in terms of five
kinematical variables:\\
i) $s_\pi$, the squared effective mass of the dipion system.\\
ii) $s_{\ell}$, the squared effective mass of the dilepton system.\\
iii) $\theta_\pi$,  the angle between the $\pi^+$ and the
  dipion line of flight with respect to the Kaon rest frame.\\
iv) $\theta_{\ell}$, the angle between the $\ell^+$ 
and the dilepton line of flight with respect to the  Kaon rest frame.\\
v) $\phi$, the angle between the $\pi-\pi$ and $e-\nu$ planes
with respect to the Kaon rest frame.

However, from the point of view of the hadronic system alone, a better
choice of variables is to use in addition to $s_\pi$,
$\cos\theta_\ell$ and $\phi$,
also
\be
t_\pi = (p_+ -p)^2 \quad \mbox{and}\quad u_\pi=(p_- -p)^2\,,
\ee
related through 
\ba
s_\pi+t_\pi+u_\pi &=& m_K^2+ 2\, m_\pi^2 + s_\ell\,,
\nonumber\\
t_\pi-u_\pi &=&-2\sigma_\pi X \cos\theta_\pi\,,
\ea
with
\ba
X &=& \frac{1}{2} \lambda^{1/2}(m_K^2,s_\pi,s_\ell)\,,
\nonumber\\
\sigma_\pi&=&\sqrt{1-4m_\pi^2/s_\pi}\,,
\nonumber\\
\lambda(m_1,m_2,m_3) &=& m_1^2+m_2^2+m_3^2 - 2\,( m_1 m_2+m_1 m_3+ m_2 m_3)\,.
\ea
With the previous notation the amplitude for the 
decay $K^+ \rightarrow \pi^+ \pi^- \ell^+ \nu_\ell$ is 
\be
\label{matrix}
T^{+-} = \frac{G_F}{\sqrt{2}} V^*_{us} \overline{u}(p_\nu) 
\gamma_\mu (1 - \gamma_5) v (p_\ell) ( V^\mu - A^\mu) \;,
\ee
with
\ba
\label{defFG}
V_\mu &=& - \frac{H}{m^3_K} \epsilon_{\mu \nu \rho \sigma} 
(p_\ell+p_\nu)^\nu (p_++p_-)^\rho (p_+-p_-)^\sigma \;,
\nonumber\\
A_\mu &=& - \frac{i}{m_K} [ (p_++p_-)_\mu \; F + 
(p_+-p_-)_\mu \; G + (p_\ell+p_\nu)_\mu \; R ]\;.
\ea
$V_{us}$ is the relevant CKM matrix-element. 
The other two amplitudes, $T^{-0}$ and $T^{00}$, are defined similarly.

Here we are interested in the $F$ and $G$ form-factors. 
The $H$ form-factor is known to ${\cal O}(p^6)$ \cite{Hanomaly} and 
the $R$ form-factor always appears with a factor of $m_\ell^2$ and 
its contribution
negligible for the electron case. $R$ is known to order $p^4$ \cite{BCG}.
The form-factors are functions of $s_\pi$, $s_\ell$ and $\cos\theta_\pi$
only, or alternatively of $s_\pi$, $t_\pi$ and $u_\pi$.

The relations between the form-factors and the intensities, easier to 
obtain from the experiment, can be found in \cite{PT} or in 
\cite{BCG}.

The amplitudes for the three processes of Eq. (\ref{processes})
are related using isospin by
\be
T^{+-} = \frac{T^{- 0}}{\sqrt{2}} + T^{0 0}
\ee
with $T^{ij}$ the matrix element defined in Eq. (\ref{matrix}).
$T^{- 0}$ is anti-symmetric under the interchange of the pion momenta while
$T^{0 0}$ is symmetric. This also implies relations between
the form-factors themselves.
Observe the different phase 
convention in the isospin states compared to \cite{BCG} where $M^{0 0}$ 
appears with a minus sign because of the Condon--Shortley phase convention.

The form-factors $F$ and $G$ can be decomposed in partial waves.
The partial wave expansion is not simply for $F$ and $G$ since the components
with a well defined $L_z=0,\pm1$ need to be expanded \cite{cabibbo,PT}.
The relevant expressions can be found in \cite{cabibbo,PT,BCG,ABT3}
and a discussion of their relative sizes and how they can be used in experiment
is in the mentioned references but more elaborated in Ref.~\cite{ABKl4}.
I will not go further into this but simply quote the experimental results.

\begin{figure}
\begin{center}
\includegraphics[width=8cm]{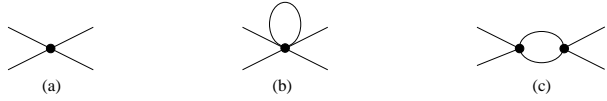}
\end{center}
\begin{center}
\begin{minipage}{16.5cm}
\caption{\label{figKl4x1loop} 
(a) One-particle irreducible tree level diagram.
(b) One-particle irreducible one-loop diagrams.
Dots refer to strong vertices or
current insertions from ${\cal L}_2$, ${\cal L}_4$ or ${\cal L}_6$.
External legs stand for
pseudoscalar or weak current. Internal lines are pseudoscalars only.
Figure from Ref.~\cite{ABT3}.}
\end{minipage}
\end{center}
\end{figure}

The lowest order calculation was done using current algebra methods
by Weinberg~\cite{weinbergKl4}.
Both the $F$ and $G$ form-factor
are equal and given by a single insertion of a ${\cal L}_2$
vertex in diagram (a) of Fig.~\ref{figKl4x1loop}. The result is
\be
F=G=\frac{m_K}{\sqrt{2} F_\pi}\,,
\ee
where $m_K$ is the physical Kaon mass. 
At higher orders more diagrams become relevant.
The tree-level and one-loop diagrams are shown in Fig.~\ref{figKl4x1loop}.
To order $p^4$ we need in addition the diagram of Fig.~\ref{figKl4x1loop}(a)
but with an order $p^4$ vertex as well as the loop diagrams (b,c) of
the same figure with only order $p^2$ vertices.
With order $p^4$ result, the experimental data at he time can be nicely fitted,
as discussed extensively in Ref.~\cite{Riggenbach} and later
in Ref.~\cite{BCG}.

\begin{figure}
\begin{center}
\includegraphics[width=10cm]{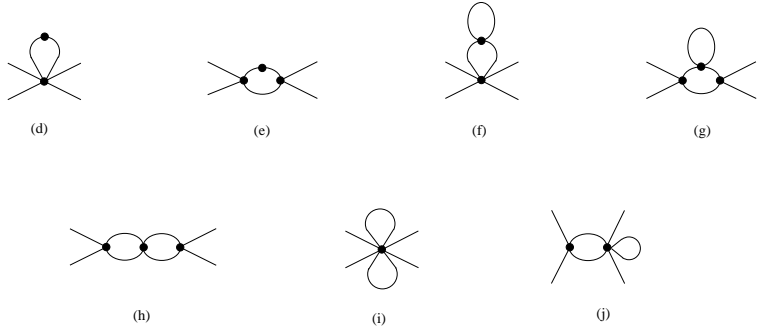}
\end{center}
\begin{center}
\begin{minipage}{16.5cm}
\caption{\label{figKl4x2loop1} 
One-particle irreducible order $p^6$ diagrams with only one-loop
integrals. The dots are order $p^2$ vertices except
the top dot in (d) and (e) which is an order $p^4$ vertex.
In addition there are the diagrams
(b) and (c) of Fig. \ref{figKl4x1loop} with one
of the dots replaced by a ${\cal O}(p^4)$ vertex.
Figure from Ref.~\cite{ABT3}.}
\end{minipage}
\end{center}
\end{figure}

The diagrams at order $p^6$ that are relevant for the calculation of the
$F$ and $G$ form-factors defined in Eq.~\ref{defFG} are shown in Figs.
\ref{figKl4x2loop1} and \ref{figKl4x2loop1} in addition to diagrams
in Fig.~\ref{figKl4x1loop} with insertions of an order $p^6$ vertex in
diagram (a) or one order $p^4$ vertex in (b,c).
The diagrams shown in Fig.~\ref{figKl4x2loop1} do not require any proper
two-loop integrals. They can be calculated using straightforward methods
and only present difficulties due to the length of the expressions.

\begin{figure}
\begin{center}
\includegraphics[width=6cm]{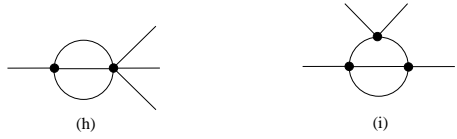}
\end{center}
\begin{center}
\begin{minipage}{16.5cm}
\caption{\label{figKl4x2loop2} 
One-particle irreducible order $p^6$ diagrams with irreducible two-loop
integrals. The dots are order $p^2$ vertices.
Figure from Ref.~\cite{ABT3}.}
\end{minipage}
\end{center}
\end{figure}

Proper two-loop integrals show up in the diagrams shown in
Fig.~\ref{figKl4x2loop2}. The left one, (a), is called the sunset-diagram
and the integrals needed are the sunsetintegrals. These were discussed
briefly in Sect.~\ref{massdecay}. The diagram on the right-hand-side, (b),
in Fig.~\ref{figKl4x2loop2} is the single most difficult diagram
in this calculation. It is called the vertex-diagram or sometimes the
fish-diagram. There have been quite a few different methods used
to evaluate this type in integrals in order $p^6$ calculations.
The one used in most calculations is based on the work of Ghinculov,
van der Bij and Yao~\cite{Ghinculov} and can be found in those references
or in Ref.~\cite{BT2}. 

In addition to the diagrams shown in 
Figs.~\ref{figKl4x1loop}-\ref{figKl4x2loop2}, there is the contribution from
wave-function renormalization from the diagrams shown in Fig.~\ref{figAA1PI}.
The order $p^6$ expressions are very long and are partly reported in a
numerical parameterization only in Ref.~\cite{ABT3}.
It was found that the parameterization based on a truncated partial wave
expansion as proposed in Ref.~\cite{ABKl4} worked very well.

\begin{figure}
\begin{center}
\includegraphics[width=8cm,angle=-90]{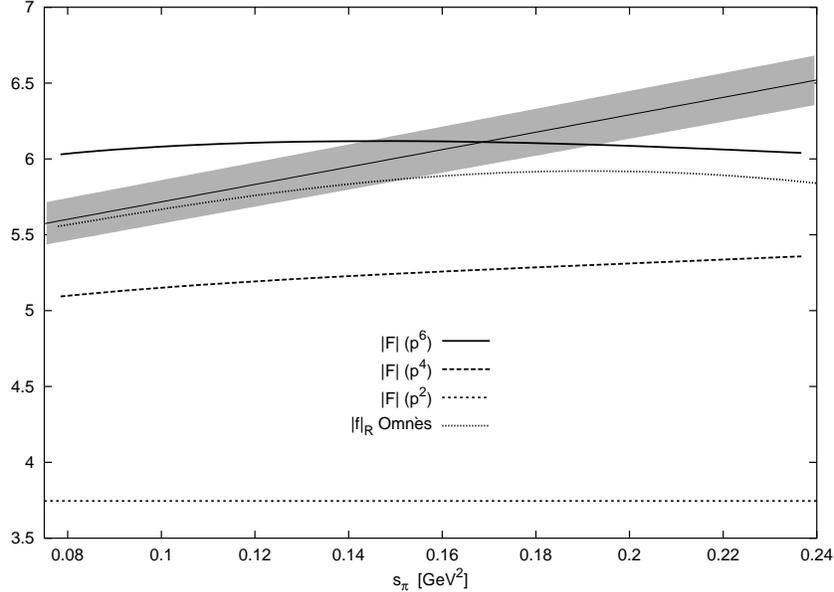}
\end{center}
\begin{center}
\begin{minipage}{16.5cm}
\caption{\label{figBCG} Comparison of the Omn\`es improved estimate
of Ref.~\cite{BCG}
with the full order $p^6$ of calculation of Ref.~\cite{ABT3}
using the same
set of values of $L_i^r$ as input together with $F_\pi=93.2$~MeV.
The shaded band is the experimental result of \cite{Rosselet}.
Figure from Ref.~\cite{ABT3}.}
\end{minipage}
\end{center}
\end{figure}

To show the results, we first compare the Omn\`es improved order $p^4$
result of Ref.~\cite{BCG} together with the data of Ref.~\cite{Rosselet}
in Fig.~\ref{figBCG}. The result was fitted to that data so there is quite
good agreement. However, when we compare the full order $p^6$ calculation we
notice that there is quite a difference with the dispersive, i.e.
the Omn\`es improved,  estimate. We also see a reasonable convergence of
$F$ going from order $p^2$ to $p^4$ to $p^6$.

\begin{figure}
\begin{center}
\begin{minipage}{8cm}
\includegraphics[angle=-90,width=0.99\textwidth]{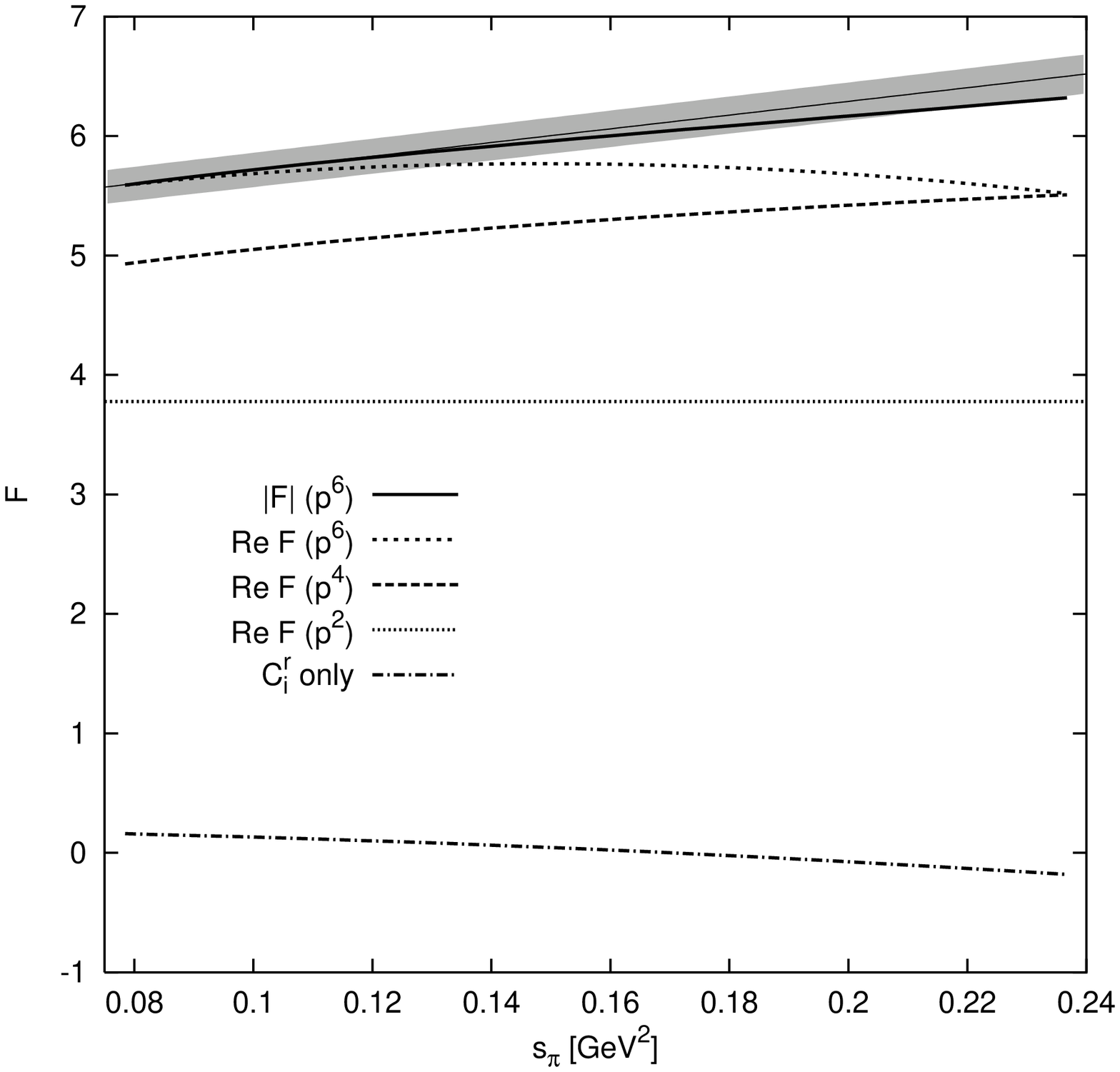}
\centerline{(a)}
\end{minipage}
\begin{minipage}{8cm}
\includegraphics[angle=-90,width=0.99\textwidth]{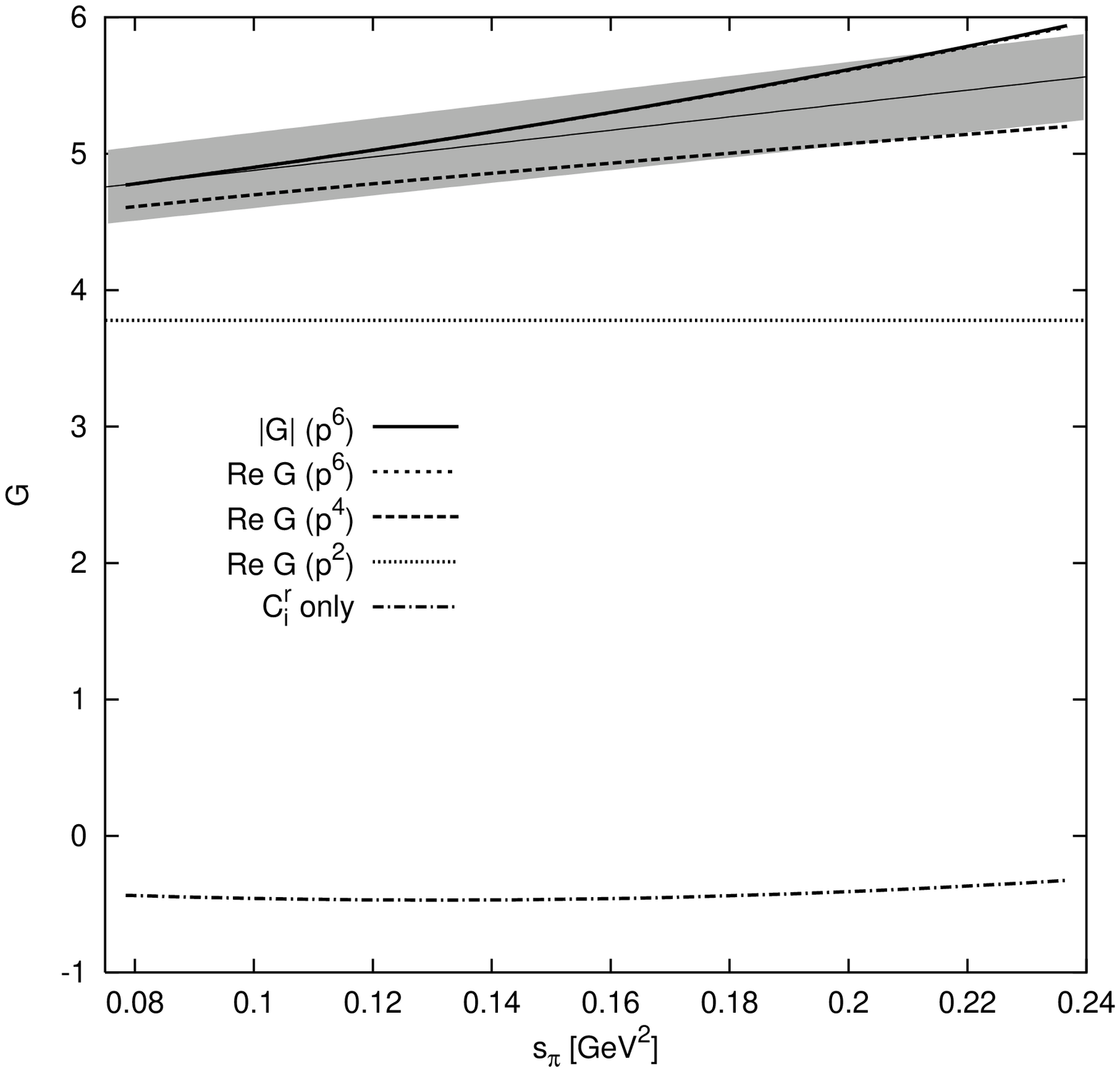}
\centerline{(a)}
\end{minipage}
\end{center}
\begin{center}
\begin{minipage}{16.5cm}
\caption{\label{figKl4FG}
The $F$ (a) and  $G$ (b) form-factor at $s_\ell =0$ and
$\cos\theta_\pi=0$.
The shaded band is the result of \cite{Rosselet}. Shown are
the full result for the absolute value, and the real part to lowest-order,
${\cal O}(p^4)$ and ${\cal O}(p^6)$. We also show the contribution from
the ${\cal O}(p^6)$ Lagrangian; this is dominated by the vector contribution.
The curves for Re~$G$ and $|G|$ are very close since Im~$G$ is very small.
Figure from Ref.~\cite{ABT3}.}
\end{minipage}
\end{center}
\end{figure}

After performing a fit to the data of Rosselet et al.~\cite{Rosselet},
the order $p^6$ result fits the data nicely.
The resulting different orders are shown in Fig.~\ref{figKl4FG}.
For both form-factors there is a good convergence.
The estimate of the contributions of the order $p^6$ Lagrangian, the curve
labeled $C_i^r$-only, is also shown. It is small for both form-factors.
The estimate for $G$ is somewhat larger, but follows from measured
resonance decays so it is fairly certain. The full description of how this
estimate has been done can be found in Refs.~\cite{BCG,ABT3}.

Another feature visible in the figures is that for $F$, which is dominated by
the $S$-wave, final state rescattering is rather important since
the contribution from the imaginary part to $|F|$ is quite visible.
In contrast, $G$ is dominated by the $P$-wave and final state rescattering
effects are fairly small.

After the work of Ref.~\cite{ABT3} was finished, new data have appeared.
The main new data from the BNL experiment~\cite{Pislak1,Pislak2}
were included in the full fit done in Ref.~\cite{ABT4}.
These data were preliminary at the time, so they were not called the main fit
in that reference. The best fit, including the newer data, is called ``fit 10''
in Ref.~\cite{ABT4} and is the one which has been used in most works following
it.

In Ref.~\cite{ABT3}, a study of the dependence on the constants $L_4^r$
and $L_6^r$ was done as well.
A main goal was to see whether good fits could be obtained with these
at values different from zero as well. It was found that such is indeed
the case, and that in order to obtain these two order $p^4$ constants,
additional experimental input was needed.
This will be discussed in the context of the scalar form-factors, pion-pion
and pion-kaon scattering below.

We also expect that new data on various $K_{\ell4}$ decays will become
available from the NA48 experiment.

\subsection{\it Vector or electromagnetic form-factors}
\label{vectorform}

The most general structure for the on-shell pseudoscalar-pseudoscalar-vector
Green function
is dictated by Lorentz
invariance. With the additional use of
charge conjugation and electromagnetic gauge invariance
one can parameterize the 
pion and kaon electromagnetic matrix elements as
\ba
\label{ff}
\langle \pi^+ (q)\vert j_\mu \vert \pi^+ (p)\rangle& =&
 (q_\mu+p_\mu) F^\pi_V(t)\,,
\nonumber \\
\langle K^+(q) \vert j_\mu \vert K^+ (p)\rangle& =& 
(q_\mu+p_\mu) F^{K^+}_V(t)\,,
\nonumber \\
\langle K^0 (q)\vert j_\mu \vert K^0 (p)\rangle& =& 
(q_\mu+p_\mu) F^{K^0}_V(t)\,,
\ea
with  $t = (q-p)^2$.
The current $j_\mu$ refers to the electromagnetic current of the light flavours
\be
j_\mu = \frac{2}{3} \left(\bar{u}\gamma_\mu u\right) -
\frac{1}{3}\left( \bar{d}\gamma_\mu d +\bar{s}\gamma_\mu s\right) \,.
\ee
The quantities $F_V^\pi, F_V^{K^0}$ and $F_V^{K^+}$
will be referred to hereafter as the vector form-factors
or simply the form-factors.
They are also defined in the crossed channel 
$ \langle 0 \vert j_\mu \vert M^a(p) M^b(-q)\rangle$. 

To lowest order the vector form-factors are constant and are simply the
charge of the relevant meson under the current under consideration.
The order $p^4$ calculation was performed by Gasser and Leutwyler in
Ref.~\cite{GL3} and confirmed for the pion form-factor in Ref.~\cite{BC}.
The full order $p^6$ calculation was performed independently by
Bijnens and Talavera~\cite{BT2} and Post and Schilcher~\cite{PS1,PS2,PS3}.
The results of Post and Schilcher for the neutral kaon form-factor
are in Ref.~\cite{PS2} and for the pion form-factor in the appendix of
Ref.~\cite{PS3}. The radius was also derived for
the charge radius of the combination of vector form-factors
which has only higher order breaking in the quark-masses as derived
by Sirlin~\cite{Sirlin1,Sirlin2} in Ref.~\cite{PS1}.
The two calculations use a very different method to calculate the
sunset and vertex integrals. In addition, Post and Schilcher have used
a slightly different subtraction scheme as well as only older values
of the order $p^4$ LECs. The analytical results which could be
compared between the two calculations agree.

\begin{figure}
\begin{center}
\includegraphics[width=10cm]{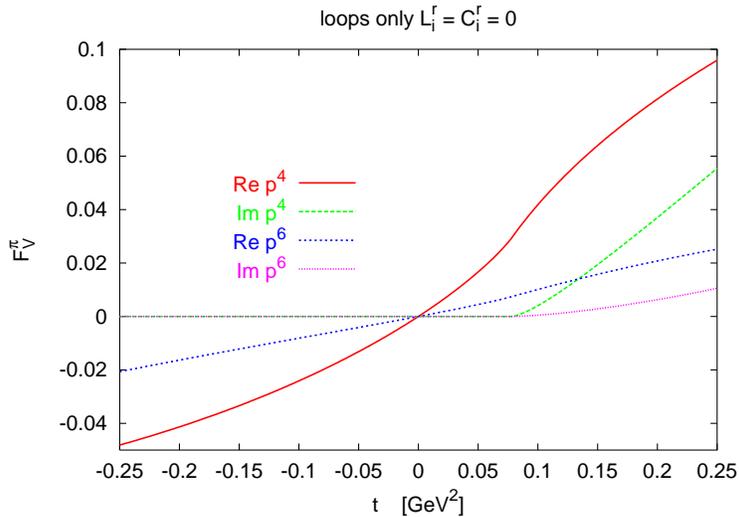}
\end{center}
\begin{center}
\begin{minipage}{16.5cm}
\caption{The real and imaginary parts of the loop diagrams at order $p^6$ and order $p^6$ with all $L_i^r=0$ and $C_i^r=0$,
for the pion form-factor. Notice that there is convergence
both for the real and the imaginary part.
Figure from Ref.~\cite{BT2}.
\label{figvectorloopsonly}}
\end{minipage}
\end{center}
\end{figure}

The numerical results for the pion vector form-factor are very similar
to those of the two-flavour calculation discussed in Sect.~\ref{piform}.
The contribution from the pure loop diagrams of order $p^4$
and order $p^6$ is shown in Fig.~\ref{figvectorloopsonly}.
They are rather small at each order and the convergence is nice. The
order $p^6$ contribution is significantly smaller than the order $p^4$
result.

\begin{figure}
\begin{center}
\includegraphics[width=10cm]{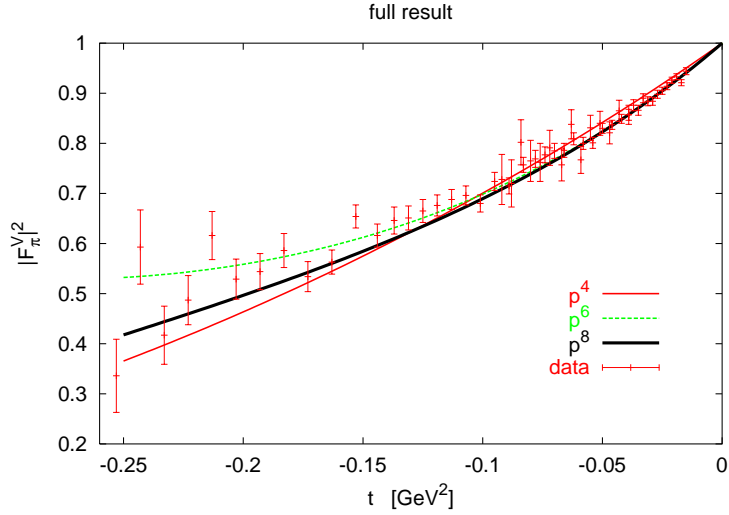}
\end{center}
\begin{center}
\begin{minipage}{16.5cm}
\caption{\label{figFVnf3} The comparison of the space-like measurements
with the ChPT calculation for the pion electromagnetic form-factor.
Notice that there is excellent convergence
over the whole kinematical range. The curve labeled $p^8$ includes
a term $c_f t^3$ in the fit in addition to the full order $p^6$ expression.
The data are from Refs.~\cite{NA7,Dallyold}.
Figure from Ref.~\cite{BT2}.}
\end{minipage}
\end{center}
\end{figure}

The calculation can now be used to fit to the available data. The fit is very
similar to the one in the two-flavour case shown in Fig.~\ref{figFV}.
The data included in the best fit as shown in Fig.~\ref{figFVnf3}.
The inclusion of time-like data does not really change the fits as discussed
in \cite{BCT} and \cite{BT2}. Figures for that part of the fits can be found
in those references. Note there seems to be some numerical problems
in the work of Ref.~\cite{PS3}. The real part of the result is compared with
the measurements in the appendix of that reference but the imaginary part
quoted there is much larger than thee one found in \cite{BT2} and way too large
to be compatible with the known pion phase-shift at that order in ChPT.

The fits of Ref.~\cite{BT2} allow to determine the parameter $L_9^r$
to order $p^6$ as
\be
\label{L9final}
L_9^r(0.77~\mbox{GeV}) = (5.93\pm0.43)\times 10^{-3}\,.
\ee
The error due to experiment only
is about half this, the remainder is mainly from the estimate of the order
$p^6$ parameters that contributes proportional to $m_\pi^2$ or $m_K^2$.
The curvature also has a contribution from an order $p^6$ constant
allowing a direct determination giving~\cite{BT2}.
\be
\label{Civectorcurvature}
R_{V2}^\pi = -4\left(C_{88}^r-C_{90}^r\right) =(0.22\pm0.02)\times10^{-3}\,.
\ee
This result is in good agreement with the various resonance estimates
discussed below.
Eqs.~(\ref{L9final}) and (\ref{Civectorcurvature})
are the direct three-flavour equivalents of the two-flavour results
of Ref.~\cite{BCT} given in Eq.~(\ref{valuel6}).

The kaon electromagnetic form-factors have also been calculated.
The data here are not so good and there seem to be some problems
with the absolute normalization. The data from
Refs.~\cite{Dallyold,Dallykaon,NA7kaon} are in agreement with the results
from ChPT within the experimental errors.

Electromagnetic form-factors are near $t=0$ often described by the
charge radius defined by
\be
\langle r^2 \rangle_V^{\pi,K^+,K^0}
= 6 \left.\frac{d}{dt} F_V^{\pi,K^+,K^0}\right|_{t=0}\,,
\ee
and the data and fits given can be used to extract/predict the various
radii. This has been done in Refs.~\cite{BCT,BT2} for the two- and
three-flavour case respectively.

The $K^0$ electromagnetic radius has been measured in a kaon regeneration
on electrons experiment~\cite{Molzon} with the result
\be
\label{rK0}
\langle r^2 \rangle_V^{K^0} = (-0.054\pm0.026)~\mbox{fm}^2\,.
\ee
The decay $K_L\to\pi^+\pi^-e^+e^-$ also has contributions
that contain the neutral kaon electromagnetic form-factor but
the extraction from the data is rather model dependent and I will not
use these data.
This radius was predicted in Refs.~\cite{PS2,BT2}, the result from
\cite{PS2} used older values of the order $p^4$ LECs. The prediction
from \cite{BT2} is
\ba
\label{rpredK0}
\langle r^2\rangle_V^{K^0} &=&\Big\{ -0.0365[\mbox{loops } p^4]
-0.0057[\mbox{loops } p^6] \Big\} ~\mbox{fm}^2
+\frac{6}{F_\pi^4} R_{V2}^{K^0} 
\nonumber\\
&=& (-0.042\pm 0.012) ~\mbox{fm}^2\,.
\ea
The error is based on {\em assuming} that the unknown contribution
is not larger than twice the order $p^6$ loop contribution.
The result is in good agreement with the measurement (\ref{rK0}). 
We can also  turn the argument around and obtain
\be
R_{V2}^{K^0} = (-0.4\pm0.9)\times 10^{-5}~{GeV}^2\,.
\ee
This result is in reasonable agreement with the resonance estimate
presented in Ref.~\cite{BT2}.

\subsection{\it $K_{\ell3}$}
\label{Kl3}

The decays considered in this subsection are
\ba
 K^+(p) &\rightarrow& \pi^0 (p') \ell^+ (p_\ell) \nu _\ell (p_\nu)\,,
 \hspace{1cm}
[K_{\ell3}^+] \label{decayKp}
\\
K^0(p) &\rightarrow &\pi^- (p') \ell^+ (p_\ell) \nu_\ell (p_\nu)\,,
\hspace{1cm}
[K_{\ell3}^0]
\label{decayKo}
\ea
and their charge conjugate modes.
$\ell$ stands for $\mu$ or $e$. The short notation for each decay is given in
the square brackets.

The matrix-element for $K_{\ell3}^+$,
neglecting scalar and tensor contributions,
has the structure
\ba
T& =& \frac{G_F} {\sqrt{2}} V_{us}^\star \ell^\mu { F_\mu}^+ (p',p)\,,
\label{s34}
\ea
with
\ba
\ell^\mu& =& \bar{u} (p_\nu)\gamma^\mu  (1- \gamma_5) v (p_\ell)\,,
\nonumber \\
{ F_\mu}^+ (p',p)& =& < \pi^0 (p') \mid V_\mu^{4-i5} (0)
\mid K^+(p)>\,,
\nonumber \\
&=& \frac{1}{\sqrt{2}} [(p'+p)_\mu f^{K^+\pi^0}_+ (t) + (p-p')_\mu
f_-^{K^+\pi^0} (t)]. \label{s35}
\ea
To obtain the $K_{\ell3}^0$ matrix-element, one replaces $F^+_\mu$ by
\ba
{F_\mu }^0 (p',p)& =& < \pi^- (p') \mid V_\mu^{4-i5} (0)
\mid K^0(p)>
\nonumber \\
&=&  (p'+p)_\mu f^{K^0\pi^-}_+ (t) + (p-p')_\mu f_-^{K^0\pi^-} (t).
\label{s35a}
\ea
 The processes (\ref{decayKp})
and (\ref{decayKo}) thus involve the four $K_{\ell3}$ form-factors
$f^{K^+\pi^0}_\pm (t)$, $f^{K^0 \pi^-}_\pm (t)$ which depend on
\be
t = (p'-p)^2 = (p_\ell + p_\nu)^2,
\label{s36}
\ee
the square of the four momentum transfer to the leptons.

Only the isospin conserving part is known to order $p^6$. The isospin breaking
corrections are one of the remaining unknowns in the determination of the
CKM matrix-element $V_{us}$.
In the isospin limit we have that
\be
f_\pm = f_\pm^{K\pi}=f_\pm^{K^+\pi^0} = f_\pm^{K^0\pi^-}\,.
\ee
$f_+^{K\pi}$ is referred to as the vector form-factor, because
it specifies the $P$-wave  projection of the crossed channel matrix-elements
 $< 0 \mid V^{4-i5}_\mu(0) \mid K^+, \pi^0 \;\mbox{in} >$.
 The $S$-wave projection is described by the scalar form-factor
\be
f_0 (t) = f_+ (t) + \frac{t}{m^2_K - m^2_\pi} f_-(t)
\,.
 \label{s37}
\ee

Analyses of $K_{\ell3}$ data frequently assume a linear dependence
\be
f_{+,0} (t) = f_+ (0) \left[ 1 + \lambda_{+,0}
\frac{t}{m^2_{\pi^+}} \right] \; \; .
\label{s38}
\ee
For an early discussion of the validity of this approximation see \cite{GL2}
and references cited therein.
As pointed out in Ref.~\cite{BT3}
and checked afterwards by the ISTRA+~\cite{ISTRA}, KTeV~\cite{KTeV}
and NA48~\cite{NA48}, the linear approximation is not sufficient at
the present level of precision.

The form-factors $f_{\pm,0} (t)$ are analytic functions in the complex
$t$-plane cut along the positive real axis. The cut starts at $t=(m_K +
m_\pi)^2$. In the phase convention used here,
the form-factors are real in the
physical region
\be
m^2_\ell \leq t \leq (m_K - m_\pi)^2.
\label{s39}
\ee
A discussion of the kinematics in $K_{\ell3}$ decays can be found
in \cite{daphnereview} and references cited therein.

The total result can be split by chiral order
\be
f_i(t) = f_i^{(2)}(t)+f_i^{(4)}(t)
+f_i^{(6)}(t)\,, \quad (i=+,-,0)\,.
\ee

The lowest order result has been known for a very long time and is fully
determined by gauge invariance.
\be
f_+^{(2)}(t) = f_0^{(2)}(t) = 1,,\quad\quad\quad
f_-^{(2)}(t) = 0\,.
\ee
The order $p^4$ contribution was first calculated within the ChPT framework
by Gasser and Leutwyler~\cite{GL3} with earlier results by Leutwyler
and Roos~\cite{LR}.
The result contains the nonanalytic dependence in the symmetry parameters
predicted by \cite{DashenLi}.
Partial studies at order $p^6$ have also been done,
the double logarithm contribution
is small as was shown in Ref.~\cite{BCE3}
and a possibly large role for terms with two
powers of quark masses has been argued for in Ref.~\cite{FKS}.
The latter reference also pointed out the strong interdependence of the
measurements of $F_K$ and $V_{us}$ as a possible solution to the CKM unitarity
problem. 

There exist two full calculations at order $p^6$.
The work of Post and Schilcher~\cite{PS3} and Bijnens and Talavera~\cite{BT3}.
The first work uses outdated values of the ChPT constants as well as an older
version of the classification of $p^6$ constants. Due to the different methods
of subtraction and splitting up the various proper two-loop integrals in an
analytical and numerical part, a full analytical comparison between the two
results has not been done. The parts which can be compared are in agreement
analytically. The numerical results of both papers are in agreement for $f_+$
when the same input values for all parameters are used. There is a disagreement
for $f_-$ but as discussed in Sect.~\ref{vectorform} there are some indications
for numerical errors in Ref.~\cite{PS3}.

Let me now discuss the main results of Ref.~\cite{BT3}. I will present
some of their numerical results later but first discuss the model-independent
relations at order $p^6$.
There are a large
number of order $p^6$ LECs contributing to both form-factors,
their contributions to the processes discussed in the previous, this one
and the next subsection satisfy several relations.
The combination which is relevant for the curvature of $f_+(t)$
is the same combination that appears in the curvature of the pion
electromagnetic form-factor. Its value has been determined from the data
and is given in Eq.~(\ref{Civectorcurvature}). That way Ref.~\cite{BT3}
predicted the curvature in $f_+(t)$.
In a similar way, the curvature of the scalar form-factor, $f_0(t)$,
is determined by the order $p^6$ constants $C_{12}^r$. Its value can in
principle also be determined from the curvature of the pion scalar form-factor
\cite{BD}, Sect. \ref{scalarform}
There are also relations between the order $p^6$ constants contributing
to the various charge radii. This relation is essentially the Sirlin relation
of \cite{Sirlin1,Sirlin2}.

A more surprising relation that connects the values of $f_+(0)$ with the slope
and curvature of the scalar form-factor $f_0(t)$, was discovered in
Ref.~\cite{BT3}.
One constructs the quantity
\be
\label{deftildef0}
\tilde f_0(t) = f_+(t)+\frac{t}{m_K^2-m_\pi^2}
\left(f_-(t)+1-F_K/F_\pi\right)
= f_0(t)+\frac{t}{m_K^2-m_\pi^2}\left(1-F_K/F_\pi\right)
\,.
\ee
This has no dependence on the $L_i^r$ at order $p^4$, only via
order $p^6$ contributions. Inspection of the dependence on the $C_i^r$ shows
that
\ba
\label{resultfp0}
\tilde f_0(t) &=& 1-\frac{8}{F_\pi^4}\left(C_{12}^r+C_{34}^r\right)
\left(m_K^2-m_\pi^2\right)^2
+8\frac{t}{F_\pi^4}\left(2C_{12}^r+C_{34}^r\right)\left(m_K^2+m_\pi^2\right)
\nonumber\\&&
-\frac{8}{F_\pi^4} t^2 C_{12}^r
+\overline\Delta(t)+\Delta(0)\,.
\ea
It should be emphasized that the quantities $\overline\Delta(t)$ and
$\Delta(0)$ can in principle be calculated
to order $p^6$ accuracy with knowledge of the $L_i^r$ to order $p^4$
accuracy. In practice, since a $p^4$ fit will include in the values of the
$L_i^r$ effects that come from the $p^6$ loops (due to the fitting to
experimental values)
we consider the $p^6$
fits to be the relevant ones to avoid double counting effects.

The definition in (\ref{deftildef0}) has essentially used the
Dashen-Weinstein
relation \cite{DashenWeinstein}
to remove the $L_i^r$ dependence at order $p^4$. It has also the
effect that it removed many of the $C_i^r$ from the scalar form-factor
as well. The corrections which appear in the Dashen-Weinstein relation
are include in the functions $\overline\Delta(t)$ and $\Delta(0)$,
these have both order $p^4$ \cite{GL3,DashenLi} and order $p^6$
contributions.

It is obvious from Eq.~(\ref{resultfp0}) that the needed combination
of $C_i^r$ can be determined from the slope and the
curvature of the scalar form-factor in $K_{\ell3}$ decays.

It seems possible that
$C_{12}^r$ can be measured from the curvature of the pion scalar form-factor
near 0 \cite{BD}. With this calculation is complete, one can use the
dispersive estimates of the pion scalar form-factor together with only a
$\lambda_0$ measurement in $K_{\mu3}$ to obtain the $p^6$ value for
$f_+(0)$. There are also some dispersive estimates for the relevant
scalar form-factor. Unfortunately, these were not in a usable form
 \cite{Pich2} when Ref.~\cite{BT3} appeared,
but have since been treated in Ref.~\cite{Pich3}.
Other estimates of the relevant constants are the one used in
\cite{BT3} coming from Ref.~\cite{LR} and the resonance chiral theory
result of \cite{Reso2}.

One feature that is visible in Eq.~\ref{resultfp0} is that the value
of $f_+(0)$ only differs from 1 by terms of order $(m_K^2-m_\pi^2)^2$.
This is not only true for the analytic contributions written out explicitly
but also for the entire expressions for $f_+(0)$. This is known as the
Ademollo-Gatto theorem~\cite{ademollo} and holds to all orders in ChPT.

\begin{figure}
\begin{center}
\includegraphics[angle=270,width=12cm]{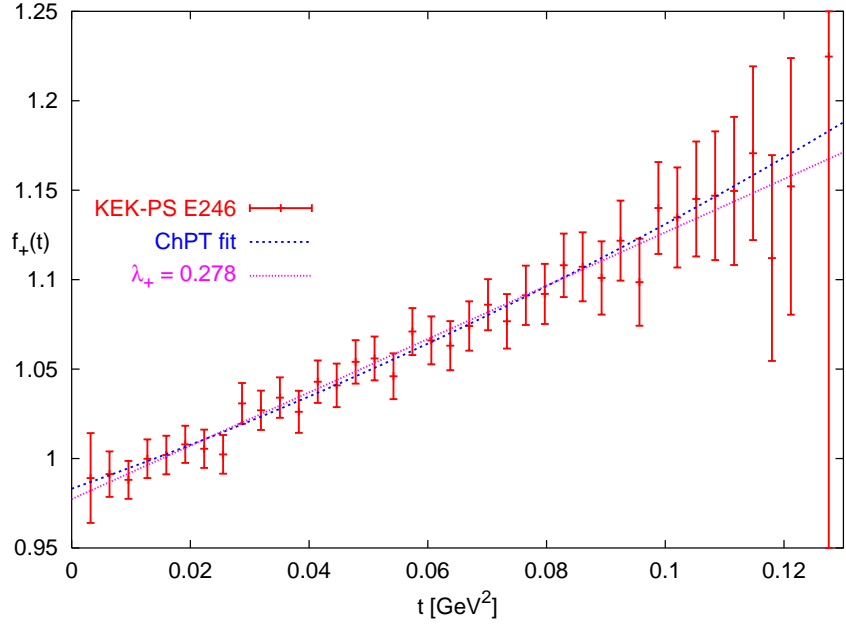}
\end{center}
\begin{center}
\begin{minipage}{16.5cm}
\caption{The KEK-PS E246 data together with the ChPT result,
with the linear coefficient fitted and curvature predicted from ChPT
as well as the linear fit of the KEK-PS collaboration.}
\label{fig:PSE246fit}
\end{minipage}
\end{center}
\end{figure}

Let us now discuss the adequacy of the linear parameterization for the
form-factors. In Ref.~\cite{BT3} it was pointed out the measured value
of a form-factor at zero and the slope are rather dependent on the curvature,
even if the data themselves do not show the presence of curvature.
The fitted value of the slope and the form-factor at zero can change
significantly outside the quoted errors due to this.
This was shown in Ref.~\cite{BT3} for the case of the then best
available data in both the neutral decay channel, CPLEAR~\cite{CPLEAR}
and charged decay channel, KEK-PS E246~\cite{KEK-E246}.
This is shown in Fig.~\ref{fig:PSE246fit}
for the KEK-PS E246 data. The discussion for the CPLEAR data can be found
in Ref.~\cite{BT3}.
The size of this effect is fairly small for the KEK-PS data but much larger
for the CPLEAR data. It is very important to get data at small $t$ to minimize
this effect as far as possible.

Since Ref.~\cite{BT3} appeared there have been newer high precisions
data from the ISTRA+~\cite{ISTRA}, KTeV~\cite{KTeV} and NA48~\cite{NA48}
collaborations. The last two have clearly demonstrated the existence of the
curvature experimentally but there are some disagreements in their results.
The question above is relevant now at a higher level
of precision. How much does the presence of a $t^3$ term affect the 
latest high precision fits?

A recent reference discussing the various results and impact
on the determination of $V_{us}$ is \cite{Reso2}. The final theoretical error
has been significantly improved since the work of Leutwyler and
Roos~\cite{LR} and is now roughly one percent. The bulk of this comes from
the uncertainty in the needed order $p^6$ constants. More theoretical
work on dispersion relations in the scalar form-factor together with
experimental studies of the scalar form-factor together with the
relation discussed above should allow to pin down this contribution to better
precision. The uncertainty on the theory side is definitely smaller than the
violation of CKM unitarity when taking $V_{ud}$ from neutron beta decay.

\subsection{\it Scalar form-factors}
\label{scalarform}

The scalar form-factors for the pions and kaons are defined as follows:
\be
\label{ffdef}
\langle M_2(p)|\bar{q}_i q_j|M_1(q) \rangle  =  F^{M_1 M_2}_{ij}(t) 
\ee
with $t=(p-q)^2$ and $i,j=u,d,s$.
$M_1$, $M_2$ are meson states with the indicated momentum.

In the case of isospin symmetry $m_u=m_d=\hat{m}$, the various pion
scalar form-factors obey
\ba
\label{isopions}
F_{S}^\pi(t) &\equiv&  
        2 F^{\pi^0\pi^0}_{uu}(t) =
        2 F^{\pi^0\pi^0}_{dd}(t) =
        2 F^{\pi^+\pi^+}_{uu}(t) =
        2 F^{\pi^+\pi^+}_{dd}(t) 
\nonumber\\
&=&
-2\sqrt{2}F^{\pi^0\pi^+}_{du}(t) =
 2\sqrt{2}F^{\pi^0\pi^-}_{ud}(t)\,,
\nonumber\\
F_{Ss}^\pi &\equiv & F^{\pi^+\pi^+}_{ss} = F^{\pi^0\pi^0}_{ss}\,.
\ea
The kaon currents are related by the rotations in flavour space
\ba
\label{isokaons}
F_{Su}^K(t) &\equiv& F_{uu}^{K^+K^+(t)} =  F_{dd}^{K^0K^0}(t)\,,
\nonumber\\
F_{Sd}^K(t) &\equiv& F_{dd}^{K^+K^+}(t) =  F_{uu}^{K^0K^0}(t)\,,
\nonumber\\
F_{Ss}^K(t) &\equiv& F_{ss}^{K^+K^+}(t) =  F_{ss}^{K^0K^0}(t)\,,
\nonumber\\
F_{S}^{K\pi}(t) &\equiv& F_{su}^{K^0\pi^-}(t)
 =  \sqrt{2} F_{su}^{K^+\pi^0}(t)\,,
\nonumber\\
F_{Sq}^K(t) & \equiv&F_{Su}^K(t)+F_{Sd}^K(t)\,.
\ea
The other scalar form-factors can be obtained from the above
using charge conjugation
and time reversal. Eqs. (\ref{isopions}) and (\ref{isokaons}) 
also show the notation used for the form-factors in the remainder of this
subsection.

The scalar form-factor $F_{S}^{K\pi}(t)$ is proportional to the form-factor
$f_0(t)$ used in $K_{\ell3}$ decays, Sect.~\ref{Kl3}.

The scalar form-factors obey a relation similar to
the Sirlin \cite{Sirlin1,Sirlin2} relation for the vector form-factor~\cite{BD}
\be
\label{eq:relation}
F_{S}^\pi(t)-2 F_{Ss}^\pi(t)-2 F_{Sd}^K(t)+2 F_{Ss}^K(t)-4 F_{S}^{K\pi}(t)
= \mathcal{O}\left((m_s-\hat m)^2\right)\,.
\ee
The proof of this relation is in App.~A of Ref.~\cite{BD}
and uses a method similar
to the one given in Ref.~\cite{Sirlin2}. The data at present do not allow
to test this result.

The values at zero momentum transfer are related to the derivatives of the
masses w.r.t. to quark masses because of the Feynman-Hellman theorem
(see \cite{GL3})
\ba
F_{S}^\pi(0) =  \frac{\partial}{\partial\hat m}m_\pi^2\,,&&
F_{Ss}^\pi(0) =  \frac{\partial}{\partial m_s}m_\pi^2\,,
\nonumber\\
F_{Su}^K(0) = \frac{\partial}{\partial m_u}m_K^2\,,&&
F_{Ss}^K(0) =  \frac{\partial}{\partial m_s}m_K^2\,,
\nonumber\\
F_{Sd}^K(0) = \frac{\partial}{\partial m_d}m_K^2\,.
\ea
The masses are known to order $p^6$ as discussed in Sect.~\ref{massdecay}
and there were sizable corrections at order $p^6$.
This also shows that large corrections at $t=0$ can be expected. The argument
goes as follows. If
\be
m_K^2 \approx B_0 m_s + \beta (B_0 m_s)^2 + \gamma (B_0 m_s)^3\,,
\ee
then
\be
F_{Ss}^K(0) \approx B_0 + 2 \beta B_0 m_s + 3 \gamma (B_0 m_s)^2\,.
\ee
So we see that in the scalar form-factors the relative $p^6$ corrections can
get enhanced by factors of order 3 compared to the masses.

\begin{figure}
\begin{minipage}{0.49\textwidth}
\includegraphics[height=0.99\textwidth,angle=-90]{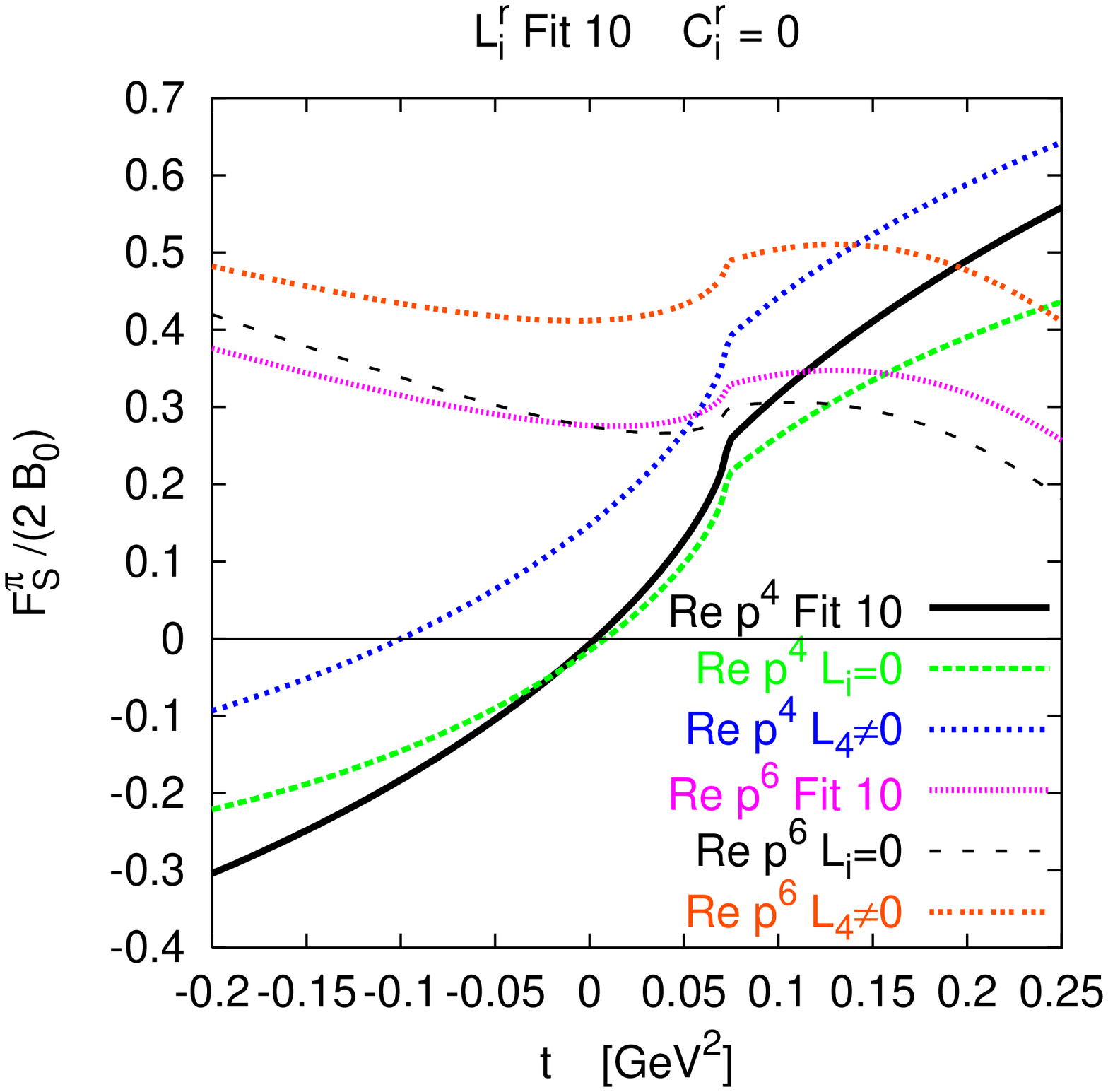}
\centerline{(a)}
\end{minipage}
\begin{minipage}{0.49\textwidth}
\includegraphics[height=0.99\textwidth,angle=-90]{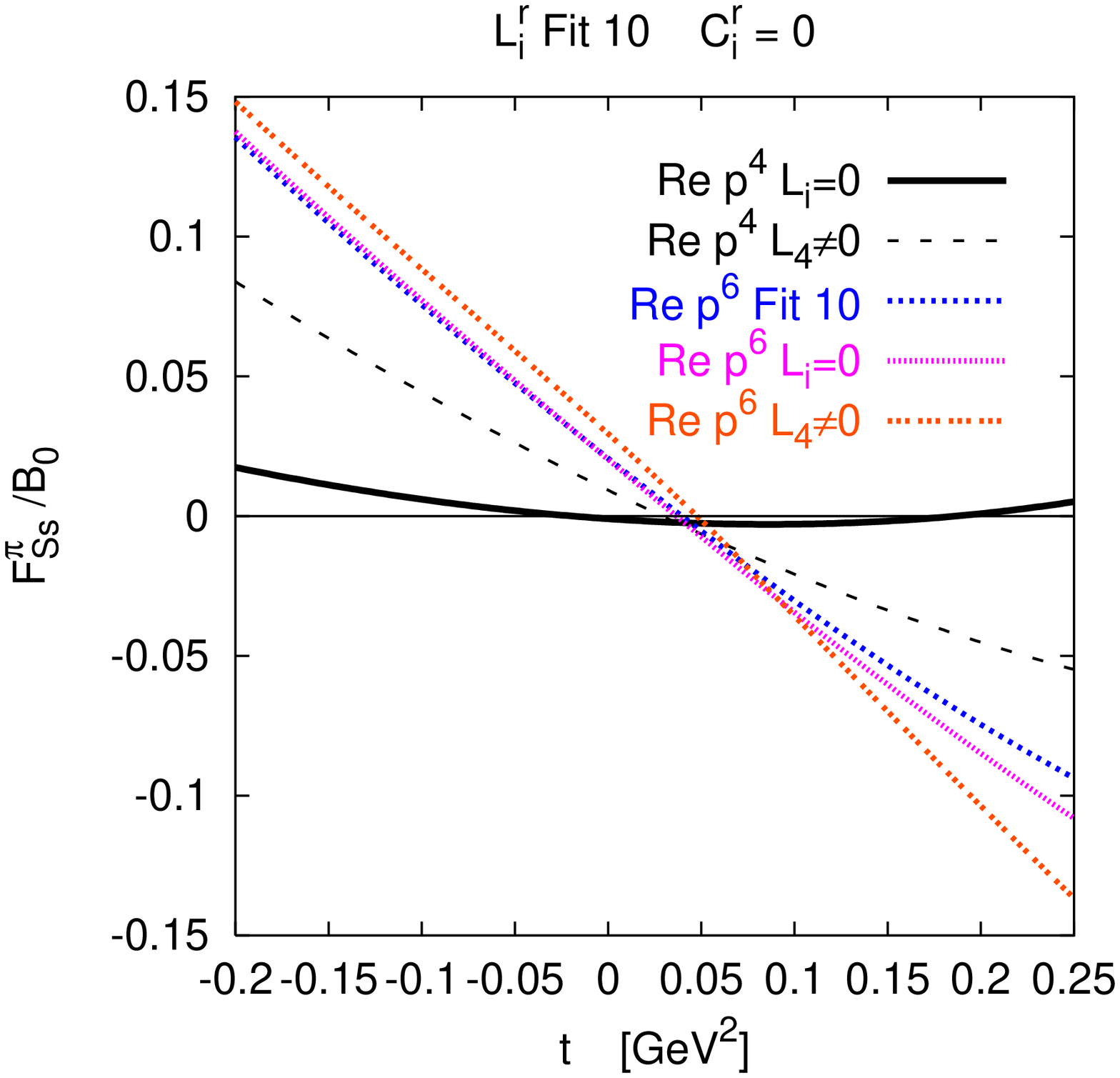}
\centerline{(b)}
\end{minipage}
\begin{center}
\begin{minipage}{16.5cm}
\caption{The effects of the $L_i^r$ on the various contributions to 
(a) $F^\pi_{S}(t)/(2B_0)$
and (b) $F^\pi_{Ss}(t)/B_0$
as a function of $t$ for the case with $C_i^r=0$. 
The curves are for the real parts
at order $p^4$ and order $p^6$. The curves labeled respectively
Fit 10, $L_4^r\ne0$ and $L_i^r= 0$ are for the
standard values of the $L_i^r$ of fit 10 in~\cite{ABT4}, the same values but 
$L_4^r = -0.003$ and with all $L_i^r=0$.
Figure from Ref.~\cite{BD}.}
\label{figpiqLi}
\end{minipage}
\end{center}
\end{figure}

The lowest order results have been long known. Order $p^4$ results were
obtained in \cite{GL3,Meissner1}. 
The actual calculations of the scalar form-factor to order $p^6$
in three-flavour ChPT were
performed by the authors of Ref.~\cite{BD}. As expected from the argument just
presented, large corrections were found to several of the form-factors
involved, especially for the kaon ones. The results are also quite sensitive
to the input values used for the LECs. In Fig.~\ref{figpiqLi}
the contributions to the two pion scalar form-factors at order $p^4$
and $p^6$ are shown. The are shown for the values of fit~10,
one with a changed value for $L_4^r$ and the pure loop parts (all $L_i^r=0$).
It is clear that while for these values the corrections are not enormous,
they nonetheless are large and show no obvious convergence
in that for many $t$ the order $p^6$ contribution is larger than order $p^4$.

\begin{figure}
\begin{center}
\includegraphics[height=10cm,angle=-90]{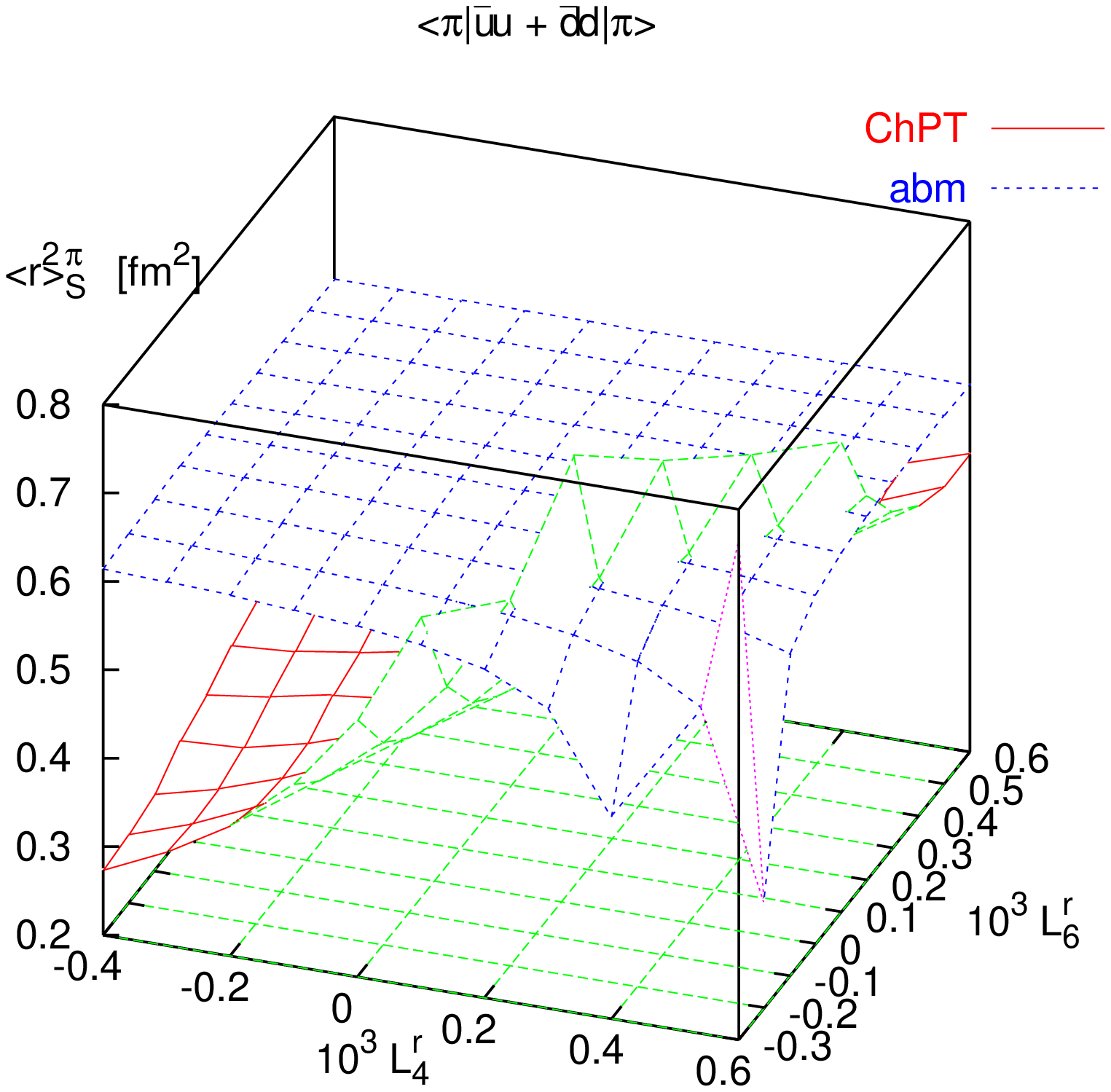}
\end{center}
\begin{center}
\begin{minipage}{16.5cm}
\caption{The result for the scalar radius $<r^2>_\pi^S$
as a function of  $L_4^r$ and $L_6^r$, with the dispersive estimate (abm)
and the ChPT calculation. Figure from Ref.~\cite{BD}.}
\label{figrpi2}
\end{minipage}
\end{center}
\end{figure}

The analysis for the pion and kaon form-factors followed
the method introduced by Ref.~\cite{DGH} as updated in
Refs.~\cite{Moussallam2,ABM}. The actual discussion of the results can be found
in Ref.~\cite{BD}. I will restrict myself to a few small comments here.
One uses the ChPT input for the form-factors at zero and then calculates
using the dispersive methods of Muskhelishvili-Omn\`es
its momentum dependence. 
In order to have a consistent set of other LECs when $L_4^r$ and $L_6^r$
were varied, Ref.~\cite{BD} redid the fits with the same assumptions
as fit~10 in Ref.~\cite{ABT4} but with range of inputs for those two $L_i^r$.
The form-factors at zero momentum were then calculated from the order $p^6$
ChPT expression and the momentum dependence calculated from those with the
dispersive method using \cite{Moussallam2,ABM}
result. The comparison of the dispersively estimated momentum dependence
with the ChPT calculated momentum dependence leads to the constraint~\cite{BD}
\be
L_6^r\approx L_4^r-0.00035.
\ee
The comparison for the scalar radius is shown in Fig.~\ref{figrpi2}.
The constraint roughly describes the line where the ChPT and dispersive radius
are in reasonable agreement.

That the actual values of $L_4^r$ and $L_6^r$ has a large impact on the
masses and scalar sector is shown in Tab.~\ref{tabmass}.
\begin{table}
\begin{center}
\begin{minipage}{16.5cm}
\caption{The results for various quantities for the $L_i^r$ of fit 10
and three other representative sets A, B and C. Adapted from Ref.~\cite{BD}.}
\label{tabmass}
\end{minipage}
\begin{tabular}{c|cccc}
\hline
\rule{0cm}{2.6ex}Set            & Fit 10 & A & B & C \\[0.25ex]
\hline
\rule{0cm}{2.6ex}$10^3\cdot L_4^r$                   & 0.0 & 0.4 & 0.5 & 0.5 \\
$10^3\cdot L_6^r$                   & 0.0 & 0.1 & 0.1 & 0.2 \\[0.25ex]
\hline
\rule{0cm}{2.6ex}$F_S^\pi(0)/B_0$ (ChPT, $C_i^r=0$)& 2.54 & 1.99 & 1.75 &2.12\\
$F_{Ss}^\pi(0)/B_0$    (ChPT, $C_i^r=0$)& 0.020 & 0.004 & $-$0.002 & 0.008\\
$F_{Sq}^K(0)/B_0$    (ChPT, $C_i^r=0$)  & 1.94 & 1.36 & 1.12 & 1.51 \\
$F_{Ss}^K(0)/B_0$   (ChPT, $C_i^r=0$)   & 1.77 & 1.35 & 1.17 & 1.45 \\[0.25ex]
\hline
\rule{0cm}{2.6ex}$<r^2>^{\pi, disp}_S$ (fm$^2$)     
     & 0.617 &  0.612 &  0.610 & 0.614\\
$<r^2>^{\pi, ChPT, C_i^r=0}_S$ (fm$^2$) & 0.384 &  0.547 &  0.625 & 0.563\\
\hline
\rule{0cm}{2.6ex}$10^5 \left(C_{12}^r+2C_{13}^r\right)\, [c_S^\pi]$
 & $-$2.6 &  $-$0.56 &   0.55 & $-$0.71 \\
$10^5 \left(C_{12}^r+2C_{13}^r\right) [\gamma_S^\pi]$
 & $-$3.3 &  $-$0.55 &   0.48 & $-$0.75 \\
$10^5 \left(C_{12}^r+2C_{13}^r\right) [\gamma_{Ss}^K]$
 & $-$0.55 &    0.11 &   0.33 &    0.15 \\
$10^5\,C_{13}^r [\gamma_{Ss}^\pi]$
 & $-$0.56&  $-$0.02 &   0.15 &    0.03 \\
$10^5\left(C_{12}^r+4C_{13}^r\right) [\gamma_{Sq}^K]$
 & $-$4.1 &  $-$0.27 &   0.99  &$-$0.08 \\
$10^5\,C_{12}^r [c_S^\pi ,\,\gamma_{Ss}^\pi]$ &
 $ -$1.5&    $-$0.52 &   0.26 & $-$0.78    \\
$10^5\,C_{12}^r [c_S^\pi ,\,\gamma_{Sq}^K]$ &
 $-$1.1 &    $-$0.84 &   0.12 & $-$1.3   \\[0.25ex]
\hline
\rule{0cm}{2.6ex}
$F_0$ (MeV)        & 87.7      & 63.5  & 70.4  & 71.0 \\
$ F_{\pi^+}^{(4)}$ & 0.136     &0.230  &0.253  &0.254 \\
 $ F_{\pi^+}^{(6)}$& $-$0.083  &0.226  &0.059  &0.048 \\
$ F_{K\pi}^{(4)} $ & 0.169     &0.157  &0.153  &0.159 \\
$  F_{K\pi}^{(6)}$ & 0.051     &0.063  &0.067  &0.061 \\[0.25ex]
\hline
$ \mu_\pi^{(2)}  $ & 0.736     &1.005   &1.129   &0.936 \\
$\mu_\pi^{(4)}   $ & 0.006     &$-$0.090&$-$0.138&$-$0.043\\
$ \mu_\pi^{(6)}  $ & 0.258     &0.085   &0.009   &0.107 \\
$\mu_K^{(2)}     $ & 0.687     &0.938   &1.055   &0.874 \\
$\mu_K^{(4)}     $ & 0.007     &$-$0.100&$-$0.149&$-$0.057\\
$\mu_K^{(6)}     $ & 0.306     &0.162   &0.094   &0.183 \\
$\mu_\eta^{(2)}  $ & 0.734     &1.001   &1.124   &0.933 \\
$\mu_\eta^{(4)}  $ & $-$0.052  &$-$0.151&$-$0.197&$-$0.104 \\
$\mu_\eta^{(6)}  $ & 0.318     &0.150   &0.073   &0.171 \\[0.25ex]
\hline
\end{tabular}
\end{center}
\end{table}
The masses and decay
constants for the sets of input parameters given
can also be found in Table~\ref{tabmass}. The various orders in
the expansion quoted there are defined as:
\ba
F_{\pi^+}/F_0  &=& 1 +  F_{\pi^+}^{(4)} +  F_{\pi^+}^{(6)}\,,
\nonumber\\
F_{K^+}/F_{\pi^+}  &=& 1 +  F_{K\pi}^{(4)} +  F_{K\pi}^{(6)}\,,
\nonumber\\
m_{\pi^\pm}^2/(m_{\pi^\pm}^2)_{\mbox{\tiny QCD}}
& = & \mu_\pi^{(2)}+\mu_\pi^{(4)}+ \mu_\pi^{(6)}
\nonumber\\
m_{K^\pm}^2/(m_{K^\pm}^2)_{\mbox{\tiny QCD}}
& = &\mu_K^{(2)}+\mu_K^{(4)}+ \mu_K^{(6)} \,,
\nonumber\\
m_{\eta}^2/(m_{\eta}^2)_{\mbox{\tiny phys}}
& = & \mu_\eta^{(2)}+\mu_\eta^{(4)}+ \mu_\eta^{(6)}\,.
\ea
Notice that in \cite{ABT4} the corresponding numbers were quoted for the
``Main Fit'' which used the old $K_{e4}$ data.
In Ref.~\cite{BD} several of the order $p^6$ constants were also determined
from the curvature in the form-factors, these are the combinations
of $C_i^r$ given in Tab.~\ref{tabmass}. The symbol in square brackets indicates
the curvature of the form-factors used in the determination.
The curvature is for the
form-factor at zero normalized to 1, for the quantities labeled $c$.
For the quantities labeled $\gamma$, the form-factors at zero
are normalized to $B_0$ or to its
lowest order value. It is obvious from the table
that there are some discrepancies still to be understood since not all
determinations are compatible.
The table shows the very large corrections to some of the scalar form-factors
at zero.

A glance at Table~\ref{tabmass} shows that the pion decay constant
in the chiral limit $F_0$ can be substantially different from the value
of about $87~$MeV for the case of fit 10. The reason is that $L_4^r$
is not that well constrained and thus allows for these different values.
As discussed in the two following subsections, the constraints
from $\pi\pi$ scattering and $\pi K$ scattering could resolve this
partly but the constraints are only marginally consistent.
The consistent area does have values of $F_0$ very close to the one
of fit 10.  
A full conclusion cannot be drawn as a comprehensive analysis of all
order $p^6$ constants is still lacking.

\subsection{\it Pion-pion scattering}
\label{pipinf3}

We have already treated pion-pion scattering in two-flavour ChPT.
In order to complete the check of going from $K_{\ell4}$ form-factors to
pion-pion scattering, the latter process also needs to be known to order
$p^6$ in three-flavour ChPT. This was accomplished in Ref.~\cite{BDT}.
All pion-pion scattering in the isospin limit can be described by the
function $A(s,t,u)$ defined in Eq.~(\ref{defAstu}).

This function was calculated in two-flavour ChPT to order $p^6$ in
Refs.~\cite{BCEGS1,BCEGS2} and is also known to order $p^6$
in three-flavour ChPT.
The expression can be found in Ref.~\cite{BDT}.

The convergence of the expansion is similar to the convergence in two-flavour
ChPT. The various isospin amplitudes can be rewritten in terms of $A(s,t,u)$
via
\ba
T^0(s,t) &=& 3 A(s,t,u) + A(t,u,s) +A(u,s,t)\,,
\nonumber\\
T^1(s,t) &=& A(t,u,s) - A(u,s,t)\,,
\nonumber\\
T^2(s,t) &=& A(t,u,s) +A(u,s,t)\,,
\ea
where the kinematical variables $t,u$ can be expressed in terms of $s$ 
and $\cos\theta$ as
\be
t = -\frac{1}{2}(s-4m_\pi^2)(1-\cos\theta)\,,\quad
u = -\frac{1}{2}(s-4m_\pi^2)(1+\cos\theta)\,.
\ee
The amplitudes are expanded in partial waves using
\be
T^I(s,t) = 
32\pi\sum_{\ell=0}^\infty (2\ell+1) P_\ell(\cos\theta) t^I_\ell(s)\,.
\ee
Near threshold these can be expanded in terms of the threshold parameters
\be
\label{defaij}
t^I_\ell = q^{2\ell}\left(a^I_\ell + b^I_\ell q^2 + {\cal O}(q^4)\right)\,,
\quad
q^2 = \frac{1}{4}\left(s-4 m_\pi^2\right)\,.
\ee
Below the inelastic threshold the partial waves satisfy
\be
\mathrm{Im} t^I_\ell(s) = \sigma(s) \left|t^I_\ell(s)\right|^2,\quad\quad
 \sigma(s) = \sqrt{1-\frac{4 m_\pi^2}{s}}\,.
\ee
In this regime, all partial waves can be written in terms of the phase-shifts.
\be
t^I_\ell(s) = \frac{1}{\sqrt{1-(4m_\pi^2/s)}}\,\frac{1}{2i}\left\{
e^{2i\delta^I_\ell(s)}-1\right\}\,.
\ee
Some results of the phase-shifts were shown already for two-flavour QCD.

The three-flavour calculation results for the threshold parameters
are shown in Tab.~\ref{tabpipi} for the sets of input parameters
fit 10, A, B and C described in earlier subsections and in Tab.~\ref{tabmass}.

\begin{table}
\begin{center}
\begin{minipage}{16.5cm}
\caption{\label{tabpipi} The values of the threshold
parameters defined in (\ref{defaij}) for the values of the input
parameters of fits 10~\cite{ABT4} and A,B,C~\cite{BD}. The lowest order
values and the contributions from the order $p^6$ LECs, $C_i^r$, are
included.
The threshold parameters are given in the corresponding power
of $m_{\pi^+}^2$. Note that $a^I_\ell$ and $b^I_\ell$ are always
given with the same
power of ten. For fit 10, the three orders are quoted separately so the
convergence can be judged.
Table adapted from \cite{BDT}.}
\end{minipage}
\begin{tabular}{c|c|ccc|c|c|c}
\hline
      &   &\multicolumn{3}{c|}{ fit 10} & fit A & fit B & fit C \\
\hline
         & $p^2$ & $p^4$ & $p^6$ & total & total & total & total \\
\hline  
$a^0_0$  & 0.159 & 0.044 & 0.016 & 0.219 & 0.220 & 0.220 & 0.221\\
$b^0_0$  & 0.182 & 0.073 & 0.025 & 0.279 & 0.282 &0.282 &0.282 \\
\hline
$10\,a^2_0$ & $-$0.454 & 0.030 & 0.013 & $-$0.410 &$-$0.427&$-$0.433&$-$0.428\\
$10\,b^2_0$ & $-$0.908 & 0.151 & 0.025 & $-$0.731  &$-$0.755&$-$0.761&$-$0.760\\\hline
$10\,a^1_1$ & 0.303 & 0.052  & 0.031 & 0.385 &0.388&0.389&0.389\\
$10\,b^1_1$ & $-$  & 0.029 & 0.038 & 0.067 &0.064&0.063&0.063\\
\hline
$10^2\,a^0_2$ & $-$ & 0.153 & 0.080 &0.233 &0.223&0.220&0.221\\
$10^2\,b^0_2$ & $-$ &$-$0.040 & 0.007 & $-$0.033  &$-$0.035&$-$0.036&$-$0.036\\
\hline
$10^3\,a^2_2$ & $-$ & 0.327  &$-$0.106  &0.221  &0.219&0.218&0.221\\
$10^3\,b^2_2$ & $-$ &$-$0.234  & $-$0.151  &$-$0.385  &$-$0.386&$-$0.385&$-$0.387\\
\hline
$10^4\,a^1_3$ & $-$ & 0.20 & 0.44  &0.64  &0.62&0.62&0.62\\
$10^4\,b^1_3$ & $-$ & $-$0.15 & $-$0.20  & $-$0.35 &$-$0.34&$-$0.34&$-$0.34\\
\hline
\end{tabular}
\end{center}
\end{table}

\begin{figure}
\begin{minipage}{0.48\textwidth}
\includegraphics[angle=270,width=0.99\textwidth]{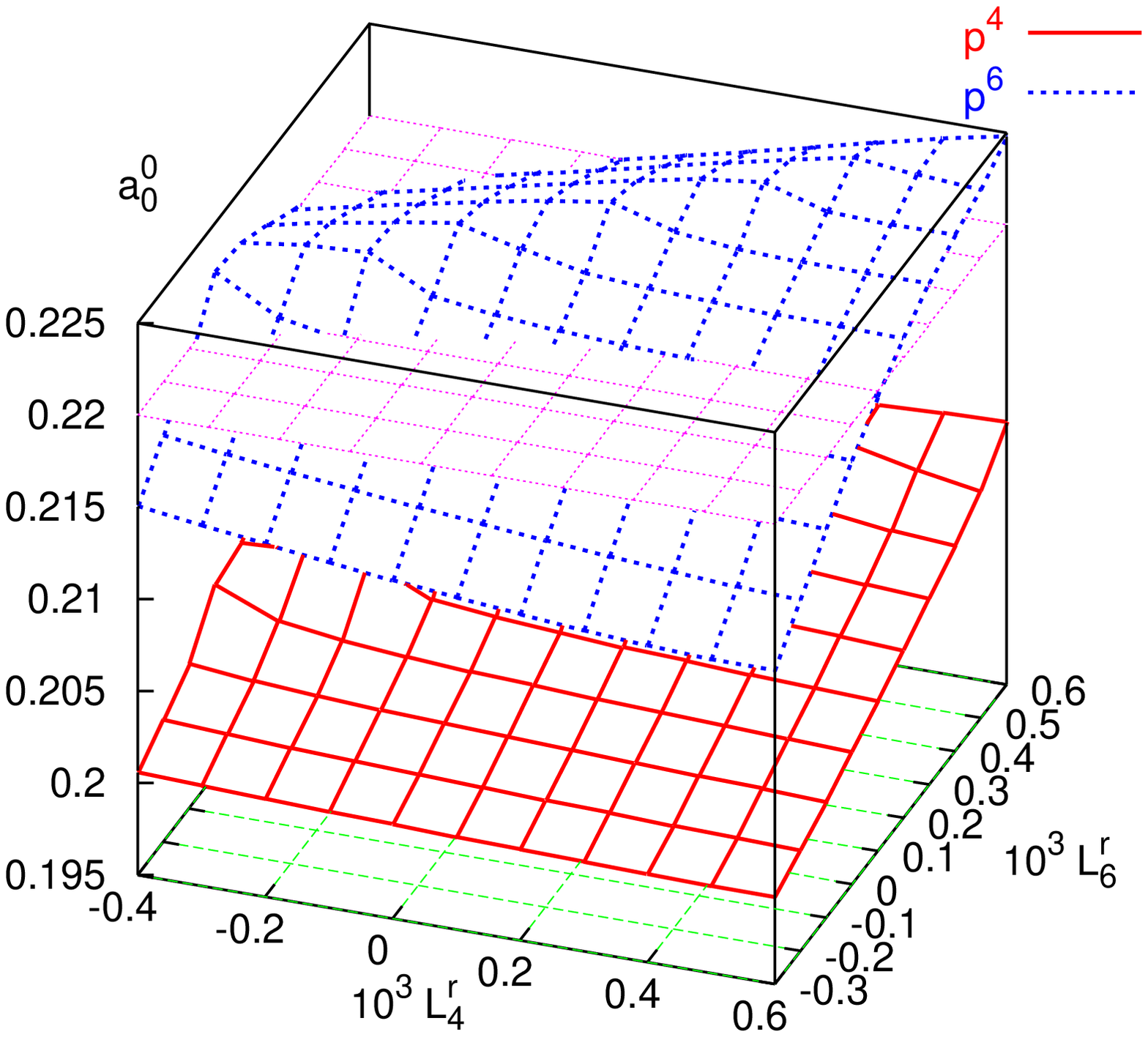}
\centerline{(a)}
\end{minipage}
\hfill
\begin{minipage}{0.48\textwidth}
\includegraphics[angle=270,width=0.99\textwidth]{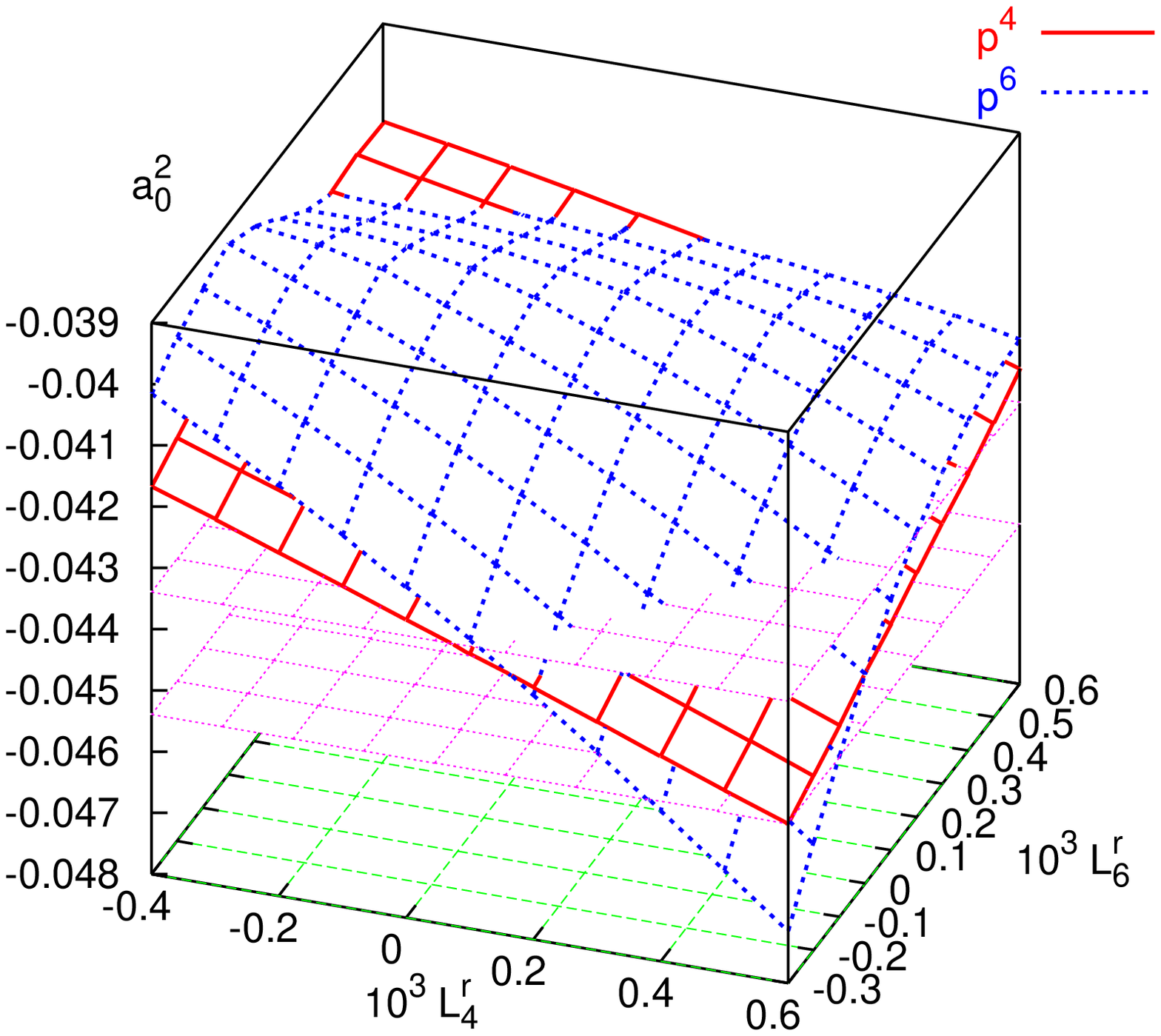}
\centerline{(b)}
\end{minipage}
\begin{center}
\begin{minipage}{16.5cm}
\caption{\label{figa00} The scattering lengths $a^0_0$ and $a^2_0$ as
a function of the input values $L_4^r$ and $L_6^r$ with the other $L_i^r$
simultaneously refitted to the $K_{e4}$ form-factors.
(a) $a^0_0$ calculated to order $p^4$ and $p^6$. The central value
of \cite{CGL3} of 0.220 is also shown. (b) $a^2_0$ at order $p^4$ and 
$p^6$. The two horizontal planes indicate the allowed region obtained
in \cite{CGL3}.
Figure from Ref.~\cite{BDT}.}
\end{minipage}
\end{center}
\end{figure}

It can already be seen from Tab.~\ref{tabpipi} that $a^0_0$ is very well
predicted for sets of input parameters but $a^2_0$ is somewhat smaller than
the result of Ref.~\cite{CGL3}. Ref.~\cite{BDT} performed a full analysis,
using the fits of \cite{BD} with the whole range of $L_4^r,L_6^r$ as input
to determine the best fits.
It was found that $a_0^0$ always fitted nicely but that the value of $a_0^2$
restricted the values to a corner of the $L_4^r,L_6^r$ only.
In Fig.~\ref{figa00} the dispersive results from Ref.~\cite{CGL3} are shown
together with the three-flavour order $p^6$ ChPT results.

Similar results can be obtained for the other threshold parameters, but
especially the higher threshold parameters have less convergence and have not
been used as constraints.
In Ref.~\cite{CGL3,BDT} one also defined the subtreshold parameters $C_1$
and $C_2$. They can similarly be used to obtain constraints on the ChPT input
parameters. The total constraints from pion-pion scattering on the
input parameters are shown in Fig.~\ref{figL4L6constraints}.

\subsection{\it Pion-kaon scattering} 
\label{piK} 

An analysis of pion-kaon scattering similar to the combination of
two-flavour ChPT at order $p^6$ and the Roy equations as done in
Ref.~\cite{CGL2,CGL3} can in principle also be done for pion-kaon scattering.
At the moment, this is not quite completed. The Roy equations analysis,
extended to the pion-kaon scattering case using the so-called Roy-Steiner
equations has been performed by the authors of Ref.~\cite{BDM}.
Some earlier sum rule work and comparisons with ChPT can be found in
Ref.~\cite{AB}.

On the ChPT side, the lowest order calculation was performed using current
algebra methods in Ref.~\cite{Griffith}. The order $p^4$ calculation
can be found in Refs.~\cite{BKM,BKM2} and the full order $p^6$ calculation was
performed by the authors of Ref.~\cite{BDT2}.

An alternative method using ChPT, is to treat the kaon as very heavy and
only the pion as a Goldstone boson. This so-called heavy-kaon approach
to pion-kaon scattering can be found in Ref.~\cite{Roessl}.

The pion-kaon scattering system can be decomposed into different isospin
amplitudes and these in turn can be expanded in partial waves.
The partial wave in turn are expanded around threshold and a full set of
threshold parameters can be defined analogous to the pion-pion scattering case.
There also exist subtreshold expansions in the pion-kaon case as well.
These quantities have been calculated using the Roy-Steiner equations
in Ref.~\cite{BDM} and to order $p^6$ in three-flavour ChPT in
Ref.~\cite{BDT2}.

\begin{figure}
\begin{center}
\begin{minipage}{7cm}
\includegraphics[width=7cm,angle=270]{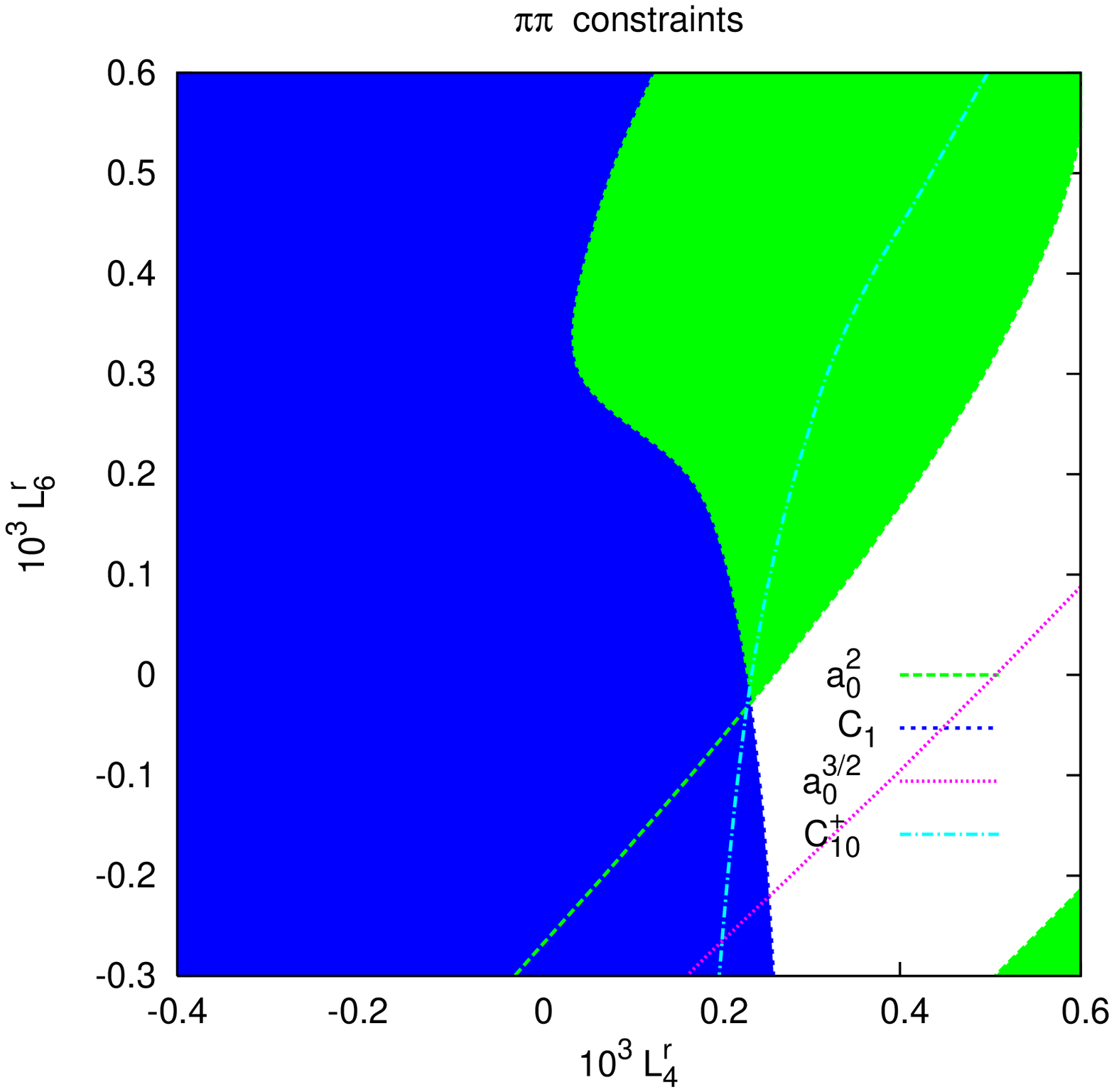}
\centerline{(a)}
\end{minipage}
\begin{minipage}{7cm}
\includegraphics[width=7cm,angle=270]{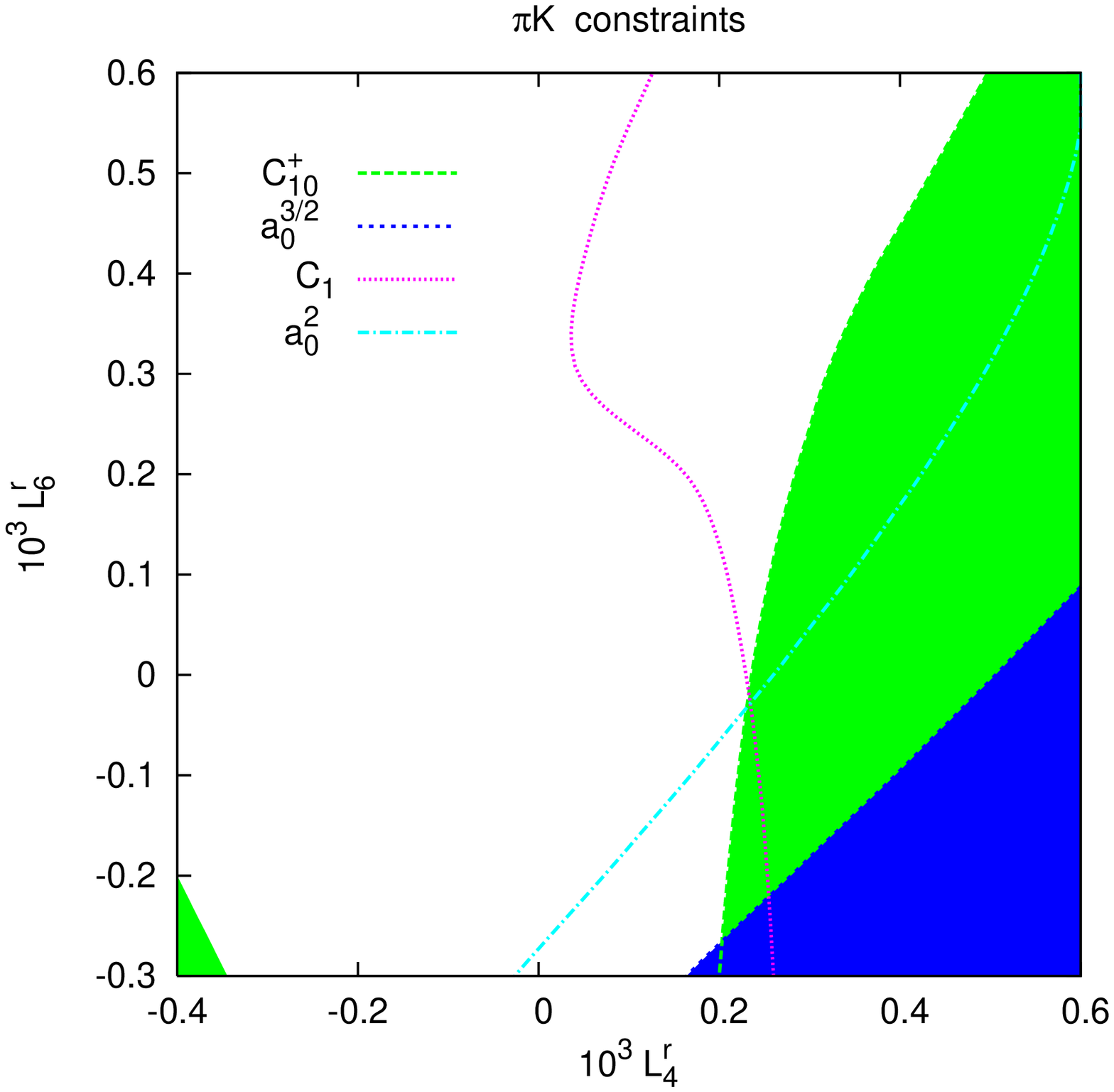}
\centerline{(b)}
\end{minipage}
\end{center}
\begin{center}
\begin{minipage}{16.5cm}
\caption{\label{figL4L6constraints}
(a) The constraints on $L_4^r $ and $L_6^r$ following from the
dispersive results in $\pi\pi$ scattering \cite{CGL2} as derived
in \cite{BDT}.
 (b) The same for the results from $\pi K$ scattering \cite{BDM}
as described in \cite{BDT2}.
The curves from the boundaries are plotted on both plots to make
comparisons easier. The shaded regions are excluded.
Figure from Ref.~\cite{BDT2}.}
\end{minipage}
\end{center}
\end{figure}

The entire comparison of inputs to ChPT, the resulting threshold and
subtreshold parameters and their comparison with the dispersive results
can be done as in the previous subsection. The results for the threshold
parameters and thee inputs of fit 10 are shown in Tab.~\ref{tabpiK}.
\begin{table}
\begin{center}
\begin{minipage}{16.5cm}
\caption{\label{tabpiK} The results
for the scattering lengths and ranges 
and the amplitude at the Cheng-Dashen point as well as the dispersive result.
The scattering lengths and ranges are given in units of $m_{\pi^+}$.
Table adapted from Ref.~\cite{BDT2}.}
\end{minipage}
\begin{tabular}{rrr}
\hline
  & Fit 10 & \cite{BDM}\\
\hline
$     a^{1/2}_0$ & 0.220 & $0.224\pm0.022$   \\
$10   a^{1/2}_1$ & 0.18  & $0.19\pm0.01$ \\  
$10   a^{3/2}_0$ & $-0.47$ & $-0.448\pm0.077$ \\
$10^2 a^{3/2}_1$ & 0.31 & $0.065\pm0.044$ \\  
$10   b^{1/2}_0$ & 1.3 & $0.85\pm0.04$ \\
$10   b^{3/2}_0$ & $-0.27$ & $-0.37\pm0.03$ \\
$T^+_{CD}$       & 2.11 & $3.90\pm1.50$ \\
\hline
\end{tabular}
\end{center}
\end{table}
As can be seen, good agreement exists for many of the threshold parameters.
A full analysis is described in Ref.~\cite{BDT2} and the constraints on the
$L_4^r,L_6^r$ plane following are shown in Fig.~\ref{figL4L6constraints}.
A region that is compatible with all results is
\be
L_4^r\approx 0.2\cdot10^{-3}\quad\mbox{and}\quad 
L_6^r\approx 0.0\cdot10^{-3}\,.
\ee
A more extensive discussion can be found in Ref.~\cite{BDT2}.

One point deserving mention, the order $p^6$ calculation naturally obeys
the chiral relations derived in the heavy kaon approach~\cite{Roessl}.
In particular, how to obtain the isospin odd low-energy theorem from
the full order $p^6$ expressions of \cite{BDT} is discussed in
Ref.~\cite{Schweizer}. There has also been some recent progress
in calculating the threshold parameters fully analytically~\cite{Schweizer2}.

\subsection{\it $\pi,K\to\ell\nu\gamma$}
\label{Klng}

\begin{figure}
\begin{center}
\includegraphics[width=16cm]{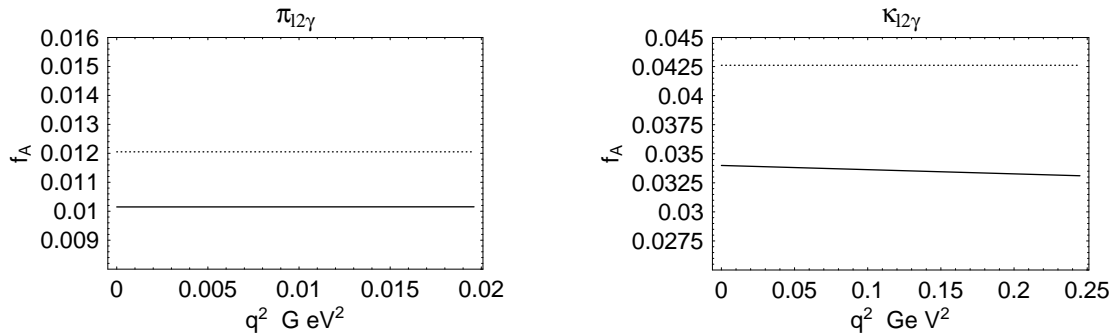}
\end{center}
\begin{center}
\begin{minipage}{16.5cm}
\caption{The axial vector form-factor in $\pi\to\ell\nu\gamma$ (left) and
$K\to\ell\nu\gamma$ at order $p^6$ in ChPT.
The dashed and solid curves represent the full results at
order $p^4$ and order $p^6$. Note that the latter includes data fitting.
Figure from Ref.~\cite{Geng}.
} \label{figpiKlng}
\end{minipage}
\end{center}
\end{figure}

To determine the parameter $L_{10}^r$ at the same order as precision
as was described in the previous subsections the processes
$\pi,K\to\ell\nu\gamma$ need to be calculated to order $p^6$.
This calculation has been done by the authors of Ref.~\cite{Geng}.
They found a reasonably converging result.
I refer to their paper for a longer discussion, but from Fig.~\ref{figpiKlng}
it can be seen that they found a nicely convergent series.

\subsection{\it Estimates of order $p^6$ constants}
\label{Ciestimates}

One of the major problems is the sheer number of order $p^6$ constants
that contributes. In some cases, they could be determined from experiment
directly. Good cases here are those responsible for kinematical quantities
at order $p^6$. These are the constants responsible for the
curvature in the various form-factors, the kinematic factors $b_5$ and $b_6$
in pion-pion scattering and similar quantities in other decays.

Especially the curvature in the vector form-factors leads to 
a well-determined combination. However, there are many more order
$p^6$ LECs that cannot be so easily determined directly from data.
Since at order $p^6$ typically many $L_i^r$ contribute to particular
processes, the determination of them becomes entangled between different
processes. As a result, an estimate of an order $p^6$ LEC used in one process
where an $L_i^r$ is determined sneaks in the determination of
the other $L_i^r$ and possibly order $p^6$ constants in the other processes.

The solution, a full comprehensive analysis of {\em all} processes
at the same time is a major undertaking which has not been done.
It is also not clear whether sufficient data exist to constrain
all involved $L_i^r$ and order $p^6$ LECs.

The solution which has mainly been used up to the present is to
{\em estimate} the values of the order $p^6$ LECs by simple resonance
saturation. This approach was started already in Ref.~\cite{GL1}.
There it was shown that the experimental values of the order $p^4$
LECs cannot be the low-energy limit of the linear sigma model.
They also found that exchange of vector mesons lead to reasonable
predictions for the largest LECs at order $p^4$.
This work was systematized in Refs.~\cite{Ecker1} where it was shown
that the values of all order $p^4$ constants can be understood on the basis
of one-resonance exchange. Ref.~\cite{Ecker2} showed how these results,
when taken together with short-distance QCD constraints, are independent
of the actual chiral representation chosen for the resonance fields.
They also found that the main contribution came from the exchange of
vector mesons.

The approach used in most order $p^6$ papers
consists in taking the simplest Lagrangian containing
vector, axial-vector, scalar and sometimes pseudoscalar and tensor
resonances, fitting its parameters as much as possible from
data, and then taking the masses large to determine the LECs of ChPT.
This is what has been done in basically all of the papers reviewed so
far, with the exception of the two-flavour pion-pion scattering work.

The main underlying theory argument is the large number of colours, $N_c$,
limit~\cite{nc}, where QCD reduces to a theory of stable resonances.
The additional assumption made in the estimates
is that the lowest resonance in each channel
will be important. 

Apart from the simplest version mainly used in the papers on
ChPT to order $p^6$ more comprehensive approaches exist.
There is the minimal hadronic approximation approach~\cite{MHA},
the ladder resummation model approach~\cite{BGLP} and the
resonance chiral model approach~\cite{Reso1,Reso2,Reso3,Reso4}.
All of these approaches are different ways of taking a certain number
of resonances into account in each channel. They differ somewhat in the
way the various couplings are treated.
One problem is that there is an ambiguity in how one deals with short
distance QCD constraints. It is known that there are incompatibilities
for a finite number of resonances~\cite{BGLP}.

Other approaches are the older quark model approach, see
\cite{ETR} and references therein, as well those based on the
Nambu-Jona-Lasinio model~\cite{NJL}, see 
\cite{ENJL,physrep,ximo,BFP} and references therein.
All these approaches give similar results in vector dominated channels
but in those cases where quark mass effects play a major role, no
full comparison with experimentally determined higher order coefficients has
been done. For predictions regarding order $p^4$ LECs, most of these
get similar results.

One problem that affects all the large $N_c$ based methods is
that the ChPT
LECs are subtraction scale dependent while the estimates are not. 
This can only be solved by going beyond
the leading terms in $1/N_c$ also in the resonance models.
Some attempts where also more references can be found are in
Refs.~\cite{loopsreso1,loopsreso2,loopsreso3}.

This area of estimating higher order $p^6$ LECs
is where most improvement is needed in the future of higher order
ChPT calculations.

\section{Other}
\label{other}

In this section I dicuss the order $p^6$ calculations which have been
done but which did not fit in the two main parts discussed in the previous
two sections. 

\subsection{Finite temperature and volume calculations}
\label{finiteT}

The field of finite temperature calculations
was started by Gasser and Leutwyler in Ref~\cite{GLfiniteT}.
It means making the time part periodic and rotating it into the
imaginary time direction. Two-loop calculations I am aware of in this
area are the density dependence of the pion propagator by
Schenk~\cite{Schenk}, as well as the pion mass and decay constant
by Toublan~\cite{Toublan}.

At finite volume, there exists also work.
Finite volume in ChPT was introduced by Gasser and Leutwyler
in Ref.~\cite{GLfiniteV}. A lot of work has been
done to higher orders in this area using  L\"uscher's method to determine
the leading finite volume correction~\cite{Luscher}.
Proper two-loop calculations at finite volume in ChPT have
only become available recently.
The mass and the decay constant in two-flavour ChPT can be found
in Ref.~\cite{CH} and the quark-antiquark condensate in three-flavour
ChPT in Ref.~\cite{KG}.

There is another regime at finite volume, the small volume or
$\epsilon$ regime. Here the lowest mode of zero momentum dominates the
higher order corrections. It therefore needs to taken into account
exactly and forms in this way a all-order calculation.
This is a large area of active research and I will not discuss it here.

\subsection{Partially quenched calculations}

A large effort goes into numerically evaluating the functional integral
of QCD. This is area is known as lattice QCD. One of the things that
appears naturally in the way those calculations are done is the different
treatment of quark lines connected to external legs, so-called valence quark
lines, and the closed quark loops, called sea-quark loops.

The effect of the latter is numerically very difficult to compute and
has therefore often been approximated. Neglecting the sea-quarks completely
is known as the quenched approximation and treating them with different
masses from the valence quarks is known as the partially quenched
approximation.

In neither of these cases is the resulting theory a bona-fide field theory
and the arguments that derive ChPT from QCD, given clearly
in Ref.~\cite{Leutwyler1} do not all hold. But both
quenched and partially-quenched are still well defined statistical models
and many of the arguments when thought of in terms of Feynman diagrams
still seem to hold. We can thus hope that a ChPT extended to the case of
quenched and partially quenched QCD still makes sense. Many of the qualitative
prediction predicted by quenched ChPT (QChPT) and partially quenched
ChPT (PQChPT) seem to be present in actual lattice QCD calculations.
Especially the quenched chiral logarithms seem to exist.

The extension of ChPT to the quenched case was started by Morel~\cite{Morel}
and properly extended by Sharpe~\cite{Sharpequenched1,Sharpequenched2}
and Bernard and Golterman~\cite{BGquenched}. 
Infinities and Lagrangians at one-loop
have been discussed in Ref.~\cite{CP}. The extension to the
partially quenched case was again done by Bernard, Golterman and Sharpe
\cite{BGPQ,SharpePQ}. Articles which provide a good overview
of the field at order $p^4$ are Refs.~\cite{SS1,SS2}.

Work on extending the work to two-loops started some years ago but
there were two difficulties to be solved. One is to find the checks
coming from the knowledge of the divergence structure and the Lagrangian
at order $p^6$ and the other is the sheer complexity of the calculations
in terms of the size of the expressions.
The solution to the former turned out to be easy.
Already in Ref.~\cite{CP} it was noted that the quenched Lagrangian and
infinity structure at order $p^4$ had a very strong resemblance
to the one of $n_F$-flavour ChPT at the same order.
More arguments for this were given in Ref.~\cite{replica}.
When Bijnens, Danielsson and L\"ahde started looking at two-loop
PQChPT, they realized that a very simple recipe allowed to obtain the
full order $p^6$ both for the Lagrangian and the divergence structure.
To understand this, one goes back to the suggestion of Morel~\cite{Morel}
how to deal with the quenched approximation systematically.
This is the approach worked out properly by Ref.~\cite{BGquenched}.
In order to systematically remove the closed quark loops, one adds for each
quark a bosonic quark, spin 1/2 but bosonic. Due to the different
statistics, in this way all closed loops are exactly canceled. Partially
quenched can be treated by leaving some of the quarks without a bosonic
partner. Closed loops from these quarks then are not canceled.
The formalism used in ChPT can be fully extended to this case by using
supertraces instead of traces and adding extra bosonic and fermionic
``Goldstone bosons.'' It was then also shown that in the partially quenched
case the equivalent of the singlet eta would be heavy and could be
systematically removed from the theory~\cite{SS2}.
With that in hand, one realizes that all manipulations done in $n_F$-flavour
ChPT to obtain Lagrangians and divergence structures go through with
the changes mentioned above and $n_F$ changed to
the number of active sea-quarks.
The PQChPT Lagrangian at order $p^6$
and the divergence structure can thus be derived immediately
from the work of Refs.~\cite{BCE1,BCE2}. One also sees immediately the origin
of the extra order $p^4$ term of \cite{SharpeVandewater}.
The number of terms in the Lagrangian this leads to is given in
Tab.~\ref{tabLEC}.

\begin{figure}
\begin{center}
\includegraphics{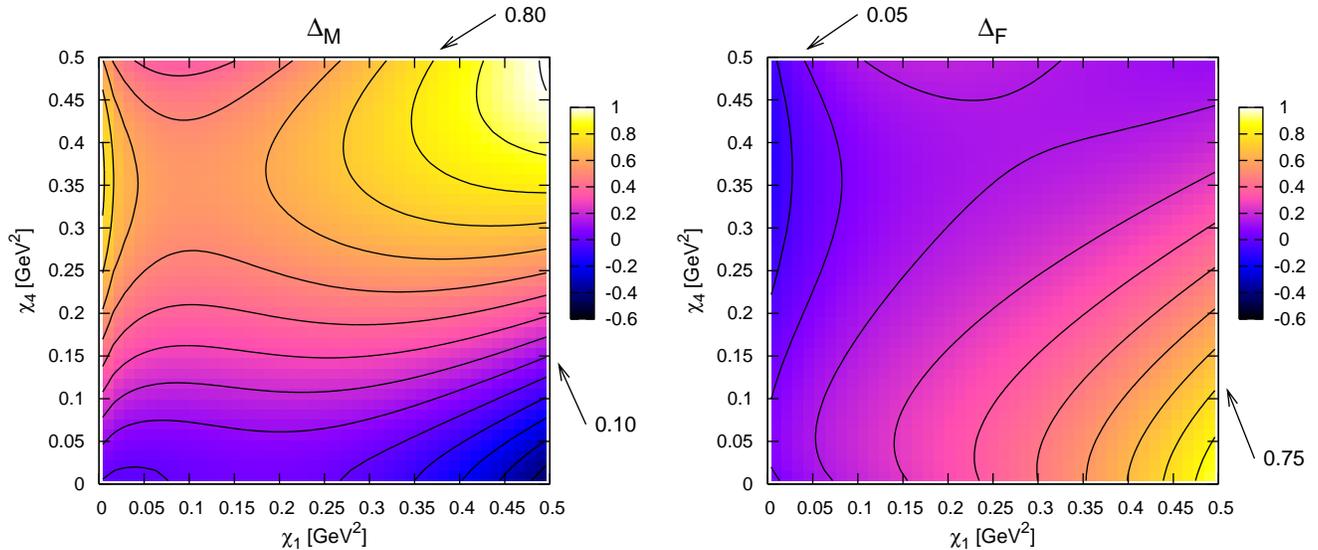}
\end{center}
\begin{center}
\begin{minipage}{16.5cm}
\caption{The relative corrections to the charged pseudoscalar meson mass, 
$\Delta_M$, and decay constant, $\Delta_F$, to NNLO for
the case of three sea-quarks. The valence quarks have quark-mass
$m_1$ and the sea-quarks mass $m_4$. The plot is as a function of
the quark-masses in the combination $\chi_1=2 B_0 m_1$
and $\chi_4 = 2 B_0 m_4$.
The quantity plotted represents 
the sum of the NLO and NNLO corrections normalized to lowest order.
The difference between two 
successive contour lines in the plots is~$0.10$. The values chosen for 
the LECs correspond to "fit~10" of Ref.~\cite{ABT4}.
Figure from Ref.~\cite{BDL2}.}
\label{figPQChPT}
\end{minipage}
\end{center}
\end{figure}

The complexity of the calculations was solved by combining the power of
FORM~\cite{FORM} with a large amount of hand-optimized simplification of the
various terms. At present no direct comparison with data are made,
so I only show a representative plot. The relative corrections to lowest order
for the case with three sea-quark flavours and valence-quarks of equal
quark-mass, $m_4$, and sea-quarks of equal quark-mass, $m_2$,
are shown in Fig.~\ref{figPQChPT}.
Some visible features are the sizable corrections
and the fact that the corrections
do not vanish when $m_1\to0$ but $m_4$ kept fixed. The latter is the
manifestation of
the quenched chiral logarithm.
More details can be found in the actual papers~\cite{BDL1,BL1,BL2,BDL2}.
The actual calculations that have been performed are
the masses for three sea-quarks~\cite{BDL1,BDL2},
the decay constants for three sea-quarks~\cite{BL1,BDL2}
and the same quantities for two sea-quark flavours as well~\cite{BL2}.
The expressions can be found on the website~\cite{formulas}.

This is clearly an area where further progress is desirable.

\section{Conclusions}

In this review I have briefly discussed the foundations of ChPT
and some of its technical aspects.
The longest parts has been concerned with the actual calculations which
have been performed at order $p^6$ and their confrontation
with experimental results.

Here the two-flavour part, Sect.~\ref{twoflavour},
is in rather good shape. Most calculations
have been done and in many cases the contributions from unknown
order $p^6$ constants are expected to be rather small.
I particularly want to emphasize again here the work on
pion-pion scattering where major progress in the theoretical
description was made.

In the three-flavour sector, Sect.~\ref{threeflavour},
many calculations have been done
and fitted to experimental results. One main highlight here is
the predictions made in the context of $K_{\ell3}$ and its relation to the
determination of $V_{us}$ but many other low-energy processes involving kaons
and etas are also known. I have only briefly discussed the main remaining block
towards a full comprehensive application of ChPT at NNLO, estimating
in a consistent fashion all needed order $p^6$ LECs. Work in this area
is going on, so I hope there will be significant progress in the 
not-too-distant future.

And last, I have mentioned the finite temperature, volume and partially
quenched work to two-loop order in ChPT. The interface between lattice
QCD and ChPT is a growing area of research and should benefit both
communities. At present only a few full two-loop order calculations in this
sector are available as discussed in Sect.~\ref{other}.

\section*{Acknowledgements}

The work reviewed here has involved many
collaborators and I wish to thank all of them for the enjoyment of working
together. I have also had many discussions regarding various aspects of ChPT
with many people. I would like to mention
G.~Amor\'os, W.A.~Bardeen, C.~Bernard, V.~Cirigliano, F.~Cornet,
G.~Colangelo, N.~Danielsson, P.~Dhonte, 
G.~Ecker, E.~Gamiz, J.~Gasser, B.~Golterman,
H.~Leutwyler, T.~L\"ahde, 
U.-G.~Mei~ss{}ner, B.~Moussallam
A.~Pich, P.~Post, J.~Prades,
E.~de Rafael, M.Sainio, M.~Savage, S.~Sharpe, J.~Stern, P.~Talavera,
but this list is certainly incomplete.
Many of the calculations reported here would have not been possible
without the algebraic manipulation program FORM from J.~Vermaseren~\cite{FORM}.

This work is supported by the Swedish Research Council,
the European Union 
TMR network, Contract No. HPRN - CT - 2002 - 00311 (EURIDICE) and the 
EU - Research Infrastructure Activity RII3 - CT - 2004 - 506078 
(HadronPhysics).

\appendix

\section{A few remarks on alternative notations}

I have tried to stick in this review to one choice of notation.
Many alternative parameterizations have been around especially for
the two-flavour case.

There are two main conventions around for the parameters $\hat F, F_0, F$
which differ by a factor of $\sqrt{2}$. Which is in use can normally be
detected by the forefactor in Eq.~(\ref{L2}). $F^2/4$ corresponds to
an $F_\pi$ of order 93~MeV, $F^2/8$ to an $F_\pi$ of order 130~MeV
while in the older literature sometimes even an $F_\pi$ of order 180~MeV
appears.

The parameters $\hat B,B_0,B$ often appear in the combination
$\hat F^2 \hat B/2$ which is frequently called $r$.

Many people use a left-right version instead of the right-left version
introduced in Ref.~\cite{GL2}. They typically use $\Sigma=U^\dagger$
as the quantity containing the Goldstone bosons.
The quantity $u$ also appears often denoted as $\xi$ and 
exists as well in a version
with left and right interchanged.

For the two-flavour case, the fact that
\be
SU(2)\times SU(2)/Z_2 = SO(4)
\ee
and
\be
SU(2)/Z_2 = SO(3)\,,
\ee
allows a parameterization inspired by the breaking of $SO(4)$ to
$SO(3)$. These are sometimes called sigma-model parameterizations.
In Ref.~\cite{GL1} a four-vector $U=(U^0,U^1,U^2,U^3)$ was used
with the restriction $U^T U =1$. Here the pion fields appear
via $U^i = \pi^i/F$ for $i=1,2,3$. All needed external fields can also
be brought into this representation. The explicit connection
between the exponential and the sigma-model representation is
\ba
U^i = -\frac{i}{2}\langle \tau^i U\rangle
\ea
where the $\tau^i$ are the three Pauli-matrices.
Weinberg often uses a three-vector
$\vec\pi$, and a covariant derivative which depends nonlinearly
on the $\vec\pi$ field. This is the form he used in his
original paper~\cite{Weinberg0}.

All of these parameterizations are equivalent, as it has
been proven that all can be brought into one standard form in
Ref.~\cite{CWZ}. However, when performing calculations, individual Feynman
diagrams can be quite different in the different formulations.
As an example, the pion wave function renormalization vanishes
at NLO in the parameterization of Ref.~\cite{GL1} but it doesn't
in the exponential parameterization used in this paper.

\end{document}